\documentclass[a4paper,12pt]{report}
\usepackage{indentfirst}
\usepackage{verbatim}
\usepackage{multicol}
\usepackage{latexsym}
\usepackage{amsmath,amssymb,amsthm}
\usepackage{mathrsfs}
\usepackage{bbm}
\usepackage{array}
\usepackage[dvips]{graphicx}

\def\bra#1{\left\langle #1\right|}
\def\ket#1{\left|#1\right\rangle}
\def\hs#1#2{\left\langle #1\right.\left| #2\right\rangle}
\def\hstar#1#2{\left\langle #1\stackrel{\star}{|} #2\right\rangle}
\def\hsstar#1#2{\left\langle #1\stackrel{\star}{\big|} #2\right\rangle}
\def\hsstarmoy#1#2{\left\langle #1\stackrel{\star_{M}}{\big|} #2\right\rangle}
\def\hsstarvor#1#2{\left\langle #1\stackrel{\star_{V}}{\big|} #2\right\rangle}

\begin{document}
%\pagenumbering{roman}
\numberwithin{equation}{section}

{\thispagestyle{empty}

\begin{center}
\begin{large}
{\Large Universit\`{a} degli Studi di Napoli Federico II}

\begin{center}
\includegraphics[scale=0.16]{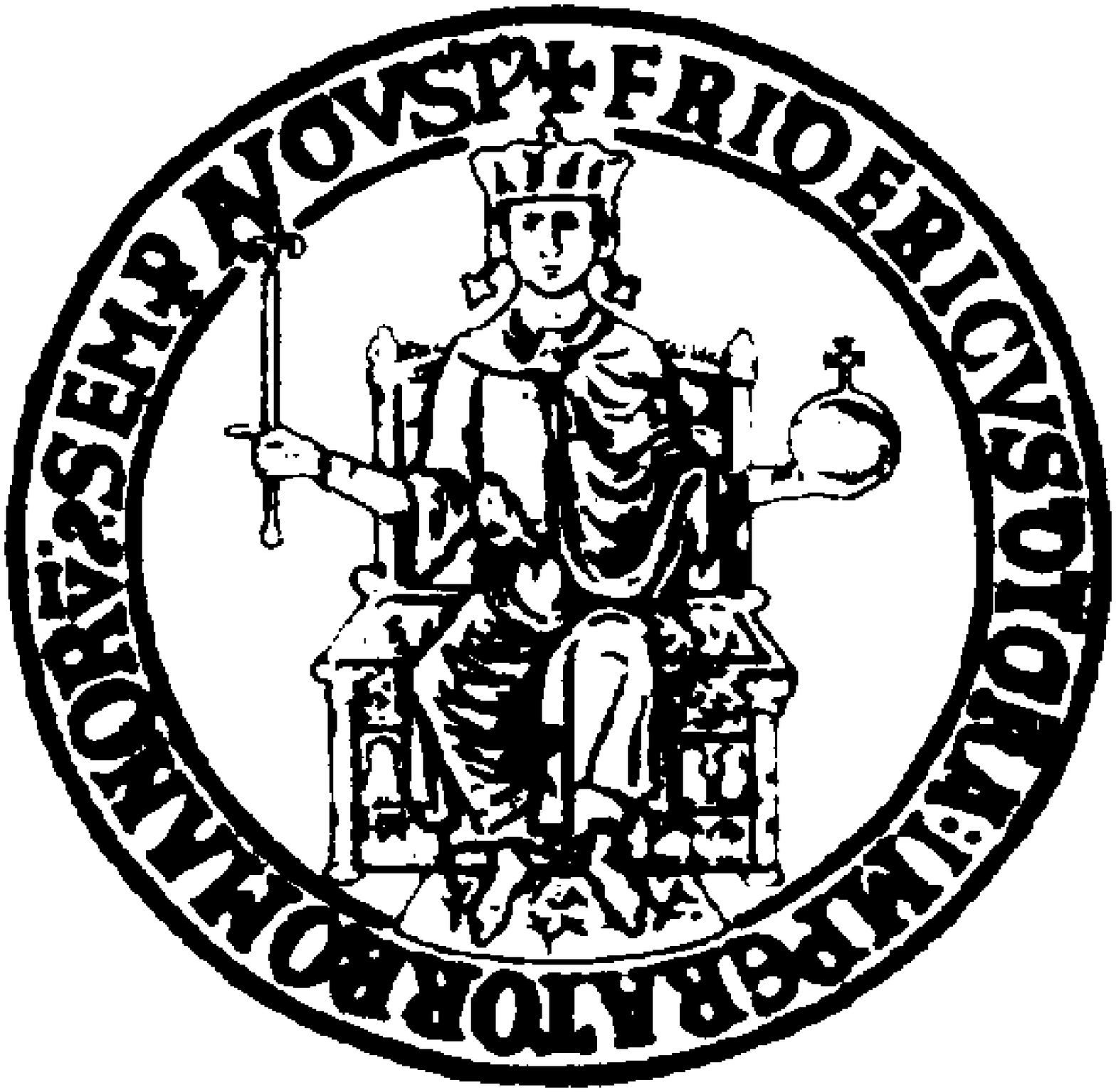}
\end{center}

Dottorato di Ricerca in Fisica Fondamentale ed Applicata XXII ciclo

\vfill

{\Large\textbf{Non-commutative Field Theory, \\Translational Invariant Products and Ultraviolet/Infrared Mixing}}

\vfill

Dissertation submitted for the degree of Philosophiae Doctor

January 13, 2010

\vfill

{\Large\textbf{Salvatore Galluccio}}

\vfill

Under the supervision of

\textbf{Prof. Fedele Lizzi} and \textbf{Dr. Patrizia Vitale}

\vfill\eject
\end{large}
\end{center}

\tableofcontents
%\listoffigures

\chapter*{Introduction}
\addcontentsline{toc}{chapter}{Introduction}

There are several motivations to consider a non-commutative structure of space-time.
One of them is that at short distances that is, at distances of the order of the Planck length:
\begin{equation*}
\ell_{P}=\sqrt{\frac{G\hslash}{c^{3}}}\approx10^{-33}\mathrm{cm}
\end{equation*}
the geometry of space-time has to be described by a dif\mbox{}ferent theory because its points become no longer localizable.
Therefore, one is forced to deal with a ``pointless'' geometry and this leads in a natural way
to the introduction of non-commutative geometry~\cite{Connes,Madore,Landi,Gracia-BondiaVarillyFigueroa}.
A historical motivation~\cite{Heisenberg, Snyder} to consider a non-commutative structure of space-time
was the hope that a modif\mbox{}ication of the short distance properties of space-time, by means of a deformation parameter, could resolve the problem of the inf\mbox{}inities of quantum f\mbox{}ield theory, like the introduction of the fundamental quantity $\hslash$ solved the so-called ultraviolet catastrophe of the black-body radiation.

The simplest kind of non-commutativity is the so-called canonical one
which is characterized by the following commutation relation between the coordinate functions on the space-time:
\begin{equation*}
[x^{\mu},x^{\nu}]=i\theta^{\mu\nu}
\end{equation*}
where $(\theta^{\mu\nu})$ is constant matrix i.e.~it does not depend on the $x$'s.
There are several reasons to consider such a kind of non-commutativity,
going from the localizability of events in space-time~\cite{DoplicherFredenhagenRoberts,Doplicher}
to the string theory~\cite{SeibergWitten}.
Moreover, f\mbox{}ield theories on a space-time equipped with the canonical non-commutativity
have interesting renormalization properties~\cite{GrosseWulkenhaar,GrosseSteinacker,Rivasseau}.
What is generally done to construct a non-commutative f\mbox{}ield theory~\cite{Szabo,DouglasNekrasov}
is to deform the ordinary pointwise commutative product among functions on space-time
with the introduction of a star product which is non-commutative
and reduces to the usual one in certain limit.
The choice of the star product compatible with the canonical non-commutativity is not unique
and throughout this thesis we discuss two dif\mbox{}ferent products, the Moyal product~\cite{Gronewold,Moyal}
and the Wick-Voros one~\cite{Bayen,Voros,BordemannWaldmann1,BordemannWaldmann2,LizziVitaleZampini1,LizziVitaleZampini2}
and investigate their ultraviolet behaviour and compare their ``physical predictions".

In the case of a f\mbox{}ield theory with the Moyal product the hope that the product resolve
the problem of the inf\mbox{}inities of quantum f\mbox{}ield theory is not fulf\mbox{}illed.
Indeed, in this case instead of the elimination (at least partial) of the ultraviolet
divergences, we encounter the phenomenon of ultraviolet/infrared mixing~\cite{MVRS},
one of the novel features of a non-commutative f\mbox{}ield theory.
Therefore, while the ultraviolet properties of the theory
are changed in the sense of a mitigation of the inf\mbox{}inities,
the price paid is the appearance of new kind of inf\mbox{}inity.
We show that the ultraviolet/infrared mixing persists in an unchanged way as well as
for a f\mbox{}ield theory with the Wick-Voros product~\cite{GalluccioLizziVitale1}
which can be seen as a variant of the Moyal one.
This is to be expected because heuristically this is consequence of commutation relation which is,
of course, the same for both products.
Our analysis is centered mainly on the one-loop correction to the propagator which is the source of all mixing
and we discuss only the scalar $\phi^{4}$ theory, but the results are more general.
Moreover, we show that the ultraviolet/infrared mixing for the Moyal product
is a generic feature of any translation invariant associative product~\cite{GalluccioLizziVitale2}.

We have to note that the two f\mbox{}ield theories with the Moyal and Wick-Voros products
are not completely equivalent because their Green's functions are dif\mbox{}ferent
and this leads to a contradiction. In fact, one can heuristically reason as follows.
What really counts is the non-commutative structure of space-time
and the star product is just a way to express such a structure so that
one can always choose the most convenient star product.
As long as one is describing the same f\mbox{}ield theory,
the results should be the same as already noted in~\cite{HammouLagraaSheikh-Jabbari}.
We see that this contradiction is only apparent. Indeed, Green's functions are not observable quantities
and what is observable is the $S$-matrix.

Discussions of the properties of the $S$-matrix naturally go together with the issue of Poincar\'{e} invariance.
The canonical non-commutativity relation is not Poincar\'{e} invariant
and this can cast doubts on its being fundamental.
However, it is possible to preserve the Poincar\'{e} symmetry at a deformed level,
as a non-commutative and non-cocommutative Hopf algebra
because both the Moyal and Wick-voros products come from a Drinfeld twist~\cite{Drinfeld1,Drinfeld2,Oeckl}.
\mbox{In other} words, the theory has a twisted Poincar\'{e} symmetry~\cite{ChaichianKulishNishijimaTureanu,Wess,AschieriBlohmannDimitrijevicMeyerSchuppWess}.

The presence of a twist forces us to reconsider all of the steps
in a f\mbox{}ield theory which has to be built in a coherent twisted way and
we show that there is equivalence between the Moyal and Wick-Voros f\mbox{}ield theories at the level of $S$-matrix
only if a consistent procedure of twisting all products is applied.
There can be some ambiguity in the issue of twisting and in an ideal context
one should let experiments resolve these ambiguities.
However, the non-commutative theory is not yet mature for a confrontation with experiments.
Thus what we do is just to use the f\mbox{}ield theories built
with the Moyal and Wick-Voros products to check each other.
This gives us the indication on the procedure to follow
for non-commutative theories coming from a twist.

\bigskip

The thesis is organized as follows.
In the f\mbox{}irst chapter we will introduce the Moyal and Wick-Voros products
and show that they are both coming from a ``Weyl map''. In particular, we will see that the Moyal product
comes from the usual Weyl map, while the Wick-Voros one comes from a generalization of Weyl map called
a weighted Weyl map.

In the second chapter we will discuss that the Moyal and Wick-Voros product can be set in a more general framework. Indeed, we will show that both products can be derived from a general quantization scheme as well.
In particular, we will see that the Moyal product derives from the so-called Weyl-Wigner quantization scheme.

In the third chapter we will investigate the ultraviolet behaviour of a non-commutative f\mbox{}ield theory
obtained from a commutative one replacing the ordinary product with the Moyal one.
To this end, we will discuss the one-loop corrections to the two- and four-point Green's functions
and see that in the non-planar cases the Moyal product softens the ultraviolet divergences,
but it is responsible for the infrared divergences. Therefore, the Moyal product presents
the phenomenon of ultraviolet/infrared mixing.

In the last three chapters we will present our original work.
In the fourth chapter we will investigate the ultraviolet behaviour
of a non-commutative f\mbox{}ield theory with the Wick-Voros product.
We will show that the ultraviolet properties in this case is the same as in the Moyal one
and in particular they present the same ultraviolet/infrared mixing as heuristically expected.
However, we will f\mbox{}ind that the two theories are not equivalent
since their Green's functions are dif\mbox{}ferent.

In the f\mbox{}ifth chapter we will proceed to the discussion of the relationship between
the translation invariance and the ultraviolet/infrared mixing
and show that the ultraviolet/infrared mixing found for the Moyal and Wick-Voros products
is not specif\mbox{}ic of the two products, but it is a generic feature
of any translation invariant associative product.

In the last chapter we will present a comparison of the non-commutative f\mbox{}ield theories
with the Moyal and Wick-Voros products in the framework of the twisted non-commutativity
and see that the two theories are equivalent at level of $S$-matrix
by means of a consistent procedure of twisting all products,
in agreement with our physical intuition, although the Green's functions are dif\mbox{}ferent.

Finally, there is an appendix in which we will recall the principal notions
of Hopf algebras that we will use throughout the thesis.

\chapter{The Moyal and Wick-Voros products from a Weyl map}
%\pagenumbering{arabic}

\emph{In this chapter we introduce the Moyal and Wick-Voros products and show that the two products
can be cast in the same general framework in that they are both coming from a ``Weyl map''.
More precisely, we show that the Moyal product comes from a map, called the Weyl map,
which associates operators to functions with symmetric ordering,
while the Wick-Voros one comes from a similar map, a weighted Weyl map,
which associates operators to functions with normal ordering.
Furthermore, we exhibit the integral form of the two products.}

\section{The Weyl map}

For the sake of simplicity, we consider the Weyl map on the plane
since its generalization to a several dimension is straightforward.
The Weyl map~\cite{Weyl} is the map which associates to a function on the plane
an operator according to \footnote{For a more modern treatment see~\cite{Zampini}.}
\begin{equation}
\hat{\Omega}_{M}(f)=\frac{1}{2\pi\theta}\int\mathrm{d}^{2}\alpha\,
\tilde{f}(\alpha)e^{i\theta_{ij}\hat{x}^{i}\alpha^{j}}
\end{equation}
where $\theta$ is a real constant parameter of dimensions of a square length,
\begin{equation}
\tilde{f}(\alpha)=\frac{1}{2\pi\theta}\int\mathrm{d}^{2}x\,
f(x)e^{-i\theta_{ij}x^{i}\alpha^{j}}
\end{equation}
is the symplectic Fourier transform of the function $f$,
\begin{equation}
\theta_{ij}=\theta^{-1}\varepsilon_{ij}\quad\mathrm{with}\quad
(\varepsilon_{ij})=
\left(\begin{array}{cc}
0 & -1\\
1 & 0
\end{array}\right)
\end{equation}
and the $\hat{x}$'s are operators which satisfy the commutation relation
\begin{equation}
[\hat{x}^{i},\hat{x}^{j}]=i\theta^{ij}
\end{equation}
where the matrix $(\theta^{ij})$ is the inverse of the matrix $(\theta_{ij})$.
In general, it is always possible to consider the operators $\hat{x}$'s in an abstract way
and def\mbox{}ine them as
\begin{align}
\nonumber
\hat{x}^{1}&=\frac{\hat{a}+\hat{a}^{\dag}}{\sqrt{2}}\\
\hat{x}^{2}&=\frac{\hat{a}-\hat{a}^{\dag}}{i\sqrt{2}}
\end{align}
where $\hat{a}$ and $\hat{a}^{\dag}$ are two operators which satisfy the commutation relation
\begin{equation}\label{CR}
[\hat{a},\hat{a}^{\dag}]=\theta.
\end{equation}
Therefore, the Weyl map can be explicitly written as
\begin{equation}
\hat{\Omega}_{M}(f)=\frac{1}{(2\pi\theta)^{2}}\int\mathrm{d}^{2}x\,\mathrm{d}^{2}\alpha\,
f(x)e^{-i\theta_{ij}x^{i}\alpha^{j}}e^{i\theta_{ij}\hat{x}^{i}\alpha^{j}}.
\end{equation}
It can be equivalently written as
\begin{equation}\label{WM}
\hat{\Omega}_{M}(f)=\frac{1}{(2\pi\theta)^{2}}\int\mathrm{d}^{2}x\,\mathrm{d}^{2}\alpha
f(x)e^{-i\theta_{ij}x^{i}\alpha^{j}}W(\alpha)
\end{equation}
where
\begin{equation}\label{WS}
W(\alpha)=e^{i\theta_{ij}\hat{x}^{i}\alpha^{j}}.
\end{equation}
This last formula has the advantage to involve the operators $W(\alpha)$
which form a Weyl system~\cite{Weyl,EspositoMarmoSudarshan}.
Indeed, by using the Baker-Campbell-Hausdorf\mbox{}f formula\footnote{
If $\hat{A}$ and $\hat{B}$ are two operators such that $[\hat{A},[\hat{A},\hat{B}]]=[\hat{B},[\hat{A},\hat{B}]]=0$, then
\begin{equation*}
e^{\hat{A}}e^{\hat{B}}=e^{\hat{A}+\hat{B}}e^{\frac{1}{2}[\hat{A},\hat{B}]}
\end{equation*}
from which follows that
\begin{equation*}
e^{\hat{A}}e^{\hat{B}}=e^{\hat{B}}e^{\hat{A}}e^{[\hat{A},\hat{B}]}.
\end{equation*}},
it is easy to verify that
\begin{equation}\label{WSR}
W(\alpha)W(\beta)=W(\alpha+\beta)e^{\frac{i}{2}\theta_{ij}\alpha^{i}\beta^{j}}.
\end{equation}
The Weyl map is linear and invertible and its inverse is given by the Wigner map
\begin{equation}
\Omega^{-1}_{M}\left(\hat{\Omega}_{M}(f)\right)=\frac{1}{2\pi\theta}\int\mathrm{d}^{2}\alpha\,
e^{i\theta_{ij}x^{i}\alpha^{j}}\operatorname{Tr}\left(\hat{\Omega}_{M}(f)W^{\dag}(\alpha)\right).
\end{equation}
In fact,
\begin{align}
\nonumber
\Omega^{-1}_{M}\left(\hat{\Omega}_{M}(f)\right)&=
\frac{1}{(2\pi\theta)^{2}}\int\mathrm{d}^{2}\alpha\,\mathrm{d}^{2}\beta\,
\tilde{f}(\alpha)e^{i\theta_{ij}x^{i}\beta^{j}}
\operatorname{Tr}\left(W(\alpha)W^{\dag}(\beta)\right)\\
\nonumber
&=\int\mathrm{d}^{2}\alpha\,\mathrm{d}^{2}\beta\,
\tilde{f}(\alpha)e^{i\theta_{ij}x^{i}\beta^{j}}\delta^{(2)}(\alpha-\beta)\\
&=\int\mathrm{d}^{2}\alpha\,
\tilde{f}(\alpha)e^{i\theta_{ij}x^{i}\alpha^{j}}=f(x)
\end{align}
since in the last line appears the symplectic Fourier antitransform of $\tilde{f}(\alpha)$.
Moreover, it can be show that~\cite{Pool} the Weyl map is an isomorphism
between $L^2(\mathbb{R}^{2})$ i.e~the Hilbert space of square-integrable functions on the plane
and $\mathcal{HS}(L^2(\mathbb{R}))$ i.e.~the Hilbert space of Hilbert-Schmidt operators on $L^2(\mathbb{R})$.

\section{The Moyal product from the Weyl map}

The Moyal product~\cite{Gronewold,Moyal}, often called the Gr\"{o}newold-Moyal product,
is def\mbox{}ined by the relation
\begin{equation}\label{M}
\hat{\Omega}_{M}(f\star_{M}g)=\hat{\Omega}_{M}(f)\hat{\Omega}_{M}(g).
\end{equation}
We can very easily obtain the integral form of the Moyal product.
Indeed, from \eqref{WM} follows that the left-hand side of \eqref{M} can be written as
\begin{equation}\label{E1}
\hat{\Omega}_{M}(f\star_{M}g)=\frac{1}{(2\pi\theta)^{2}}\int\mathrm{d}^{2}x\,\mathrm{d}^{2}\alpha\,
(f\star_{M}g)(x)e^{-i\theta_{ij}x^{i}\alpha^{j}}W(\alpha)
\end{equation}
and the right-hand side of \eqref{M} as
\begin{multline}
\hat{\Omega}_{M}(f)\hat{\Omega}_{M}(g)=
\frac{1}{(2\pi\theta)^{4}}\int\mathrm{d}^{2}y\,\mathrm{d}^{2}\beta\,
\mathrm{d}^{2}z\,\mathrm{d}^{2}\gamma\,f(y)g(z)\\
e^{-i\theta_{ij}y^{i}\beta^{j}}e^{-i\theta_{ij}z^{i}\gamma^{j}}W(\beta)W(\gamma)
\end{multline}
which can be written because of \eqref{WSR} as
\begin{multline}
\hat{\Omega}_{M}(f)\hat{\Omega}_{M}(g)=
\frac{1}{(2\pi\theta)^{4}}\int\mathrm{d}^{2}y\,\mathrm{d}^{2}\beta\,
\mathrm{d}^{2}z\,\mathrm{d}^{2}\gamma\,f(y)g(z)\\
e^{-i\theta_{ij}y^{i}\beta^{j}}e^{-i\theta_{ij}z^{i}\gamma^{j}}
e^{\frac{i}{2}\theta_{ij}\beta^{i}\gamma^{j}}W(\beta+\gamma).
\end{multline}
By means of the linear transformation
\begin{align}
\nonumber
\beta=&\alpha-2x+2y\\
\gamma=&2x-2y
\end{align}
with $y$ constant, it takes the form
\begin{multline}\label{E2}
\hat{\Omega}_{M}(f)\hat{\Omega}_{M}(g)=
\frac{2^{2}}{(2\pi\theta)^{4}}\int\mathrm{d}^{2}x\,\mathrm{d}^{2}\alpha\,
\mathrm{d}^{2}y\,\mathrm{d}^{2}z\,f(y)g(z)\\
e^{-2i\theta_{ij}(x^{i}-y^{i})(x^{j}-z^{j})}e^{-i\theta_{ij}x^{i}\alpha^{j}}W(\alpha).
\end{multline}
By confronting \eqref{E1} with \eqref{E2} we obtain the integral form of the Moyal product
\begin{equation}\label{IFMP}
(f\star_{M}g)(x)=
\frac{1}{(\pi\theta)^{2}}\int\mathrm{d}^{2}y\,\mathrm{d}^{2}z\,
f(y)g(z)e^{-2i\theta_{ij}(x^{i}-y^{i})(x^{j}-z^{j})}.
\end{equation}
Other integral expressions are possible
some of which can be found in the appendix of~\cite{Gracia-BondiaLizziMarmoVitale}.
Note that the Moyal product can be expressed as well as in a dif\mbox{}ferential form which is
an asymptotic expansion of the integral one~\cite{EstradaVarillyGracia-Bondia}.
However, the integral form has the advantage to be def\mbox{}ined
on a set wider than the one on which is def\mbox{}ined the dif\mbox{}ferential form.

\section{The Wick-Voros product from a weighted Weyl map}

A weighted Weyl map is a generalization of the Weyl map def\mbox{}ined as
\begin{equation}
\hat{\Omega}(f)=\frac{1}{(2\pi\theta)^{2}}\int\mathrm{d}^{2}x\,\mathrm{d}^{2}\alpha
f(x)w(\alpha)e^{-i\theta_{ij}x^{i}\alpha^{j}}W(\alpha)
\end{equation}
where $w(\alpha)$ is an invertible function, called weighted function.
A general weighted Weyl map is linear and invertible and its inverse is
\begin{equation}
\Omega^{-1}\left(\hat{\Omega}(f)\right)=\frac{1}{2\pi\theta}\int\mathrm{d}^{2}\alpha\,w^{-1}(\alpha)
e^{i\theta_{ij}x^{i}\alpha^{j}}\operatorname{Tr}\left(\hat{\Omega}(f)W^{\dag}(\alpha)\right).
\end{equation}
Here we are interesting to the weighted Weyl map given by
\begin{equation}
\hat{\Omega}_{V}(f)=\frac{1}{(2\pi\theta)^{2}}\int\mathrm{d}^{2}x\,\mathrm{d}^{2}\alpha
f(x)e^{-\frac{1}{4\theta}\boldsymbol{\alpha}^{2}}e^{-i\theta_{ij}x^{i}\alpha^{j}}W(\alpha)
\end{equation}
which leads to the Wick-Voros product. This weighted Weyl map can be written in complex coordinates:
\begin{equation}\label{ComplexCoordinates}
x^{\pm}=\frac{x^{1}\pm ix^{2}}{\sqrt{2}}
\end{equation}
as
\begin{equation}\label{WVWM}
\hat{\Omega}_{V}(f)=\frac{1}{(2\pi\theta)^{2}}\int\mathrm{d}^{2}x\,\mathrm{d}^{2}\alpha
f(x)e^{-\frac{1}{2\theta}\alpha^{+}\alpha^{-}}
e^{-\frac{1}{\theta}\left(\alpha^{+}x^{-}-\alpha^{-}x^{+}\right)}W(\alpha)
\end{equation}
where
\begin{equation}
\alpha^{\pm}=\frac{\alpha^{1}\pm i\alpha^{2}}{\sqrt{2}}.
\end{equation}
In these coordinates the Weyl system \eqref{WS} takes the form
\begin{equation}
W(\alpha)=e^{\frac{1}{\theta}\left(\alpha^{-}\hat{a}-\alpha^{+}\hat{a}^{\dag}\right)}
\end{equation}
and the relation \eqref{WSR} is given by
\begin{equation}
W(\alpha)W(\beta)=W(\alpha+\beta)e^{\frac{1}{2\theta}(\alpha^{+}\beta^{-}-\alpha^{-}\beta^{+})}.
\end{equation}
The Wick-Voros product is then def\mbox{}ined by the relation\footnote{See also~\cite{Daoud}.}
\begin{equation}
\hat{\Omega}_{V}(f\star_{V}g)=\hat{\Omega}_{V}(f)\hat{\Omega}_{V}(g).
\end{equation}
and it is possible to show that it reads
\begin{equation}\label{IFVWP}
(f\star_{V}g)(x)=\int\frac{\mathrm{d}^{2}y}{\pi\theta}f(x^{-},y^{+})g(y^{-},x^{+})
e^{-\frac{1}{\theta}(x^{-}-y^{-})(x^{+}-y^{+})}.
\end{equation}
This product, like the Moyal one, can be expressed as well as in
a dif\mbox{}ferential form which is an asymptotic expansion of the integral one as we will see in the following.

\chapter{The Moyal and Wick-Voros products from a quantization scheme}

\emph{In this second chapter we show how the Moyal and Wick-Voros products can be derived
from a general quantization scheme. To this end, we f\mbox{}irst review a general quantization scheme
for associating operators with functions and vice versa and producing new star products.
In particular, we describe the duality symmetry of a quantization scheme and the notion of dual star product.
We f\mbox{}inally introduce the Weyl-Wigner and Wick-Voros quantization schemes for
the Moyal and Wick-Voros products respectively.}

\section{Quantization schemes and star products}

We begin with a review of a general scheme to associate operators
with functions and vice versa~\cite{CahillGlauber1} and produce new star products~\cite{MankoMankoMarmo}
for operator symbols.
In this scheme the symbols of the operators are def\mbox{}ined in terms of a family of operators, called dequantizers,
while the reconstruction of operators in terms of their symbols
is determined using another family of operators, called quantizers.

Let us consider a Hilbert space $\mathcal{H}$ and two sets of
operators $\hat{U}(x)$ and $\hat{V}(x)$ on $\mathcal{H}$
parameterized by an $n$-dimensional vector $x=(x_{1},x_{2},...,x_{n})$
and suppose they satisfy the consistency condition
\begin{equation}\label{CI}
\operatorname{Tr}\left(\hat{U}(x)\hat{V}(x')\right)=\delta^{(n)}(x-x').
\end{equation}
With these two families of operators we can construct an invertible map
which associates to each operator $\hat{A}$ on $\mathcal{H}$
a function $f_{\hat{A}}(x)$, called the symbol of the operator $\hat{A}$, def\mbox{}ined by
\begin{equation}\label{S}
f_{\hat{A}}(x)=\operatorname{Tr}\left(\hat{A}\hat{V}(x)\right)
\end{equation}
and to each function $f_{\hat{A}}(x)$ an operator $\hat{A}$ on $\mathcal{H}$ def\mbox{}ined by
\begin{equation}\label{RF}
\hat{A}=\int f_{\hat{A}}(x)\hat{U}(x)\mathrm{d}^{n}x.
\end{equation}
Indeed, multiplying both sides of equation \eqref{RF} by the
operator $\hat{V}(x')$ and taking the trace, we have
\begin{equation}
\operatorname{Tr}\left(\hat{A}\hat{V}(x')\right)=
\operatorname{Tr}\int f_{\hat{A}}(x)\hat{U}(x)\hat{V}(x')\mathrm{d}^{n}x
\end{equation}
and, assuming it is possible to exchange the trace with the integral, we have
\begin{equation}
\operatorname{Tr}\left(\hat{A}\hat{V}(x')\right)=\int
f_{\hat{A}}(x)\operatorname{Tr}\left(\hat{U}(x)\hat{V}(x')\right)\mathrm{d}^{n}x=f_{\hat{A}}(x')
\end{equation}
where we have used \eqref{CI}.
Therefore, the operators $\hat{V}(x)$ associate to the operator $\hat{A}$ (quantum observable)
a function $f_{\hat{A}}(x)$ (classical observable) i.e.~they ``dequantize'' the quantum observable,
while the role of the other operators $\hat{U}(x)$ is opposite; they associate to the function $f_{\hat{A}}(x)$
an operator $\hat{A}$ i.e.~they ``quantize'' the classical observable.
For this reason we call the operators $\hat{U}(x)$ and $\hat{V}(x)$ quantizers and dequantizers respectively.
However, there is an ambiguity in def\mbox{}ining the operators $\hat{U}(x)$ and $\hat{V}(x)$.
Indeed, we can  make a scaling transformation of the operators $\hat{U}(x)$ and $\hat{V}(x)$
without violating the consistency of the quantization scheme i.e.~the condition \eqref{CI}.
Moreover, if we require that the symbol of identity operator $\mathbbm{1}$ is equal to the constant function 1,
this ambiguity is removed because the operators $\hat{V}(x)$ have to satisfy the condition
\begin{equation}
\operatorname{Tr}\hat{V}(x)=1
\end{equation}
and the operators $\hat{U}(x)$ the condition
\begin{equation}
\int\hat{U}(x)\mathrm{d}^{n}x=\mathbbm{1}.
\end{equation}

Now we can introduce the star product of the symbols $f_{\hat{A}}(x)$ and $f_{\hat{B}}(x)$
of two operators $\hat{A}$ and $\hat{B}$ on $\mathcal{H}$ by the relationships
\begin{equation}
f_{\hat{A}}(x)*f_{\hat{B}}(x)=f_{\hat{A}\hat{B}}(x)
\end{equation}
that is,
\begin{equation}
f_{\hat{A}}(x)*f_{\hat{B}}(x)=
\operatorname{Tr}\left(\hat{A}\hat{B}\hat{V}(x)\right).
\end{equation}
In other words, the star product of the symbols $f_{\hat{A}}(x)$ and $f_{\hat{B}}(x)$
of two operators $\hat{A}$ and $\hat{B}$ on $\mathcal{H}$ is the symbol of the their product.
The star product is associative due to the associativity of the operator product. Indeed,
\begin{align}
\nonumber
(f_{\hat{A}}(x)\star f_{\hat{B}}(x))\star f_{\hat{C}}(x)&=f_{\hat{A}\hat{B}}(x)\star f_{\hat{C}}(x)=
f_{(\hat{A}\hat{B})\hat{C}}(x)=f_{\hat{A}(\hat{B}\hat{C})}(x)\\
&=f_{\hat{A}}(x)\star f_{\hat{B}\hat{C}}(x)=f_{\hat{A}}(x)\star(f_{\hat{B}}(x)\star f_{\hat{C}}(x)).
\end{align}
However, it is not commutative. From the def\mbox{}inition of the star product follows that
\begin{equation}
f_{\hat{A}}(x)*f_{\hat{B}}(x)=\operatorname{Tr}\int
f_{\hat{A}}(x')f_{\hat{B}}(x'')\hat{U}(x')\hat{U}(x'')\hat{V}(x)\mathrm{d}^{n}x'\mathrm{d}^{n}x''
\end{equation}
and, assuming once again it is possible to exchange the trace with the integral, we have
\begin{equation}
f_{\hat{A}}(x)*f_{\hat{B}}(x)=\int f_{\hat{A}}(x')f_{\hat{B}}(x'')
\operatorname{Tr}\left(\hat{U}(x')\hat{U}(x'')\hat{V}(x)\right)\mathrm{d}^{n}x'\mathrm{d}^{n}x''.
\end{equation}
Therefore, we can rewritten the star product as
\begin{equation}
f_{\hat{A}}(x)*f_{\hat{B}}(x)=\int K(x',x'',x)
f_{\hat{A}}(x')f_{\hat{B}}(x'')\mathrm{d}^{n}x'\mathrm{d}^{n}x''
\end{equation}
where the kernel is given by
\begin{equation}\label{Kernel}
K(x',x'',x)=\operatorname{Tr}\left(\hat{U}(x')\hat{U}(x'')\hat{V}(x)\right).
\end{equation}
Note that the usual product is also of this kind for
\begin{equation}
K(x',x'',x)=\delta^{(n)}(x'-x)\delta^{(n)}(x''-x).
\end{equation}
Moreover, the expression \eqref{Kernel} is quadratic with respect to $\hat{U}(x)$
and linear with respect to $\hat{V}(x)$
and then there is an asymmetry in the kernel with respect to quantizers and dequantizers.
Furthermore, thanks to the cyclical property of the trace,
the kernel and then the star product is invariant under the transformation
\begin{align}
\hat{U}'(x)&=\hat{S}\hat{U}(x)\hat{S}^{-1}\\
\hat{V}'(x)&=\hat{S}\hat{V}(x)\hat{S}^{-1}
\end{align}
where $\hat{S}$ is an invertible operator.
Finally, the associativity condition for the operator symbols
imposes a strong constrain on the kernel $K(x',x'',x)$. Indeed, for associativity
\begin{multline}
(f_{\hat{A}}(x)\star f_{\hat{B}}(x))\star f_{\hat{C}}(x)=
\int K(x_{1},x_{2},x)(f_{\hat{A}}(x_{1})\star f_{\hat{B}}(x_{1}))f_{\hat{C}}(x_{2})
\mathrm{d}^{n}x_{1}\mathrm{d}^{n}x_{2}\\
=\int K(x_{1},x_{2},x)K(x_{3},x_{4},x_{1})f_{\hat{A}}(x_{3})f_{\hat{B}}(x_{4})f_{\hat{C}}(x_{2})
\mathrm{d}^{n}x_{1}\mathrm{d}^{n}x_{2}\mathrm{d}^{n}x_{3}\mathrm{d}^{n}x_{4}
\end{multline}
must be equal to
\begin{multline}
f_{\hat{A}}(x)\star(f_{\hat{B}}(x)\star f_{\hat{C}}(x))=
\int K(x_{1},x_{2},x)f_{\hat{A}}(x_{1})(f_{\hat{B}}(x_{2})\star f_{\hat{C}}(x_{2}))
\mathrm{d}^{n}x_{1}\mathrm{d}^{n}x_{2}\\
=\int K(x_{1},x_{2},x)K(x_{3},x_{4},x_{2})f_{\hat{A}}(x_{1})f_{\hat{B}}(x_{3})f_{\hat{C}}(x_{4})
\mathrm{d}^{n}x_{1}\mathrm{d}^{n}x_{2}\mathrm{d}^{n}x_{3}\mathrm{d}^{n}x_{4}\\
=\int K(x_{3},x_{1},x)K(x_{4},x_{2},x_{1})f_{\hat{A}}(x_{3})f_{\hat{B}}(x_{4})f_{\hat{C}}(x_{2})
\mathrm{d}^{n}x_{1}\mathrm{d}^{n}x_{2}\mathrm{d}^{n}x_{3}\mathrm{d}^{n}x_{4}.
\end{multline}
Therefore, the kernel $K(x',x'',x)$ must satisfy the equation
\begin{equation}\label{ACFK}
\int K(x_{1},x_{2},x)K(x_{3},x_{4},x_{1})\mathrm{d}^{n}x_{1}=
\int K(x_{3},x_{1},x)K(x_{4},x_{2},x_{1})\mathrm{d}^{n}x_{1}
\end{equation}
which is, of course, satisf\mbox{}ied by \eqref{Kernel}.
Observe that this equation has like symmetry a scaling transform.
That is, given a solution $K(x',x'',x)$ of the equation \eqref{ACFK},
\begin{equation}
K'(x',x'',x)=\lambda K(x',x'',x)
\end{equation}
is still a solution of \eqref{ACFK}, where $\lambda$ is a non-vanishing complex number.
Note that the scaling transform of the kernel can be induced
transforming the quantizers and dequantizers as
\begin{align}
\hat{U}'(x)&=\lambda\hat{U}(x)\\
\hat{V}'(x)&=\lambda^{-1}\hat{V}(x).
\end{align}
In the next section we will introduce the duality symmetry
namely we will show that the role of $\hat{U}(x)$ and $\hat{V}(x)$ can be
exchanged without violating the consistency of the quantization scheme.
Furthermore, we will def\mbox{}ine dual star products.

\section{Dual quantization schemes}

As we have already said, the duality symmetry is due to the fact
that the role of $\hat{U}(x)$ and $\hat{V}(x)$ can be exchanged
without violating the consistency of the quantization scheme.
The dual quantization scheme~\cite{MankoMarmoVitale,MankoMankoMarmoVitale} is def\mbox{}ined by
\begin{align}
\hat{U}^{(d)}(x)&=\hat{V}(x)\\
\hat{V}^{(d)}(x)&=\hat{U}(x)
\end{align}
and condition \eqref{CI} is obviously satisf\mbox{}ied.
In the dual scheme, the symbol $f^{(d)}_{A}(x)$ of an operator $\hat{A}$ on $\mathcal{H}$,
called the dual symbol of the operator $\hat{A}$, is given by
\begin{equation}
f^{(d)}_{\hat{A}}(x)=\operatorname{Tr}\left(\hat{A}\hat{U}(x)\right)
\end{equation}
and the reconstruction formula for the operator $\hat{A}$ is given by
\begin{equation}
\hat{A}=\int f^{(d)}_{\hat{A}}(x)\hat{V}(x)\mathrm{d}^{n}x.
\end{equation}
Therefore, in the dual scheme the operators $\hat{U}(x)$ are used to dequantize,
while the operators $\hat{V}(x)$ to quantize.
In other words, the dual dequantizers correspond to old quantizers,
while the dual quantizers correspond to old dequantizers.

In the dual scheme, the star product of the dual symbols
$f^{(d)}_{\hat{A}}(x)$ and $f^{(d)}_{\hat{B}}(x)$ of two operators
$\hat{A}$ and $\hat{B}$ on $\mathcal{H}$, called dual star product, is given by
\begin{equation}
f^{(d)}_{\hat{A}}(x)*f^{(d)}_{\hat{B}}(x)=f^{(d)}_{\hat{A}\hat{B}}(x),
\end{equation}
that is,
\begin{equation}
f^{(d)}_{\hat{A}}(x)*f^{(d)}_{\hat{B}}(x)=
\operatorname{Tr}\left(\hat{A}\hat{B}\hat{U}(x)\right).
\end{equation}
Equivalently, we can rewritten the dual star product as
\begin{equation}
f^{(d)}_{\hat{A}}(x)*f^{(d)}_{\hat{B}}(x)=\int K^{(d)}(x',x'',x)
f^{(d)}_{\hat{A}}(x')f^{(d)}_{\hat{B}}(x'')\mathrm{d}^{n}x'\mathrm{d}^{n}x''
\end{equation}
where the kernel, called the dual kernel, is given by
\begin{equation}\label{DK}
K^{(d)}(x',x'',x)=\operatorname{Tr}\left(\hat{V}(x')\hat{V}(x'')\hat{U}(x)\right).
\end{equation}
The dual kernel $K^{(d)}(x',x'',x)$, unlike the kernel $K(x',x'',x)$,
is linear with respect to $\hat{U}(x)$ and quadratic with respect to $\hat{V}(x)$.
So, in general, the star product and its dual are dif\mbox{}ferent to each other.
We will call the quantization scheme self-dual if
\begin{equation}
K^{(d)}(x',x'',x)=K(x',x'',x).
\end{equation}
In other words, a quantization scheme is self-dual if it and its dual scheme produce the same star product.

\section{The Moyal and Wick-Voros products from a quantization scheme}

The main quantization scheme known is the Weyl-Wigner quantization scheme.
This scheme is self-dual and the star product associated with it is the Moyal product.

In order to introduce this scheme, let us consider a system with a single degree of freedom.
In this case, the Hilbert space $\mathcal{H}$ is identif\mbox{}ied with $L^{2}(\mathbb{R})$,
the Hilbert space of square-integrable functions on $\mathbb{R}$, and the
quantizers and dequantizers~\cite{MankoMankoMarmoVitale} are given respectively by
\begin{align}
\hat{U}(\lambda)&=\frac{1}{2\pi}\hat{V}(\lambda)\\
\hat{V}(\lambda)&=2\hat{D}(\lambda)(-1)^{a^{\dag}a}\hat{D}(-\lambda)
\end{align}
where $\lambda=x+ip$, with $x$ and $p$ representing respectively
the position and momentum, $(-1)^{a^{\dag}a}$ is the parity operator, with
the annihilation and creation operators $a$ and $a^{\dag}$ given respectively by
\begin{align}
a&=\frac{1}{\sqrt{2}}(\hat{x}+i\hat{p})\\
a^{\dag}&=\frac{1}{\sqrt{2}}(\hat{x}-i\hat{p})
\end{align}
which satisfy the commutation relation \eqref{CR} and
the displacement operator $\hat{D}(\lambda)$ given by
\begin{equation}
\hat{D}(\lambda)=e^{\lambda a^{\dag}-\lambda^{*}a}
\end{equation}
which is unitary and obeys the relation
\begin{equation}
D^{\dag}(\lambda)=D^{-1}(\lambda)=D(-\lambda).
\end{equation}
As well-known, the displacement operator creates the coherent states namely,
the eigenstates of the annihilation operator $a$, from the vacuum state:
\begin{equation}
D(\lambda)|0\rangle=|\lambda\rangle
\end{equation}
for every complex number $\lambda$ where the vacuum state is as usual def\mbox{}ined by
\begin{equation}
a|0\rangle=0.
\end{equation}
It is not dif\mbox{}f\mbox{}icult to show that~\cite{MankoMankoMarmo,MankoMankoMarmoVitale}
the star product associated with the Weyl-Wigner quantization scheme is the Moyal product \eqref{IFMP}.
Moreover, this scheme is self-dual and from \eqref{S} follows that:
\begin{align}
f_{\mathbbm{1}}(x,p)&=1\\
f_{\hat{q}}(x,p)&=x\\
f_{\hat{p}}(x,p)&=p.
\end{align}

Eventually, consider the following quantization scheme~\cite{CahillGlauber1,MankoMankoMarmoVitale}
described by the two families of operators
\begin{align}
\hat{U}_{s}(\lambda)&=\frac{1}{2\pi}\hat{V}_{-s}(\lambda)\\
\hat{V}_{s}(\lambda)&=\frac{2}{1-s}\hat{D}(\lambda)\left(\frac{s+1}{s-1}\right)^{a^{\dag}a}\hat{D}(\lambda)
\end{align}
where $s$ is a real parameter. It is possible to show that the Wick-Voros product \eqref{IFVWP}
is obtained in the limit $s=1$.
Moreover, it is easy to see that the case $s=0$ corresponds to the Weyl-Wigner scheme described above.

\chapter{Non-commutative Moyal f\mbox{}ield theory}

\emph{In this chapter we brief\mbox{}ly review a non-commutative f\mbox{}ield theory
obtained from a commutative one replacing the ordinary product with the Moyal one.
We show that the free case is the same as the ordinary one, but the interacting case does not.
In particular, we show that in the interacting case the Moyal product
softens the ultraviolet divergence, but it is responsible for the so-called ultraviolet/infrared mixing.}

\section{The Moyal product}

For the sake of simplicity, we consider a $(1+2)$-dimensional space-time
and consider exclusively spatial non-commutativity.
Therefore, the canonical non-commutative relation between
the coordinate functions on the space-time takes the form
\begin{equation}\label{CNC}
[x^{i},x^{j}]=i\theta^{ij}
\end{equation}
where
\begin{equation}
\theta^{ij}=\theta\varepsilon^{ij}
\quad\mathrm{with}\quad
(\varepsilon^{ij})=
\left(\begin{array}{cc}
0 & 1\\
-1 & 0
\end{array}\right)
\end{equation}
and $\theta$ is a real constant parameter of dimensions of a square length
which can be seen as a deformation parameter.
Consider now the dif\mbox{}ferential form of the Moyal product~\cite{EstradaVarillyGracia-Bondia} which is given by
\begin{equation}\label{MP}
f\star_{M}g=fe^{\frac{i}{2}\theta^{ij}
\stackrel{\leftarrow}{\partial}_{i}\stackrel{\rightarrow}{\partial}_{j}}g
\end{equation}
where
$\stackrel{\leftarrow}{\partial}_{i}$ and $\stackrel{\rightarrow}{\partial}_{j}$
act respectively on the left and on the right.
More explicitly, it can be written as
\begin{equation}
f\star_{M}g=fe^{\frac{i}{2}\theta\left(
\stackrel{\leftarrow}{\partial}_{1}\stackrel{\rightarrow}{\partial}_{2}-
\stackrel{\leftarrow}{\partial}_{2}\stackrel{\rightarrow}{\partial}_{1}\right)}g
\end{equation}
which reduces to the commutative product in the limit when $\theta$ goes to $0$,
namely it is a deformation of the commutative one. The Moyal product is associative, but not commutative.
In particular, we have
\begin{align}
x^{1}\star_{M}x^{2}&=x^{1}x^{2}+\frac{i}{2}\theta\\
x^{2}\star_{M}x^{1}&=x^{1}x^{2}-\frac{i}{2}\theta.
\end{align}
Hence the Moyal bracket of $x^{1}$ and $x^{2}$ reads
\begin{equation}
[x^{1},x^{2}]_{\star_{M}}=x^{1}\star_{M}x^{2}-x^{2}\star_{M}x^{1}=i\theta
\end{equation}
in agreement with the canonical non-commutative relation \eqref{CNC}.
More in general, the Moyal bracket of two functions%
\footnote{Like every commutator, the Moyal bracket is bilinear and antisymmetric.
Moreover, it satisf\mbox{}ies the Jacobi identity
\begin{equation*}
[f,[g,h]_{\star_{M}}]_{\star_{M}}+[g,[h,f]_{\star_{M}}]_{\star_{M}}+[h,[f,g]_{\star_{M}}]_{\star_{M}}=0
\end{equation*}
and the Leibniz rule
\begin{equation*}
[f,g\star_{M}h]_{\star_{M}}=[f,g]_{\star_{M}}\star_{M}h+g\star_{M}[f,h]_{\star_{M}}.
\end{equation*}}
\begin{equation}
[f,g]_{\star_{M}}=f\star_{M}g-g\star_{M}f
\end{equation}
to f\mbox{}irst order in $\theta$ reads
\begin{equation}
[f,g]_{\star_{M}}=i\theta\{f,g\}+\ldots
\end{equation}
where as usual
\begin{equation}
\{f,g\}=(\partial_{1}f)\partial_{2}g-(\partial_{2}f)\partial_{1}g.
\end{equation}
Thus the Moyal bracket of two functions to f\mbox{}irst order in the deformation parameter $\theta$
is proportional to the Poisson bracket of the two functions.

It is very useful to write the Moyal product in momentum space. Since in Fourier transform
\begin{equation}
f(x)=\int\frac{\mathrm{d}^{3}p}{(2\pi)^{3}}\tilde{f}(p)e^{-ip\cdot x}
\end{equation}
from \eqref{MP} we have
\begin{align}\label{mp}
\nonumber
(f\star_{M}g)(x)=&\int\frac{\mathrm{d}^{3}p}{(2\pi)^{3}}\frac{\mathrm{d}^{3}q}{(2\pi)^{3}}
\tilde{f}(p)\tilde{g}(q)e^{-\frac{i}{2}\theta^{ij}p_{i}q_{j}}e^{i(p+q)\cdot x}\\
=&\int\frac{\mathrm{d}^{3}p}{(2\pi)^{3}}\frac{\mathrm{d}^{3}q}{(2\pi)^{3}}
\tilde{f}(p)\tilde{g}(q)e^{-\frac{i}{2}\theta\boldsymbol{p}\wedge\boldsymbol{q}}e^{i(p+q)\cdot x}
\end{align}
where we have set
\begin{equation}
\boldsymbol{p}\wedge\boldsymbol{q}=\varepsilon^{ij}p_{i}q_{j}
\end{equation}
which is, of course, antisymmetric for the exchange of $\boldsymbol{p}$ and $\boldsymbol{q}$.
Therefore, the Moyal product in momentum space is the standard convolution of Fourier
transforms twisted by a phase. For example, we can calculate the Moyal product of two exponentials.
Indeed, by using \eqref{mp} we easily get
\begin{align}
\nonumber
e^{-ip\cdot x}\star_{M}e^{-iq\cdot x}&=
\int\frac{\mathrm{d}^{3}r}{(2\pi)^{3}}\frac{\mathrm{d}^{3}s}{(2\pi)^{3}}\,
\delta^{(3)}(r-p)\delta^{(3)}(s-q)e^{-\frac{i}{2}\theta\boldsymbol{r}\wedge\boldsymbol{s}}e^{i(r+s)\cdot x}\\
&=e^{-\frac{i}{2}\theta\boldsymbol{p}\wedge\boldsymbol{q}}e^{i(p+q)\cdot x}.
\end{align}
It is now easy to see that from \eqref{mp} follows that the integral of the Moyal product
of two functions is equal to the integral of the ordinary product of the two functions. Indeed,
\begin{align}\label{imp}
\nonumber
\int\mathrm{d}^{3}x\,f\star_{M}g&=
\int\mathrm{d}^{3}x\frac{\mathrm{d}^{3}p}{(2\pi)^{3}}\frac{\mathrm{d}^{3}q}{(2\pi)^{3}}
\tilde{f}(p)\tilde{g}(q)e^{-\frac{i}{2}\theta\boldsymbol{p}\wedge\boldsymbol{q}}e^{i(p+q)\cdot x}\\
\nonumber
&=\int\frac{\mathrm{d}^{3}p}{(2\pi)^{3}}\tilde{f}(p)\tilde{g}(-p)
=\int\mathrm{d}^{3}x\,\mathrm{d}^{3}y\frac{\mathrm{d}^{3}p}{(2\pi)^{3}}f(x)g(y)e^{ip\cdot(x-y)}\\
&=\int\mathrm{d}^{3}x\,f(x)g(x).
\end{align}
This property is very important in non-commutative f\mbox{}ield theory since,
as we will show in the next section, it allows to state that the free non-commutative f\mbox{}ield theory
with the Moyal product is the same as the commutative one.
Finally, note that from \eqref{imp} follows that the Moyal product has the trace property
\begin{equation}
\int\mathrm{d}^{3}x\,f\star_{M}g=\int\mathrm{d}^{3}x\,g\star_{M}f.
\end{equation}

\section{The Moyal f\mbox{}ield theory}

At this point we proceed to the discussion of a non-commutative f\mbox{}ield theory
obtained from a commutative one replacing the ordinary product with the Moyal one.
To this end, consider the commutative f\mbox{}ield theory described by the action
\begin{equation}\label{OA}
S^{(0)}=S_{0}^{(0)}+S_{\mathrm{int}}^{(0)},
\end{equation}
where $S_{0}^{(0)}$ is the free Klein-Gordon action given by
\begin{equation}\label{FOA}
S_{0}^{(0)}=\int\mathrm{d}^{3}x\,
\frac{1}{2}\left(\partial_{\mu}\phi\,\partial^{\mu}\phi-m^{2}\phi^{2}\right)
\end{equation}
and $S_{\mathrm{int}}^{(0)}$ is the interacting action given by
\begin{equation}
S_{\mathrm{int}}^{(0)}=\frac{g}{4!}\int\mathrm{d}^{3}x\,\phi^{4}.
\end{equation}
In order to construct a non-commutative f\mbox{}ield theory, we replace the ordinary product with the Moyal one.
This procedure is part of a general framework called deformation quantization which consists
in a modif\mbox{}ication of a theory in such a way that it reduces to the undeformed one in a certain limit.
So the free non-commutative action is given by
\begin{equation}
S_{0_{M}}=\int\mathrm{d}^{3}x\,
\frac{1}{2}\left(\partial_{\mu}\phi\star_{M}\partial^{\mu}\phi-m^{2}\phi\star_{M}\phi\right)
\end{equation}
which, of course, is equal to the ordinary one \eqref{FOA} because of \eqref{imp}.
Then the free non-commutative f\mbox{}ield theory with the Moyal product is the same as the commutative one.
Instead, the interacting non-commutative action is given by
\begin{equation}\label{MIA}
S_{\mathrm{int}_{M}}=\frac{g}{4!}\int\mathrm{d}^{3}x\,
\phi\star_{M}\phi\star_{M}\phi\star_{M}\phi
\end{equation}
which is dif\mbox{}ferent from the commutative one.

We now move on to the quantum case, we calculate the Green's functions up to one-loop
for the two- and four-point cases and discuss the ultraviolet behaviour of the theory.
Since the free non-commutative f\mbox{}ield theory with the Moyal product
is the same as the commutative one, the propagator that is, the two-point Green's function
is as the usual one
\begin{equation}\label{MTPGF}
\tilde{G}^{(2)}_{M}(p)=\frac{1}{p^{2}-m^{2}}.
\end{equation}
To calculate the four-point Green's function to the tree level in the Moyal case,
we f\mbox{}irst have to determine the vertex. In this case, the vertex is dif\mbox{}ferent
with respect to the usual one since it acquires a phase~\cite{Filk}. To determine the vertex,
let us write down the interacting action \eqref{MIA} in momentum space.
By using the relations \eqref{mp} and \eqref{imp} we have
\begin{align}
\nonumber
S_{\mathrm{int}_{M}}=\frac{g}{4!}\int\mathrm{d}x\,
\frac{\mathrm{d}k_{1}}{(2\pi)^{3}}\frac{\mathrm{d}k_{2}}{(2\pi)^{3}}
\frac{\mathrm{d}k_{3}}{(2\pi)^{3}}\frac{\mathrm{d}k_{4}}{(2\pi)^{3}}
\tilde{\phi}(k_{1})\tilde{\phi}(k_{2})\tilde{\phi}(k_{3})\tilde{\phi}(k_{4})\\
\nonumber
e^{-\frac{i}{2}\theta(\boldsymbol{k}_{1}\wedge\boldsymbol{k}_{2}+
\boldsymbol{k}_{3}\wedge\boldsymbol{k}_{4})}
e^{i(k_{1}+k_{2}+k_{3}+k_{4})\cdot x}\\
\nonumber
=\frac{g}{4!}(2\pi)^{3}\!\int\frac{\mathrm{d}k_{1}}{(2\pi)^{3}}\frac{\mathrm{d}k_{2}}{(2\pi)^{3}}
\frac{\mathrm{d}k_{3}}{(2\pi)^{3}}\frac{\mathrm{d}k_{4}}{(2\pi)^{3}}
\tilde{\phi}(k_{1})\tilde{\phi}(k_{2})\tilde{\phi}(k_{3})\tilde{\phi}(k_{4})\\
e^{-\frac{i}{2}\theta[\boldsymbol{k}_{1}\wedge\boldsymbol{k}_{2}+
\boldsymbol{k}_{3}\wedge\boldsymbol{k}_{4}+
(\boldsymbol{k}_{1}+\boldsymbol{k}_{2})\wedge(\boldsymbol{k}_{3}+\boldsymbol{k}_{4})]}
\delta^{(3)}(k_{1}+k_{2}+k_{3}+k_{4})
\end{align}
which can be rewritten as
\begin{equation}
S_{\mathrm{int}_{M}}=i\int\prod_{a=1}^{4}\frac{\mathrm{d}^{3}k_{a}}{(2\pi)^{3}}
\tilde{\phi}(k_{a})V_{\star_{M}}
\end{equation}
where
\begin{equation}\label{MV}
V_{\star_{M}}=Ve^{\sum_{a<b}-\frac{i}{2}\theta\boldsymbol{k}_{a}\wedge\boldsymbol{k}_{b}}=
Ve^{\sum_{a<b}-\frac{i}{2}\theta^{ij}k_{ai} k_{bj}}
\end{equation}
is the Moyal vertex and
\begin{equation}\label{OV}
V=-i\frac{g}{4!}(2\pi)^{3}\delta^{(3)}\!\left(\sum_{a=1}^{4}k_{a}\right)
\end{equation}
is the usual vertex which is proportional to the coupling constant
multiplying the $\delta$ of momentum conservation.
Note that the presence of the phase in the vertex \eqref{MV}
makes it non-invariant for a generic exchange of the momenta.
This is a consequence of non-commutativity and of the fact that the integral of Moyal product
of more than two functions is not invariant for an exchange of the functions.
However, it is invariant for a cyclic exchange of the factors.
Eventually, to determine the four-point Green's function to the tree level,
we must attach to the vertex \eqref{MV} four propagators \eqref{MTPGF}. We have
\begin{equation}
\tilde{G}_{M}^{(4)}=-ig(2\pi)^{3}\frac{e^{\sum_{a<b}-\frac{i}{2}\theta\boldsymbol{k}_{a}\wedge\boldsymbol{k}_{b}}}
{\prod_{a=1}^{4}(k_{a}^{2}-m^{2})}\delta^{(3)}\!\left(\sum_{a=1}^{4}k_{a}\right)
\end{equation}
which is dif\mbox{}ferent with respect the usual one.

\section{UV/IR mixing for the Moyal product}

In order to discuss the ultraviolet behaviour of the theory in the Moyal case,
we calculate the one-loop corrections to the two- and four-point Green's functions.
Because the vertex \eqref{MV} is not invariant for a generic exchange of the momenta,
we must consider both the planar and non-planar diagrams in the calculation of the corrections.
To obtain  the one-loop corrections to the propagator,
consider f\mbox{}irst the planar case in f\mbox{}igure~\ref{TPD}(a).
The amplitude is obtained using three propagators \eqref{MTPGF}, two with momentum $p$,
one with momentum $q$, and the vertex \eqref{MV} with assignments
\begin{equation}\label{PA}
k_{1}=-k_{4}=p\quad\mathrm{and}\quad k_{2}=-k_{3}=q
\end{equation}
and the proper symmetry factor~\cite{Kaku}.
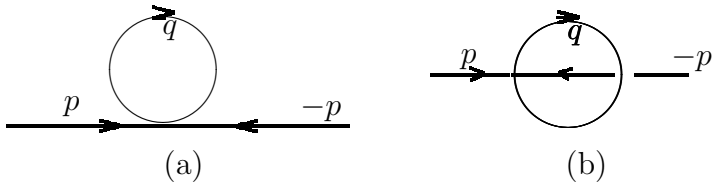
\begin{figure}[h]
\begin{center}
\unitlength 1.5mm
\linethickness{1pt}
\ifx\plotpoint\undefined\newsavebox{\plotpoint}\fi
\begin{picture}(66.733,16.902)(0,0)
\put(6.983,6.412){\line(1,0){1}} \put(7.983,6.412){\line(1,0){1}}
\put(8.983,6.412){\line(1,0){1}} \put(9.983,6.412){\line(1,0){1}}
\put(10.983,6.412){\line(1,0){1}}
\put(11.983,6.412){\line(1,0){1}}
\put(12.983,6.412){\line(1,0){1}}
\put(13.983,6.412){\line(1,0){1}}
\put(14.983,6.412){\line(1,0){1}}
\put(15.983,6.412){\line(1,0){1}}
\multiput(16.983,6.412)(-.1293333,-.0326667){15}{\line(-1,0){.1293333}}
\multiput(15.043,6.902)(.1293333,-.0326667){15}{\line(1,0){.1293333}}
\put(36.983,6.412){\line(-1,0){1}}
\put(35.983,6.412){\line(-1,0){1}}
\put(34.983,6.412){\line(-1,0){1}}
\put(33.983,6.412){\line(-1,0){1}}
\put(32.983,6.412){\line(-1,0){1}}
\put(31.983,6.412){\line(-1,0){1}}
\put(30.983,6.412){\line(-1,0){1}}
\put(29.983,6.412){\line(-1,0){1}}
\put(28.983,6.412){\line(-1,0){1}}
\put(27.983,6.412){\line(-1,0){1}}
\multiput(26.983,6.412)(.1293333,.0326667){15}{\line(1,0){.1293333}}
\multiput(28.923,5.922)(-.1293333,.0326667){15}{\line(-1,0){.1293333}}
\put(16.983,6.412){\line(1,0){1}}
\put(17.983,6.412){\line(1,0){1}}
\put(18.983,6.412){\line(1,0){1}}
\put(19.983,6.412){\line(1,0){1}}
\put(20.983,6.412){\line(1,0){1}}
\put(21.983,6.412){\line(1,0){1}}
\put(22.983,6.412){\line(1,0){1}}
\put(23.983,6.412){\line(1,0){1}}
\put(24.983,6.412){\line(1,0){1}}
\put(25.983,6.412){\line(1,0){1}}
\put(19.862,15.97){\line(1,0){1}}
\multiput(21.657,16.412)(-.1293333,-.0326667){15}{\line(-1,0){.1293333}}
\multiput(19.717,16.902)(.1293333,-.0326667){15}{\line(1,0){.1293333}}
\put(20.657,11.412){\circle{10}} \put(11.932,7.867){$p$}
\put(20.594,14.496){$q$} \put(32.704,7.425){$-p$}
\put(55.177,16.01){\line(1,0){1}}
\put(55.177,16.01){\line(1,0){1}}
\put(55.177,16.01){\line(1,0){1}}
\put(55.177,16.01){\line(1,0){1}}
\put(56.177,16.01){\line(1,0){1}}
\put(56.177,16.01){\line(1,0){1}}
\put(56.177,16.01){\line(1,0){1}}
\put(56.177,16.01){\line(1,0){1}}
\multiput(57.177,16.01)(-.1293333,-.0326667){15}{\line(-1,0){.1293333}}
\multiput(57.177,16.01)(-.1293333,-.0326667){15}{\line(-1,0){.1293333}}
\multiput(57.177,16.01)(-.1293333,-.0326667){15}{\line(-1,0){.1293333}}
\multiput(57.177,16.01)(-.1293333,-.0326667){15}{\line(-1,0){.1293333}}
\multiput(55.237,16.5)(.1293333,-.0326667){15}{\line(1,0){.1293333}}
\multiput(55.237,16.5)(.1293333,-.0326667){15}{\line(1,0){.1293333}}
\multiput(55.237,16.5)(.1293333,-.0326667){15}{\line(1,0){.1293333}}
\multiput(55.237,16.5)(.1293333,-.0326667){15}{\line(1,0){.1293333}}
\put(56.177,11.01){\circle{10}} \put(56.177,11.01){\circle{10}}
\put(56.177,11.01){\circle{10}} \put(56.177,11.01){\circle{10}}
\put(56.114,14.094){$q$} \put(56.114,14.094){$q$}
\put(56.114,14.094){$q$} \put(56.114,14.094){$q$}
\put(51,10.96){\line(-1,0){6.894}}
\multiput(48.879,11.049)(-.0828641,.0331457){16}{\line(-1,0){.0828641}}
\multiput(48.879,10.96)(-.0773399,-.0331457){16}{\line(-1,0){.0773399}}
\put(51.177,10.96){\line(1,0){8.927}}
\put(60.104,10.96){\line(0,1){0}}
\put(60.104,10.96){\line(1,0){.088}}
\multiput(55.331,11.137)(.112494,.032141){11}{\line(1,0){.112494}}
\multiput(55.42,11.049)(.088388,-.029463){6}{\line(1,0){.088388}}
\put(55.95,10.872){\line(0,1){0}}
\multiput(55.42,10.96)(.0568211,-.0315673){14}{\line(1,0){.0568211}}
\put(56.215,10.518){\line(0,1){0}}
\put(61.96,10.96){\line(1,0){4.773}} \put(46.757,12.021){$p$}
\put(65.231,11.756){$-p$} \put(20.683,2.21){(a)}
\put(55.95,2.21){(b)}
\end{picture}
\end{center}
\caption[The planar and non-planar one-loop two-point diagrams.]
{The planar (a) and non-planar (b) one-loop two-point diagrams.}
\label{TPD}
\end{figure}
We have
\begin{equation}
\tilde{G}_{\mathrm{P}}^{(2)}=-i\frac{g}{3}\int\frac{\mathrm{d}^{3}q}{(2\pi)^{3}}
\frac{1}{(p^{2}-m^{2})^{2}(q^{2}-m^{2})}
\end{equation}
which is the same as the ordinary one. Consider now the non-planar case in f\mbox{}igure~\ref{TPD}(b).
The structure is the same as in the planar case, but this time the assignments are
\begin{equation}\label{NPA}
k_{1}=-k_{3}=p\quad\mathrm{and}\quad k_{2}=-k_{4}=q.
\end{equation}
We have
\begin{equation}
\tilde{G}_{\mathrm{NP}}^{(2)}=-i\frac{g}{6}\int\frac{\mathrm{d}^{3}q}{(2\pi)^{3}}
\frac{e^{-i\theta\boldsymbol{p}\wedge\boldsymbol{q}}}{(p^{2}-m^{2})^{2}(q^{2}-m^{2})}.
\end{equation}
Therefore, the $q$-contribution does not cancel completely
and the oscillating factor in the integral softens the ultraviolet divergence
because it dampens the functions for high $q$. However, it is responsible for the infrared divergence.
Notice that the persistence of some divergences is more general than the present calculation
and was noted in~\cite{Gracia-BondiaVarilly} in the general framework of Connes' non-commutative
geometry\footnote{See also~\cite{MartinGracia-BondiaVarilly}.},
while in~\cite{ChaichianDemichevPresnajder} it is shown that not all divergences can be
eliminated in the presence of the commutation relation~\eqref{CNC}.

Finally, to get the one-loop correction to the four-point Green's function,
consider f\mbox{}irst the planar case of f\mbox{}igure~\ref{PFPD}.
The one-loop correction to the four-point Green's function
can easily be calculated by properly joining two vertices \eqref{MV}. We have
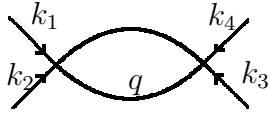
\begin{figure}[h]
\begin{center}
\unitlength 1.5mm
\linethickness{1pt}
\ifx\plotpoint\undefined\newsavebox{\plotpoint}\fi
\begin{picture}(28.838,17.169)(0,0)
\qbezier(11.966,10.852)(18.433,17.169)(24.899,11)
\qbezier(12.041,10.777)(18.581,4.905)(24.825,10.926)
\multiput(11.892,10.852)(-.033656805,.035760355){106}{\line(0,1){.035760355}}
\multiput(8.324,14.642)(-.0371627,.0318538){14}{\line(-1,0){.0371627}}
\put(7.804,15.088){\line(0,1){0}}
\multiput(11.966,10.926)(-.033545112,-.03420286){113}{\line(0,-1){.03420286}}
\multiput(24.899,11.074)(.034860607,.033545112){113}{\line(1,0){.034860607}}
\multiput(24.899,11)(.033668791,-.033668791){117}{\line(1,0){.033668791}}
\put(9.737,14.493){$k_1$} \put(7.581,9.068){$k_2$}
\put(28.169,9.291){$k_3$} \put(25.271,14.419){$k_4$}
\put(18.284,8.696){$q$} \put(10.108,11.966){\line(1,0){.818}}
\put(10.926,11.966){\line(0,1){.743}}
\put(10.926,12.71){\line(0,-1){.074}}
\put(10.108,9.885){\line(1,0){.743}}
\put(10.852,9.885){\line(0,-1){.595}}
\put(26.014,12.189){\line(1,0){.743}}
\put(26.014,9.142){\line(0,1){.818}}
\put(26.014,9.96){\line(1,0){.595}}
\put(26.014,12.189){\line(0,1){.372}}
\put(26.014,12.264){\line(0,1){.446}}
\end{picture}
\end{center}
\caption{The planar one-loop four-point diagram.}
\label{PFPD}
\end{figure}
\begin{equation}
\tilde{G}^{(4)}_{\mathrm{P}}=\frac{(-ig)^{2}}{8}(2\pi)^{3}\int\frac{\mathrm{d}^{3}q}{(2\pi)^{3}}
\frac{e^{\sum_{a<b}-\frac{i}{2}\theta\boldsymbol{k}_{a}\wedge\boldsymbol{k}_{b}}
\delta^{(3)}\!\left(\sum_{a=1}^{4}k_{a}\right)}
{(q^{2}-m^{2})\left[(k_{1}+k_{2}-q)^{2}-m^{2}\right]\prod_{a=1}^{4}(k_{a}^{2}-m^{2})}.
\end{equation}
Therefore, the internal momentum $q$ appears only in the denominator
so that also in this case the planar diagram has the same ultraviolet behaviour as the ordinary one.
Consider now the non-planar diagrams shown in f\mbox{}igure~\ref{NPFPD}.
\begin{figure}[h]
\begin{center}
\unitlength 1.5mm
\linethickness{1pt}
\ifx\plotpoint\undefined\newsavebox{\plotpoint}\fi
\begin{picture}(83.917,18.959)(0,0)
\qbezier(10.034,10.777)(16.575,4.905)(22.818,10.926)
\multiput(9.885,10.852)(-.033650943,.035754717){106}{\line(0,1){.035754717}}
\multiput(39.16,10.826)(-.033650943,.035764151){106}{\line(0,1){.035764151}}
\multiput(64.589,10.852)(-.033660377,.035754717){106}{\line(0,1){.035754717}}
\multiput(6.318,14.642)(-.0372143,.0318571){14}{\line(-1,0){.0372143}}
\multiput(35.593,14.617)(-.0372143,.0318571){14}{\line(-1,0){.0372143}}
\multiput(61.021,14.642)(-.0371429,.0318571){14}{\line(-1,0){.0371429}}
\put(5.797,15.088){\line(0,1){0}}
\put(35.072,15.063){\line(0,1){0}}
\put(60.501,15.088){\line(0,1){0}}
\multiput(9.96,10.926)(-.033548673,-.03420354){113}{\line(0,-1){.03420354}}
\multiput(64.663,10.926)(-.033539823,-.03420354){113}{\line(0,-1){.03420354}}
\multiput(22.892,11)(.033666667,-.033666667){117}{\line(1,0){.033666667}}
\multiput(52.167,10.975)(.033666667,-.033675214){117}{\line(0,-1){.033675214}}
\put(7.73,14.493){$k_1$}
\put(37.005,14.468){$k_1$}
\put(62.433,14.493){$k_1$}
\put(3.806,9.156){$k_2$}
\put(34.496,9.042){$k_2$}
\put(59.924,9.068){$k_2$}
\put(26.163,9.291){$k_3$}
\put(55.438,9.265){$k_3$}
\put(23.264,14.419){$k_4$}
\put(52.539,14.394){$k_4$}
\put(16.277,8.696){$q$}
\put(45.552,8.671){$q$}
\put(8.101,11.966){\line(1,0){.818}}
\put(37.376,11.941){\line(1,0){.818}}
\put(62.805,11.966){\line(1,0){.818}}
\put(8.919,11.966){\line(0,1){.743}}
\put(38.194,11.941){\line(0,1){.743}}
\put(63.623,11.966){\line(0,1){.743}}
\put(8.919,12.71){\line(0,-1){.074}}
\put(38.194,12.684){\line(0,-1){.074}}
\put(63.623,12.71){\line(0,-1){.074}}
\put(8.101,9.885){\line(1,0){.743}}
\put(62.805,9.885){\line(1,0){.743}}
\put(8.845,9.885){\line(0,-1){.595}}
\put(63.548,9.885){\line(0,-1){.595}}
\put(24.007,9.142){\line(0,1){.818}}
\put(53.282,9.117){\line(0,1){.818}}
\put(24.007,9.96){\line(1,0){.595}}
\put(53.282,9.934){\line(1,0){.595}}
\qbezier(9.96,10.852)(36.234,14.977)(22.818,10.926)
\qbezier(39.235,10.826)(65.509,14.951)(52.093,10.9)
\multiput(22.818,10.926)(.03361905,.03716667){42}{\line(0,1){.03716667}}
\multiput(52.093,10.9)(.03361905,.03716667){42}{\line(0,1){.03716667}}
\put(25.122,13.453){\line(1,0){.595}}
\put(54.397,13.427){\line(1,0){.595}}
\put(25.122,13.527){\line(0,1){.52}}
\put(54.397,13.502){\line(0,1){.52}}
\qbezier(52.213,10.912)(29.806,7.193)(39.275,10.85)
\put(39.275,10.85){\line(-1,0){.125}}
\multiput(39.15,10.85)(-.03289474,-.03947368){38}{\line(0,-1){.03947368}}
\multiput(37.588,8.725)(-.03341379,-.03556897){58}{\line(0,-1){.03556897}}
\put(36.4,8.412){\line(1,0){.875}}
\put(37.275,8.412){\line(0,-1){.625}}
\multiput(24.875,13.188)(.03333333,.04026667){45}{\line(0,1){.04026667}}
\multiput(54.088,13.1)(.03355556,.03703704){54}{\line(0,1){.03703704}}
\qbezier(64.7,10.872)(68.987,18.959)(73.097,10.96)
\qbezier(73.186,10.872)(77.384,5.171)(79.991,10.96)
\qbezier(73.362,11.579)(77.119,18.429)(79.991,10.96)
\multiput(79.977,11.187)(.033675214,-.033666667){117}{\line(1,0){.033675214}}
\put(83.248,9.478){$k_3$}
\put(80.349,14.606){$k_4$}
\put(81.092,9.329){\line(0,1){.818}}
\put(81.092,10.147){\line(1,0){.595}}
\multiput(79.903,11.113)(.03361905,.03716667){42}{\line(0,1){.03716667}}
\put(82.207,13.714){\line(0,1){.52}}
\multiput(81.898,13.312)(.03355556,.03703704){54}{\line(0,1){.03703704}}
\qbezier(64.612,10.872)(69.208,5.171)(72.39,10.607)
\put(16.529,4.95){(1)}
\put(45.874,5.038){(2)}
\put(73.186,5.215){(3)}
\end{picture}
\end{center}
\caption{The non-planar one-loop four-point diagrams.}
\label{NPFPD}
\end{figure}
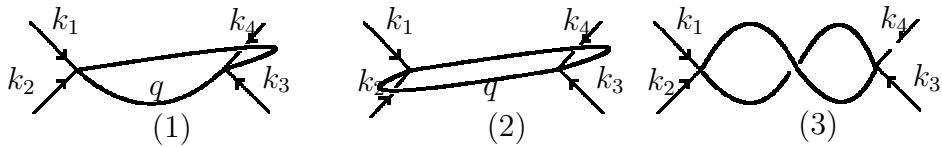
We have
\begin{equation}
\tilde{G}^{(4)}_{\mathrm{NP}_{a}}=\frac{(-ig)^{2}}{8}(2\pi)^{3}\int\frac{\mathrm{d}^{3}q}{(2\pi)^{3}}
\frac{e^{\sum_{a<b}-\frac{i}{2}\theta\boldsymbol{k}_{a}\wedge\boldsymbol{k}_{b}+E_{a}}
\delta^{(3)}\!\left(\sum_{a=1}^{4}k_{a}\right)}
{(q^{2}-m^{2})\left[(k_{1}+k_{2}-q)^{2}-m^{2}\right]\prod_{a=1}^{4}(k_{a}^{2}-m^{2})}
\end{equation}
with
\begin{align}\label{NPFPT}
\nonumber
E_{1}&=i\theta\boldsymbol{k}_{1}\wedge\boldsymbol{q}\\
\nonumber
E_{2}&=-i\theta\left(\boldsymbol{k}_{2}\wedge\boldsymbol{q}+
\boldsymbol{k}_{3}\wedge\boldsymbol{q}\right)\\
E_{3}&=-i\theta\left(\boldsymbol{k}_{1}\wedge\boldsymbol{q}+
\boldsymbol{k}_{2}\wedge\boldsymbol{q}\right).
\end{align}
So in non-planar cases the one-loop correction to the two- and four-point Green's function
have the same oscillating factor and then in both cases hold similar considerations.
In the next chapter we will introduce the Wick-Voros product that is,
a variant of the more studied Moyal product and compare the Wick-Voros f\mbox{}ield theory with the Moyal one.

\chapter{Non-commutative Wick-Voros f\mbox{}ield theory}

\emph{In this chapter we describe another non-commutative f\mbox{}ield theory
obtained from the same commutative one of the previous chapter
replacing the ordinary product with the Wick-Voros one.
We show that the free case is the same as the Moyal one (and the ordinary one),
while the interacting case is dif\mbox{}ferent and, in fact, we f\mbox{}ind dif\mbox{}ferent Green's functions.
However, the interacting case present the same kind of ultraviolet/infrared mixing as the Moyal one.}

\section{The Wick-Voros product}

Let us consider the dif\mbox{}ferential form of Wick-Voros product which is given by
\begin{equation}
f\star_{V}g=f e^{\frac{i}{2}\theta
\left[\stackrel{\leftarrow}{\partial_{1}}\stackrel{\rightarrow}{\partial}_{2}-
\stackrel{\leftarrow}{\partial_{2}}\stackrel{\rightarrow}{\partial}_{1}
-i\left(\stackrel{\leftarrow}{\partial_{1}}\stackrel{\rightarrow}{\partial}_{1}+
\stackrel{\leftarrow}{\partial_{2}}\stackrel{\rightarrow}{\partial}_{2}\right)\right]}g
\end{equation}
where $\theta$ is still a real constant parameter of dimensions of a square length.
In this case is more natural working with complex coordinates than real ones and
in complex coordinates \eqref{ComplexCoordinates} the Wick-Voros product takes the form
\begin{equation}\label{WVP}
f\star_{V}g=f e^{\theta\stackrel{\leftarrow}{\partial_{+}}
\stackrel{\rightarrow}{\partial}_{-}}g
\end{equation}
where
\begin{equation}
\partial_{\pm}=\frac{\partial}{\partial x^{\pm}}=
\frac{1}{\sqrt{2}}\left(\frac{\partial}{\partial x^{1}}\mp
i\frac{\partial}{\partial x^{2}} \right)=
\frac{1}{\sqrt{2}}(\partial_{1}\mp i\partial_{2}).
\end{equation}
Note that the Moyal product \eqref{MP} can be rewritten in these coordinates as
\begin{equation}
f\star_{M}g=f
e^{\frac{\theta}{2}\left(\stackrel{\leftarrow}{\partial_{+}}
\stackrel{\rightarrow}{\partial}_{-}-\stackrel{\leftarrow}{\partial_{-}}
\stackrel{\rightarrow}{\partial}_{+}\right)}g.
\end{equation}
Moreover, the Laplacian is
\begin{equation}
\nabla^{2}=2\partial_{+}\partial_{-}
\end{equation}
and the d'Alembertian is as usual
\begin{equation}
\Box=\partial_{0}^{2}-\nabla^{2}.
\end{equation}
The Wick-Voros product is associative and non-commutative. In particular, we have
\begin{align}
x^{+}\star_{V}x^{-}&=x^{+}x^{-}+\theta\\
x^{-}\star_{V}x^{+}&=x^{+}x^{-}.
\end{align}
Hence the Wick-Voros bracket of $x^{+}$ and $x^{-}$ reads
\begin{equation}
[x^{+},x^{-}]_{\star_{V}}=x^{+}\star_{V}x^{-}-x^{-}\star_{V}x^{+}=\theta
\end{equation}
which gives the canonical non-commutative relation \eqref{CNC} going back to the $x$'s.
It is very easy to see that the Wick-Voros bracket, like the Moyal one, of two functions:
\begin{equation}
[f,g]_{\star_{V}}=f\star_{V} g-g\star_{V} f
\end{equation}
is proportional to the Poisson bracket of the two functions to f\mbox{}irst order in $\theta$:
\begin{equation}
[f,g]_{\star_{V}}=i\theta\{f,g\}+\ldots
\end{equation}
Therefore, both products are a deformation of the ordinary one
which reduce to the ordinary one in the limit $\theta\to0$
and reproduce to f\mbox{}irst order in the deformation parameter $\theta$ the Poisson strutture.

Even in the Wick-Voros case, it is very useful to write the product in momentum space.
Since in complex coordinates the Fourier transform can be expressed as
\begin{equation}
f(x)=\int\frac{\mathrm{d}^{3}p}{(2\pi)^{3}}
\tilde{f}(p)e^{-i\left(p_{+}x^{-}+p_{-}x^{+}\right)}
\end{equation}
with
\begin{equation}
p_{\pm}=\frac{p_{1}\pm ip_{2}}{\sqrt{2}}
\end{equation}
from \eqref{WVP} we have
\begin{align}\label{wvp}
\nonumber
(f\star_{V}g)(x)&=\int\frac{\mathrm{d}^{3}p}{(2\pi)^{3}}\frac{\mathrm{d}^{3}q}{(2\pi)^{3}}
\tilde{f}(p)\tilde{g}(q)e^{-\theta p_{-}q_{+}}e^{i(p+q)\cdot x}\\
&=\int\frac{\mathrm{d}^{3}p}{(2\pi)^{3}}\frac{\mathrm{d}^{3}q}{(2\pi)^{3}}
\tilde{f}(p)\tilde{g}(q)e^{-\frac{\theta}{2}(\boldsymbol{p}\cdot\boldsymbol{q}
+i\boldsymbol{p}\wedge\boldsymbol{q})}e^{i(p+q)\cdot x}
\end{align}
since\footnote{
In complex coordinates
\begin{equation*}
\boldsymbol{p}\cdot\boldsymbol{q}=p_{-}q_{+}+q_{-}p_{+}.
\end{equation*}}
\begin{equation}
p_{-}q_{+}=\frac{1}{2}\left(\boldsymbol{p}\cdot\boldsymbol{q}+
i\boldsymbol{p}\wedge\boldsymbol{q}\right).
\end{equation}
So the Wick-Voros product in momentum space
is the standard convolution of Fourier transforms twisted by a factor
which is not just a phase like in the Moyal case, but a phase multiplied by a real exponential.
Furthermore, from \eqref{wvp} follows that
\begin{align}\label{iwvp}
\nonumber
\int\mathrm{d}^{3}x\,f\star_{V}g&=
\int\mathrm{d}^{3}x\,\frac{\mathrm{d}^{3}p}{(2\pi)^{3}}\frac{\mathrm{d}^{3}q}{(2\pi)^{3}}
\tilde{f}(p)\tilde{g}(q)e^{-\theta p_{-}q_{+}}e^{i(p+q)\cdot x}\\
&=\int\frac{\mathrm{d}^{3}p}{(2\pi)^{3}}\tilde{f}(p)\tilde{g}(-p)e^{\theta p_{-}p_{+}}=
\int\frac{\mathrm{d}^{3}p}{(2\pi)^{3}}\tilde{f}(p)\tilde{g}(-p)e^{\theta\boldsymbol{p}^{2}}.
\end{align}
Therefore, unlike for the Moyal case, the integral of the Wick-Voros product of two functions
is not equal to the integral of the ordinary product of the two functions
\begin{equation}
\int\mathrm{d}^{3}x\,f\star_{V}g\neq\int\mathrm{d}^{3}x\,fg.
\end{equation}
This has a precise interpretation in non-commutative f\mbox{}ield theory.
It means that also the free non-commutative f\mbox{}ield theory
with the Wick-Voros product could be dif\mbox{}ferent from the ordinary one.
However, it has the trace property
\begin{equation}
\int\mathrm{d}^{3}x\,f\star_{V}g=\int\mathrm{d}^{3}x\,g\star_{V}f.
\end{equation}
Notice that the integral of the Wick-Voros product of two functions can be written as well as
\begin{multline}\label{IWVP}
\int\mathrm{d}^{3}x\,f\star_{V}g=
\int\mathrm{d}^{3}x\,\sum_{n=0}^{\infty}\frac{\theta^{n}}{n!}(\partial_{+}^{n}f)(\partial_{-}^{n}g)=
\int\mathrm{d}^{3}x\,\sum_{n=0}^{\infty}\frac{(-\theta)^{n}}{n!}f\partial_{+}^{n}\partial_{-}^{n}g=\\
\int\mathrm{d}^{3}x\,f e^{-\theta\partial_{+}\partial_{-}}g=
\int\mathrm{d}^{3}x\,f e^{-\frac{\theta}{2}\nabla^{2}}g
\end{multline}
where we have integrated by parts and neglected all the boundary terms.

We conclude this section by noting that at the algebraic level the Moyal and Wick-Voros products
are equivalent in sense that they def\mbox{}ine the same deformed algebra,
namely there exists an invertible map~\cite{Zachos,AlexanianPinzulStern} $T$ such that
\begin{equation}\label{ER}
T(f\star_{M}g)=T(f)\star_{V}T(g)
\end{equation}
where
\begin{equation}\label{EM}
T=e^{\frac{\theta}{4}\nabla^{2}}.
\end{equation}
Note that from invertibility of $T$ follows that
\begin{equation}
T^{-1}(f\star_{V}g)=T^{-1}(f)\star_{M}T^{-1}(g).
\end{equation}
We can easily show that the Moyal and Wick-Voros products are algebraically equivalent.
In fact, by using \eqref{wvp} the left-hand side of \eqref{ER} can be written as
\begin{align}
\nonumber
T(f\star_{M}g)&=
e^{\frac{\theta}{4}\nabla^{2}}\int\frac{\mathrm{d}^{3}p}{(2\pi)^{3}}\frac{\mathrm{d}^{3}q}{(2\pi)^{3}}
\tilde{f}(p)\tilde{g}(q)e^{-\frac{i}{2}\theta\boldsymbol{p}\wedge\boldsymbol{q}}e^{i(p+q)\cdot x}\\
&=\int\frac{\mathrm{d}^{3}p}{(2\pi)^{3}}\frac{\mathrm{d}^{3}q}{(2\pi)^{3}}
\tilde{f}(p)\tilde{g}(q)e^{-\frac{\theta}{4}(\boldsymbol{p}+\boldsymbol{q})^{2}}
e^{-\frac{i}{2}\theta\boldsymbol{p}\wedge\boldsymbol{q}}
e^{i(p+q)\cdot x}
\end{align}
and the right-hand side of \eqref{ER} can be written in the same way
\begin{align}
\nonumber
T(f)\star_{V}T(g)&=
e^{\frac{\theta}{4}\nabla^{2}}\int\frac{\mathrm{d}^{3}p}{(2\pi)^{3}}\tilde{f}(p)e^{-ip\cdot x}\star_{V}
e^{\frac{\theta}{4}\nabla^{2}}\int\frac{\mathrm{d}^{3}q}{(2\pi)^{3}}\tilde{f}(q)e^{-iq\cdot x}\\
\nonumber
&=\int\frac{\mathrm{d}^{3}p}{(2\pi)^{3}}\frac{\mathrm{d}^{3}q}{(2\pi)^{3}}
\tilde{f}(p)\tilde{g}(q)e^{-\frac{\theta}{4}(\boldsymbol{p}^{2}+\boldsymbol{q}^{2})}
e^{-ip\cdot x}\star_{V}e^{-iq\cdot x}\\
\nonumber
&=\int\frac{\mathrm{d}^{3}p}{(2\pi)^{3}}\frac{\mathrm{d}^{3}q}{(2\pi)^{3}}
\tilde{f}(p)\tilde{g}(q)e^{-\frac{\theta}{4}(\boldsymbol{p}^{2}+\boldsymbol{q}^{2})}
e^{-\frac{\theta}{2}(\boldsymbol{p}\cdot\boldsymbol{q}
+i\boldsymbol{p}\wedge\boldsymbol{q})}e^{i(p+q)\cdot x}\\
&=\int\frac{\mathrm{d}^{3}p}{(2\pi)^{3}}\frac{\mathrm{d}^{3}q}{(2\pi)^{3}}
\tilde{f}(p)\tilde{g}(q)e^{-\frac{\theta}{4}(\boldsymbol{p}+\boldsymbol{q})^{2}}
e^{-\frac{i}{2}\theta\boldsymbol{p}\wedge\boldsymbol{q}}e^{i(p+q)\cdot x}
\end{align}
since from \eqref{wvp} follows immediately that the Wick-Voros product of the two exponentials
$e^{-ip\cdot x}$ and $e^{-iq\cdot x}$ is given by
\begin{align}
\nonumber
e^{-ip\cdot x}\star_{V}e^{-iq\cdot x}&=\int\mathrm{d}^{3}r\,\mathrm{d}^{3}s\,
\delta^{(3)}(r-p)\delta^{(3)}(s-q)e^{-\frac{\theta}{2}(\boldsymbol{r}\cdot\boldsymbol{s}
+i\boldsymbol{r}\wedge\boldsymbol{s})}e^{i(r+s)\cdot x}\\
&=e^{-\frac{\theta}{2}(\boldsymbol{p}\cdot\boldsymbol{q}
+i\boldsymbol{p}\wedge\boldsymbol{q})}e^{i(p+q)\cdot x}.
\end{align}

\section{The Wick-Voros classical f\mbox{}ield theory}

We now study a classical f\mbox{}ield theory with the Wick-Voros product and
we describe both the Lagrangian and the Hamiltonian formalisms.

\subsection{Lagrangian formalism}

At this point we proceed to the discussion of another non-commutative f\mbox{}ield theory
obtained from the commutative one described by the action \eqref{OA}
substituting the ordinary product with the Wick-Voros one.
So the free non-commutative action (as well as the Lagrangian and the Lagrangian density) is given by
\begin{equation}\label{FWVA}
S_{0_{V}}=\int\mathrm{d}t\,L_{0_{V}}=\int\mathrm{d}^{3}x\,\mathcal{L}_{0_{V}}
=\int\mathrm{d}^{3}x\,\frac{1}{2}\left(\partial_{\mu}\phi\star_{V}\partial^{\mu}\phi
-m^{2}\phi\star_{V}\phi\right)
\end{equation}
which, unlike the Moyal one, does not reduce to the commutative one
and the interacting non-commutative action is given by
\begin{equation}\label{WVIA}
S_{\mathrm{int}_{V}}=\frac{g}{4!}\int\mathrm{d}^{3}x\,
\phi\star_{V}\phi\star_{V}\phi\star_{V}\phi.
\end{equation}
As we have already seen, the Moyal product and the Wick-Voros one are algebraically equivalent.
However, this does not mean that any deformation of an action with the two products are the same.
Indeed, mapping the free action with the Wick-Voros product \eqref{FWVA}
to the corresponding action with the Moyal one by means of \eqref{EM}
\begin{equation}
S_{0_{V}}\to\int\mathrm{d}^{3}x\,T^{-1}\mathcal{L}_{0_{V}}
\end{equation}
we have
\begin{align}
\nonumber
S_{0_{V}}&\to\int\mathrm{d}^{3}x\,
\frac{1}{2}\left[\left(e^{-\frac{\theta}{4}\nabla^{2}}\partial_{\mu}\phi\right)\star_{M}
\left(e^{-\frac{\theta}{4}\nabla^{2}}\partial^{\mu}\phi\right)
-m^{2}\left(e^{-\frac{\theta}{4}\nabla^{2}}\phi\right)\star_{M}
\left(e^{-\frac{\theta}{4}\nabla^{2}}\phi\right)\right]\\
\nonumber
&=\int\mathrm{d}^{3}x\,
\frac{1}{2}\left[\left(e^{-\frac{\theta}{4}\nabla^{2}}\partial_{\mu}\phi\right)
\left(e^{-\frac{\theta}{4}\nabla^{2}}\partial^{\mu}\phi\right)
-m^{2}\left(e^{-\frac{\theta}{4}\nabla^{2}}\phi\right)
\left(e^{-\frac{\theta}{4}\nabla^{2}}\phi\right)\right]\\
&=\int\mathrm{d}^{3}x\,
\frac{1}{2}\left(\partial_{\mu}\phi\,
e^{-\frac{\theta}{2}\nabla^{2}}\partial^{\mu}\phi
-m^{2}\phi\,e^{-\frac{\theta}{2}\nabla^{2}}\phi\right)
\end{align}
which is not the free action with the Moyal product namely the ordinary one.

We begin with the discussion of the free case
since the free action \eqref{FWVA} is dif\mbox{}ferent from the ordinary one.
As well-known, the dynamical behaviour of a dynamical system
is determined by the action principle
which af\mbox{}f\mbox{}irms that the equation of motion, namely the f\mbox{}ield equation,
is obtained demanding that the variation of the action is vanishing
under any inf\mbox{}initesimal variation of the f\mbox{}ield.
So given an inf\mbox{}initesimal variation of $\phi$:
\begin{equation}
\phi\to\phi+\delta\phi
\end{equation}
the corresponding inf\mbox{}initesimal variation of the action $S_{0_{V}}$ is given by
\begin{equation}
\delta S_{0_{V}}=\int\mathrm{d}^{3}x\,
\left(\partial_{\mu}\phi\star_{V}\partial^{\mu}\delta\phi-m^{2}\phi\star_{V}\delta\phi\right).
\end{equation}
By integrating by parts we obtain, up to boundary terms,
\begin{equation}
\delta S_{0_{V}}=-\int\mathrm{d}^{3}x\,\delta\phi\star_{V}\left(\Box+m^{2}\right)\phi
\end{equation}
and using the relation \eqref{IWVP} we have
\begin{equation}
\delta S_{0_{V}}=-\int\mathrm{d}^{3}x\,
\delta\phi\,e^{-\frac{\theta}{2}\nabla^{2}}\left(\Box+m^{2}\right)\phi.
\end{equation}
Since the variation of the action $\delta S_{0}$ has to vanish
for any variation of the f\mbox{}ield $\delta\phi$, we get the equation of motion
\begin{equation}\label{WVEOM}
e^{-\frac{\theta}{2}\nabla^{2}}\left(\Box+m^{2}\right)\phi=0
\end{equation}
which is dif\mbox{}ferent from the ordinary Klein-Gordon equation given by
\begin{equation}
\left(\Box+m^{2}\right)\phi=0.
\end{equation}
However, the two equations of motion have exactly the same solutions due to
the invertibility of the operator $e^{-\frac{\theta}{2}\nabla^{2}}$.
Moreover, the on shell condition is not deformed. That is, the
dispersion relation is the same as the ordinary one.
In fact, in Fourier transform the equation of motion \eqref{WVEOM} becomes
\begin{align}
\nonumber
e^{-\frac{\theta}{2}\nabla^{2}}\left(\Box+m^{2}\right)\phi(x)&=
e^{-\frac{\theta}{2}\nabla^{2}}\left(\Box+m^{2}\right)
\int\frac{\mathrm{d}^{3}k}{(2\pi)^{3}}\tilde{\phi}(k)e^{-ik\cdot x}\\
%\nonumber
%&=e^{-\frac{\theta}{2}\nabla^{2}}\int\frac{\mathrm{d}^{3}k}{(2\pi)^{3}}
%\left(-k^{2}+m^{2}\right)\tilde{\phi}(k)e^{-ik\cdot x}\\
&=\int\frac{\mathrm{d}^{3}k}{(2\pi)^{3}}
e^{\frac{\theta}{2}\boldsymbol{k}^{2}}\left(-k^{2}+m^{2}\right)
\tilde{\phi}(k)e^{-ik\cdot x}=0.
\end{align}
Hence on shell condition is given by
\begin{equation}
e^{\frac{\theta}{2}\boldsymbol{k}^{2}}\left(k^{2}-m^{2}\right)\tilde{\phi}(k)=0
\end{equation}
which reduces to the ordinary one
\begin{equation}
\left(k^{2}-m^{2}\right)\tilde{\phi}(k)=0
\end{equation}
because the exponential never vanishes. Therefore, at classical level the free non-commutative f\mbox{}ield theory
with the Wick-Voros product, like that with the Moyal product, is the same as the commutative one.

\subsection{Hamiltonian formalism}

Before proceeding further in our analysis, we investigate the Hamiltonian formalism in the free case.
To begin with, let us f\mbox{}ind the f\mbox{}ield conjugate to $\phi$. As well-known, it is def\mbox{}ined by
\begin{equation}
\pi=\frac{\delta L_{0_{V}}}{\delta\dot{\phi}}
\end{equation}
where we have set $\dot{\phi}=\partial_{0}\phi$.
To determine the functional derivative of the Lagrangian $L_{0_{V}}$
with respect $\dot{\phi}$, consider an inf\mbox{}initesimal variation of $\dot{\phi}$:
\begin{equation}
\dot{\phi}\to\dot{\phi}+\delta\dot{\phi}.
\end{equation}
The corresponding inf\mbox{}initesimal variation of the Lagrangian $L_{0_{V}}$ is given by
\begin{equation}
\delta L_{0_{V}}=\int\mathrm{d}^{2}x\,\delta\dot{\phi}\star_{V}\dot{\phi}
\end{equation}
and using the relation \eqref{IWVP} we have
\begin{equation}
\delta L_{0_{V}}=\int\mathrm{d}^{2}x\,
\delta\dot{\phi}\,e^{-\frac{\theta}{2}\nabla^{2}}\dot{\phi}.
\end{equation}
So the f\mbox{}ield conjugate to $\phi$ is given by
\begin{equation}\label{FC}
\pi=e^{-\frac{\theta}{2}\nabla^{2}}\dot{\phi}.
\end{equation}
We now assume that the free non-commutative Hamiltonian in the Wick-Voros case is given by
\begin{equation}
H_{0_{V}}=\int\mathrm{d}^{2}x\,\mathcal{H}_{0_{V}}=
\int\mathrm{d}^{2}x\,\left(\pi\dot{\phi}-\mathcal{L}_{0_{V}}\right)
\end{equation}
expressing $\dot{\phi}$ as a function of the conjugate f\mbox{}ield $\pi$.
The choice of Hamiltonian can be justif\mbox{}ied thinking of our theory as just as a theory
with inf\mbox{}initely many numbers of derivatives without any consideration
about non-commutative geometry. Therefore, by using the relation \eqref{IWVP}
the free Hamiltonian can be written as
\begin{align}
\nonumber
%
%H_{0_{V}}&=\int\mathrm{d}^{2}x\left\{\pi e^{\frac{\theta}{2}\nabla^{2}}\pi
%-\frac{1}{2}\left[\left(e^{\frac{\theta}{2}\nabla^{2}}\pi\right)\star_{V}
%\left(e^{\frac{\theta}{2}\nabla^{2}}\pi\right)+
%\partial_{i}\phi\star_{V}\partial^{i}\phi-m^{2}\phi\star_{V}\phi\right]\right\}\\
%
H_{0_{V}}&=\int\mathrm{d}^{2}x\left[\pi e^{\frac{\theta}{2}\nabla^{2}}\pi
-\frac{1}{2}\left(e^{\frac{\theta}{2}\nabla^{2}}\pi\star_{V}e^{\frac{\theta}{2}\nabla^{2}}\pi+
\partial_{i}\phi\star_{V}\partial^{i}\phi-m^{2}\phi\star_{V}\phi\right)\right]\\
&=\int\mathrm{d}^{2}x\,\frac{1}{2}\left(\pi e^{\frac{\theta}{2}\nabla^{2}}\pi- \partial_{i}\phi\star_{V}\partial^{i}\phi+m^{2}\phi\star_{V}\phi\right)
\end{align}
which is dif\mbox{}ferent from the one that we would write
if we started directly from Hamiltonian formalism rather Lagrangian one%
\footnote{Indeed, in such a case we would write the free non-commutative Hamiltonian as
\begin{equation*}
H_{0_{V}}=\int\mathrm{d}^{2}x\,\frac{1}{2}\left(\pi\star_{V}\pi-
\partial_{i}\phi\star_{V}\partial^{i}\phi+m^{2}\phi\star_{V}\phi\right)
\end{equation*}
where the f\mbox{}ield conjugate to $\phi$ is easily seen to be
$\pi=e^{\frac{\theta}{2}\nabla^{2}}\dot{\phi}$.}.
We now are ready to derive the Hamilton equations:
\begin{align}
\dot{\phi}&=\frac{\delta H}{\delta\pi}\\
\dot{\pi}&=-\frac{\delta H}{\delta\phi}.
\end{align}
To evaluate the time evolution of $\phi$, consider an inf\mbox{}initesimal variation of $\pi$:
\begin{equation}
\pi\to\pi+\delta\pi.
\end{equation}
The corresponding inf\mbox{}initesimal variation of the Hamiltonian is given by
\begin{equation}
\delta H_{0_{V}}=\int\mathrm{d}^{2}x\,
\frac{1}{2}\left(\pi\,e^{\frac{\theta}{2}\nabla^{2}}\delta\pi+
\delta\pi\,e^{\frac{\theta}{2}\nabla^{2}}\pi\right).
\end{equation}
That is,
\begin{equation}
\delta H_{0_{V}}=\int\mathrm{d}^{2}x\,\delta\pi\,e^{\frac{\theta}{2}\nabla^{2}}\pi
\end{equation}
where we have integrated by parts and neglected all the boundary terms.
So the time evolution of $\phi$ is given by
\begin{equation}\label{TEOphi}
\dot{\phi}=e^{\frac{\theta}{2}\nabla^{2}}\pi
\end{equation}
which is, of course, consistent with \eqref{FC}.
Instead to calculate the time evolution of $\pi$,
consider an inf\mbox{}initesimal variation of $\phi$:
\begin{equation}
\phi\to\phi+\delta\phi.
\end{equation}
The corresponding inf\mbox{}initesimal variation of the Hamiltonian is given by
\begin{equation}
\delta H_{0_{V}}=-\int\mathrm{d}^{2}x\,\left(\partial_{i}\phi\star_{V}\partial^{i}\delta\phi
-m^{2}\phi\star_{V}\delta\phi\right)
\end{equation}
By integrating by parts we get, up to boundary terms,
\begin{equation}
\delta H_{0_{V}}=-\int\mathrm{d}^{2}x\,\delta\phi\star_{V}\left(\nabla^{2}-m^{2}\right)\phi
\end{equation}
and using the relation \eqref{IWVP} we have
\begin{equation}
\delta H_{0_{V}}=-\int\mathrm{d}^{2}x\,
\delta\phi\,e^{-\frac{\theta}{2}\nabla^{2}}\left(\nabla^{2}-m^{2}\right)\phi.
\end{equation}
So the time evolution of $\pi$ is given by
\begin{equation}\label{TEOpi}
\dot{\pi}=e^{-\frac{\theta}{2}\nabla^{2}}
\left(\nabla^{2}-m^{2}\right)\phi.
\end{equation}
It is now easy to show that combining \eqref{TEOphi} with \eqref{TEOpi}
we obtain the ordinary equations of motion for $\phi$ and $\pi$ f\mbox{}ields.
This result conf\mbox{}irms that at classical level the free non-commutative f\mbox{}ield theory
with the Wick-Voros product is the same as the commutative one.

\section{The Wick-Voros quantum f\mbox{}ield theory}

We now move on to the quantum case and proceed to the determination of Green's functions in the Wick-Voros case.
To calculate the propagator, we can start from its general def\mbox{}inition:
\begin{equation}
e^{-\frac{\theta}{2}\nabla^{2}}\left(\Box+m^{2}\right)G_{V}^{(2)}(x-x')=-\delta^{(3)}(x-x').
\end{equation}
In Fourier transform it becomes
\begin{align}
\nonumber
e^{-\frac{\theta}{2}\nabla^{2}}\left(\Box+m^{2}\right)G_{V}^{(2)}(x-x')&=
e^{-\frac{\theta}{2}\nabla^{2}}\left(\Box+m^{2}\right)
\int\frac{\mathrm{d}^{3}p}{(2\pi)^{3}}\tilde{G}_{V}^{(2)}(p)e^{-ip\cdot(x-x')}\\
&=\int\frac{\mathrm{d}^{3}p}{(2\pi)^{3}}
\nonumber
e^{\frac{\theta}{2}\boldsymbol{p}^{2}}\left(-p^{2}+m^{2}\right)\tilde{G}_{V}^{(2)}(p)e^{-ip\cdot(x-x')}\\
&=-\int\frac{\mathrm{d}^{3}p}{(2\pi)^{3}}e^{-ip\cdot(x-x')}
\end{align}
from which follows that
\begin{equation}
e^{\frac{\theta}{2}\boldsymbol{p}^{2}}\left(-p^{2}+m^{2}\right)\tilde{G}_{V}^{(2)}(p)=-1
\end{equation}
and then
\begin{equation}\label{WVTPGF}
\tilde{G}_{V}^{(2)}(p)=\frac{e^{-\frac{\theta}{2}\boldsymbol{p}^{2}}}{p^{2}-m^{2}}.
\end{equation}
Note that to get the propagator, we can proceed as well as follows.
The free action \eqref{FWVA} can be written as
\begin{align}
\nonumber
S_{0_{V}}&=\int\mathrm{d}^{3}x\,\frac{1}{2}
\left(\partial_{\mu}\phi\,e^{-\frac{\theta}{2}\nabla^{2}}\partial_{\mu}\phi
-m^{2}\phi\,e^{-\frac{\theta}{2}\nabla^{2}}\phi\right)\\
&=\int\mathrm{d}^{3}x\,\frac{1}{2}\phi\,e^{-\frac{\theta}{2}\nabla^{2}}
\left(-\partial_{\mu}^{2}-m^{2}\right)\phi
\end{align}
where we have used \eqref{IWVP}, integrated by parts and neglected the boundary terms.
So it can be rewritten as
\begin{equation}
S_{0_{V}}=\int\mathrm{d}^{3}x\,\mathrm{d}^{3}x'\,\phi(x)K(x,x')\phi(x')
\end{equation}
with
\begin{equation}
K(x,x')=e^{-\frac{\theta}{2}\nabla^{2}}(-\partial_{\mu}^{2}-m^{2})\delta^{(3)}(x-x')
\end{equation}
or equivalently
\begin{align}
\nonumber
K(x,x')&=e^{-\frac{\theta}{2}\nabla^{2}}(-\partial_{\mu}^{2}-m^{2})
\int\frac{\mathrm{d}^{3}p}{(2\pi)^{3}}e^{-ip\cdot(x-x')}\\
&=\int\frac{\mathrm{d}^{3}p}{(2\pi)^{3}}e^{\frac{\theta}{2}\boldsymbol{p^{2}}}
\left(p^{2}-m^{2}\right)e^{-ip\cdot(x-x')}.
\end{align}
Therefore, the propagator, the inverse of the operator $K(x,x')$, is
\begin{equation}
G_{V}^{(2)}(x,x')=\int\frac{\mathrm{d}^{3}p}{(2\pi)^{3}}
\frac{e^{-\frac{\theta}{2}\boldsymbol{p^{2}}}}{p^{2}-m^{2}}e^{-ip\cdot(x-x')}.
\end{equation}
from which we can read of\mbox{}f the propagator in momentum space \eqref{WVTPGF}.
Since the poles in the propagator \eqref{WVTPGF} are the same as in the Moyal and ordinary cases \eqref{MTPGF},
the free f\mbox{}ield theory in the Wick-Voros case is the same as in the two cases at the quantum level as well.
Nevertheless, the two propagators are not identical. Moreover, for inf\mbox{}inite momentum there is
an essential singularity or a zero of the propagator according to the sign of $\theta$.
The meaning of the essential singularity is not clear,
but the oddity is that the sign of $\theta$ has no physical meaning since
it can be changed by the exchange of the two coordinates.

To calculate the four-point Green's function to the tree level in the Wick-Voros case,
we f\mbox{}irst must determine the vertex. To this end, let us write down the interacting action \eqref{WVIA}
in momentum space. By using the relations \eqref{wvp} and \eqref{iwvp} we have
\begin{multline}
S_{\mathrm{int}_{V}}=\frac{g}{4!}\int\mathrm{d}x\,
\frac{\mathrm{d}k_{1}}{(2\pi)^{3}}\frac{\mathrm{d}k_{2}}{(2\pi)^{3}}
\frac{\mathrm{d}k_{3}}{(2\pi)^{3}}\frac{\mathrm{d}k_{4}}{(2\pi)^{3}}
\tilde{\phi}(k_{1})\tilde{\phi}(k_{2})\tilde{\phi}(k_{3})\tilde{\phi}(k_{4})\\
e^{-\theta(k_{1-}k_{2+}+k_{3-}k_{4+})}e^{i(k_{1}+k_{2})\cdot x}\star_{V}e^{i(k_{3}+k_{4})\cdot x}\\
=\frac{g}{4!}(2\pi)^{3}\!\int\frac{\mathrm{d}k_{1}}{(2\pi)^{3}}\frac{\mathrm{d}k_{2}}{(2\pi)^{3}}
\frac{\mathrm{d}k_{3}}{(2\pi)^{3}}\frac{\mathrm{d}k_{4}}{(2\pi)^{3}}
\tilde{\phi}(k_{1})\tilde{\phi}(k_{2})\tilde{\phi}(k_{3})\tilde{\phi}(k_{4})\\
e^{-\theta(k_{1-}k_{2+}+k_{3-}k_{4+}-k_{-}k_{+})}\delta^{(3)}(k_{1}+k_{2}+k)\delta^{(3)}(k_{3}+k_{4}-k)\\
=\frac{g}{4!}(2\pi)^{3}\!\int\frac{\mathrm{d}k_{1}}{(2\pi)^{3}}\frac{\mathrm{d}k_{2}}{(2\pi)^{3}}
\frac{\mathrm{d}k_{3}}{(2\pi)^{3}}\frac{\mathrm{d}k_{4}}{(2\pi)^{3}}
\tilde{\phi}(k_{1})\tilde{\phi}(k_{2})\tilde{\phi}(k_{3})\tilde{\phi}(k_{4})\\
e^{-\theta[k_{1-}k_{2+}+k_{3-}k_{4+}+(k_{1-}+k_{2-})(k_{3+}+k_{4+})]}
\delta^{(3)}(k_{1}+k_{2}+k_{3}+k_{4})
\end{multline}
where
\begin{equation}\label{WVV}
V_{\star_{V}}=Ve^{\sum_{a<b}-\theta k_{a-}k_{b+}}=
Ve^{\sum_{a<b}-\frac{\theta}{2}(\boldsymbol{k}_{a}\cdot\boldsymbol{k}_{b}+
i\boldsymbol{k}_{a}\wedge\boldsymbol{k}_{b})}
\end{equation}
is the Wick-Voros vertex and $V$ is the ordinary vertex \eqref{OV}.
To calculate the four-point Green's function at the tree level,
we have to attach to the vertex \eqref{WVV} four propagators \eqref{WVTPGF}.
We have
\begin{align}
\nonumber
\tilde{G}_{V}^{(4)}&=-ig(2\pi)^{3}
\frac{e^{-\theta\left(\sum_{a=1}^{4}k_{a-}k_{a+}+\sum_{a<b}k_{a-}k_{b+}\right)}}
{\prod_{a=1}^{4}\left(k_{a}^{2}-m^{2}\right)}\delta^{(3)}\!\left(\sum_{a=1}^{4}k_{a}\right)\\
\nonumber
&=-ig(2\pi)^{3}
\frac{e^{-\frac{\theta}{2}\left[\sum_{a=1}^{4}\boldsymbol{k}_{a}^{2}+
\sum_{a<b}\left(\boldsymbol{k}_{a}\cdot\boldsymbol{k}_{b}+
i\boldsymbol{k}_{a}\wedge\boldsymbol{k}_{b}\right)\right]}}
{\prod_{a=1}^{4}\left(k_{a}^{2}-m^{2}\right)}\delta^{(3)}\!\left(\sum_{a=1}^{4}k_{a}\right)\\
\nonumber
&=-ig(2\pi)^{3}
\frac{e^{-\frac{\theta}{4}\left[\sum_{a=1}^{4}\boldsymbol{k}_{a}^{2}+
\left(\sum_{a=1}^{4}\boldsymbol{k}_{a}\right)^{2}
+2i\sum_{a<b}\boldsymbol{k}_{a}\wedge\boldsymbol{k}_{b}\right]}}
{\prod_{a=1}^{4}\left(k_{a}^{2}-m^{2}\right)}\delta^{(3)}\!\left(\sum_{a=1}^{4}k_{a}\right)\\
&=-ig(2\pi)^{3}
\frac{e^{-\frac{\theta}{4}\left(\sum_{a=1}^{4}\boldsymbol{k}_{a}^{2}+
2i\sum_{a<b}\boldsymbol{k}_{a}\wedge\boldsymbol{k}_{b}\right)}}
{\prod_{a=1}^{4}\left(k_{a}^{2}-m^{2}\right)}\delta^{(3)}\!\left(\sum_{a=1}^{4}k_{a}\right)
\end{align}
since the $\delta$ of conservation of momentum kills the mid term in the \mbox{exponential}.
The presence of a real exponent in the Green's functions signif\mbox{}ies
that the ultraviolet behaviour of the theory could be dif\mbox{}ferent from the Moyal and ordinary ones.
In order to investigate ultraviolet behaviour of the theory in the Wick-Voros case,
we will calculate in the next section the one-loop correction to the two- and four-point Green's functions.

\section{UV/IR mixing for the Wick-Voros product}

We now proceed to the calculation of the one-loop corrections
to the two- and four-point Green's functions in the Wick-Voros case.
In order to get the one-loop corrections to the propagator,
consider f\mbox{}irst the planar case in f\mbox{}igure~\ref{TPD}(a).
The amplitude is then obtained using three propagators \eqref{WVTPGF},
two with momentum $p$, one with momentum $q$, \mbox{the vertex} \eqref{WVV} with assignments \eqref{PA}
and the proper symmetry factor.
We have
\begin{align}
\nonumber
\tilde{G}_{\mathrm{P}}^{(2)}&=-i\frac{g}{3}\int\frac{\mathrm{d}^{3}q}{(2\pi)^{3}}
\frac{e^{-\theta(2p_{-}p_{+}+q_{-}q_{+})}
e^{-\theta(p_{-}q_{+}-p_{-}q_{+}-p_{-}p_{+}-q_{-}q_{+}-q_{-}p_{+}+q_{-}p_{+})}}
{(p^{2}-m^{2})^{2}(q^{2}-m^{2})}\\
&=-i\frac{g}{3}\int\frac{\mathrm{d}^{3}q}{(2\pi)^{3}}
\frac{e^{-\theta p_{-}p_{+}}}{(p^{2}-m^{2})^{2}(q^{2}-m^{2})}
\end{align}
where the f\mbox{}irst exponential is due to the propagators and the second one to the vertex.
Therefore, the $q$-contribution cancels
completely that is, there is no change in the convergence of the integral
with respect the ordinary case.
Consider now the non-planar case in f\mbox{}igure~\ref{TPD}(b).
The structure is the same as in the planar case, but with assignments given by \eqref{NPA}.
We have
\begin{align}
\nonumber
\tilde{G}_{\mathrm{NP}}^{(2)}&=-i\frac{g}{6}\int\frac{\mathrm{d}^{3}q}{(2\pi)^{3}}
\frac{e^{-\theta(2p_{-}p_{+}+q_{-}q_{+})}
e^{-\theta(p_{-}q_{+}-p_{-}p_{+}-p_{-}q_{+}-q_{-}p_{+}-q_{-}q_{+}+p_{-}q_{+})}}
{(p^{2}-m^{2})^{2}(q^{2}-m^{2})}\\
\nonumber
&=-i\frac{g}{6}\int\frac{\mathrm{d}^{3}q}{(2\pi)^{3}}
\frac{e^{-\theta(p_{-}p_{+}+i\boldsymbol{p}\wedge\boldsymbol{q})}}{(p^{2}-m^{2})^{2}(q^{2}-m^{2})}
\end{align}
since
\begin{equation}
p_{-}q_{+}-q_{-}p_{+}=i\boldsymbol{p}\wedge\boldsymbol{q}.
\end{equation}
Therefore, the $q$-contribution does not cancel completely.
We can conclude that the ultraviolet divergence of the planar diagram
in both the Moyal and Wick-Voros cases is unchanged with respect to the ordinary one.
Instead, in the non-planar diagram the presence of the oscillating factor
in the integral softens the ultraviolet divergence, because it dampens
the functions for high values of $q$. However, it is responsible for infrared divergences.
Therefore, the ultraviolet behaviour in both cases is exactly the same
and this can be seen like a consequence of the canonical non-commutative relation \eqref{CNC}
which is unchanged between the Moyal and Wick-Voros cases.
Nevertheless, the Green's functions for the two cases are not the same.

Finally, to get the one-loop correction to the four-point Green's function,
consider f\mbox{}irst the planar case of f\mbox{}igure~\ref{PFPD}. We have
\begin{equation}
\tilde{G}_{P}^{(4)}=\frac{(-ig)^{2}}{8}(2\pi)^{3}\!\int\frac{\mathrm{d}^{3}q}{(2\pi)^{3}}
\frac{e^{-\theta\left(\sum_{a=1}^{4}k_{a-}k_{a+}+\sum_{a<b}k_{a-}k_{b+}\right)}
\delta^{(3)}\!\left(\sum_{a=1}^{4}k_{a}\right)}
{(q^{2}-m^{2})\left[(k_{1}+k_{2}-q)^{2}-m^{2}\right]\prod_{a=1}^{4}(k_{a}^{2}-m^{2})}\\
\end{equation}
Therefore, the internal momentum $q$ appears only in the denominator, like in the Moyal case,
that is, the planar diagram has the same ultraviolet behaviour as the ordinary one.
In the determination of the one-loop correction to the four-point Green's function
we must consider the non-planar diagrams shown in f\mbox{}igure~\ref{NPFPD} as well. We have
\begin{equation}
\tilde{G}_{\mathrm{NP}_{a}}^{(4)}=\frac{(-ig)^{2}}{8}(2\pi)^{3}\!\int\frac{\mathrm{d}^{3}q}{(2\pi)^{3}}
\frac{e^{-\theta\left(\sum_{a=1}^{4}k_{a-}k_{a+}+\sum_{a<b}k_{a-}k_{b+}+E_{a}\right)}
\delta^{(3)}\!\left(\sum_{a=1}^{4}k_{a}\right)}
{(q^{2}-m^{2})\left[(k_{1}+k_{2}-q)^{2}-m^{2}\right]\prod_{a=1}^{4}(k_{a}^{2}-m^{2})}
\end{equation}
where $E_{a}$'s are still given by \eqref{NPFPT}.
We can conclude that the ultraviolet properties in the Wick-Voros case
remain unchanged with respect to the Moyal case and, in particular, the ultraviolet/infrared mixing is unchanged.
This is to be expected since heuristically this is consequence of commutation relation
which is, of course, the same in both theories. However, the two theories are not equivalent,
despite the physical intuition, since the Green's functions dif\mbox{}fer
for a factor can be considered as a momentum dependent coupling constant.
In the last chapter we will discuss this problem and show that
the Moyal and Wick-Voros f\mbox{}ield theories describe the same physics at the level of $S$-matrix
and in the framework of twisted non-commutativity. Instead, in the next chapter we will introduce
a general translation invariant associative product and show that
the ultraviolet/infrared mixing is still the some as for the Moyal and Wick-Voros products.

\chapter{UV/IR mixing for a general translation invariant product}

\emph{In this chapter we introduce a general translation invariant associative product
and describe a non-commutative f\mbox{}ield theory with such a product in order to investigate
the relationship between the translation invariance and the ultraviolet/infrared mixing.
We show that the ultraviolet/infrared mixing for the Moyal and Wick-Voros products
is a generic feature of any translation invariant associative product.}

\section{A general translation invariant product}

In this section we introduce a general associative star product\footnote{
General star products were f\mbox{}irst introduced in~\cite{BFFLS1,BFFLS2}
in the framework of deformation quantization of Poisson manifolds.}
between functions on $\mathbb{R}^{d}$
and then we discuss the condition of translational invariance in order to investigate
the relationship between the translation invariance and the ultraviolet/infrared mixing.
Notice that we contemplate the possibility that the star product be commutative
although in general it is not so.

Consider a generalization of the Moyal product \eqref{mp} given by
\begin{equation}\label{GP}
(f\star g)(x)=
\int\frac{\mathrm{d}^{d}p}{(2\pi)^{d}}\frac{\mathrm{d}^{d}q}{(2\pi)^{d}}\frac{\mathrm{d}^{d}r}{(2\pi)^{d}}\,
\tilde{f}(q)\tilde{g}(r)K(p,q,r)e^{ip\cdot x}
\end{equation}
where $K$ is in general a distribution. Note that the ordinary product is also of this kind for
\begin{equation}
K(p,q,r)=\delta^{(d)}(r-p+q).
\end{equation}
In order to have an associative product, we have to impose
\begin{multline}
((f\star g)\star h)(x)=
\int\frac{\mathrm{d}^{d}p}{(2\pi)^{d}}\frac{\mathrm{d}^{d}q}{(2\pi)^{d}}\frac{\mathrm{d}^{d}r}{(2\pi)^{d}}\,
(\widetilde{f\star g})(q)\tilde{h}(r)K(p,q,r)e^{ip\cdot x}\\
=\int\frac{\mathrm{d}^{d}p}{(2\pi)^{d}}\frac{\mathrm{d}^{d}q}{(2\pi)^{d}}\frac{\mathrm{d}^{d}r}{(2\pi)^{d}}
\frac{\mathrm{d}^{d}s}{(2\pi)^{d}}\frac{\mathrm{d}^{d}t}{(2\pi)^{d}}\,
\tilde{f}(s)\tilde{g}(t)\tilde{h}(r)K(p,q,r)K(q,s,t)e^{ip\cdot x}\\
=\int\frac{\mathrm{d}^{d}p}{(2\pi)^{d}}\frac{\mathrm{d}^{d}q}{(2\pi)^{d}}\frac{\mathrm{d}^{d}r}{(2\pi)^{d}}
\frac{\mathrm{d}^{d}s}{(2\pi)^{d}}\frac{\mathrm{d}^{d}t}{(2\pi)^{d}}\,
\tilde{f}(r)\tilde{g}(s)\tilde{h}(t)K(p,q,t)K(q,r,s)e^{ip\cdot x}
\end{multline}
is equal to
\begin{multline}
(f\star(g\star h))(x)=
\int\frac{\mathrm{d}^{d}p}{(2\pi)^{d}}\frac{\mathrm{d}^{d}q}{(2\pi)^{d}}\frac{\mathrm{d}^{d}r}{(2\pi)^{d}}\,
\tilde{f}(q)(\widetilde{g\star h})(r)K(p,q,r)e^{ip\cdot x}\\
=\int\frac{\mathrm{d}^{d}p}{(2\pi)^{d}}\frac{\mathrm{d}^{d}q}{(2\pi)^{d}}\frac{\mathrm{d}^{d}r}{(2\pi)^{d}}
\frac{\mathrm{d}^{d}s}{(2\pi)^{d}}\frac{\mathrm{d}^{d}t}{(2\pi)^{d}}\,
\tilde{f}(q)\tilde{g}(s)\tilde{h}(t)K(p,q,r)K(r,s,t)e^{ip\cdot x}\\
=\int\frac{\mathrm{d}^{d}p}{(2\pi)^{d}}\frac{\mathrm{d}^{d}q}{(2\pi)^{d}}\frac{\mathrm{d}^{d}r}{(2\pi)^{d}}
\frac{\mathrm{d}^{d}s}{(2\pi)^{d}}\frac{\mathrm{d}^{d}t}{(2\pi)^{d}}\,
\tilde{f}(r)\tilde{g}(s)\tilde{h}(t)K(p,r,q)K(q,s,t)e^{ip\cdot x}.
\end{multline}
In other words, we have to impose
\begin{equation}\label{AC}
\int\mathrm{d}^{d}q\,K(p,q,t)K(q,r,s)=
\int\mathrm{d}^{d}q\,K(p,r,q)K(q,s,t).
\end{equation}
We can show that this  condition is nothing but
the usual cocycle condition in the Hochschild cohomology.
To this end, we recall that the $2$-cochain $c\in C^{2}(\mathcal{A})$ is the map
\begin{equation}
c:\mathcal{A}\otimes\mathcal{A}\to\mathcal{A}
\end{equation}
def\mbox{}ined by
\begin{equation}\label{2-cochain}
c(f,g)=f\star g
\end{equation}
where $\mathcal{A}$ is the non-commutative algebra of functions with the product \eqref{GP}
and the coboundary operator is the map
\begin{equation}
\partial:C^{k}(\mathcal{A})\to C^{k+1}(\mathcal{A})
\end{equation}
which transforms the $k$-cochain $c(f_{1},\ldots f_{k})$ in the $(k+1)$-cochain def\mbox{}ined by
\begin{align}
\notag
\partial c(f_{0},\ldots,f_{k})&=f_{0}\star c(f_{1},\ldots,f_{k})+
\sum_{i=0}^{k-1}(-1)^{i+1}c(f_{0},\ldots,f_{i}\star f_{i+1},\ldots,f_k)\\
&+(-1)^{k+1}c(f_{0},\ldots,f_{k-1})\star f_{k}.
\end{align}
In order for the $2$-cochain \eqref{2-cochain} to be a $2$-cocycle, it has to be
\begin{align}
\notag
0=\partial c(f,g,h)&=f\star c(g,h)-c(f\star g,h)+c(f,g\star h)-c(f,g)\star h\\
&=2\left(f\star(g\star h)-(f\star g)\star h\right)
\end{align}
which gives \eqref{AC}. In general, the product \eqref{GTIP} is non-commutative.
However, it becomes commutative if we impose a constraint on $K$. Indeed, imposing
\begin{align}
\nonumber
(f\star g)(x)&=
\int\frac{\mathrm{d}^{d}p}{(2\pi)^{d}}\frac{\mathrm{d}^{d}q}{(2\pi)^{d}}\frac{\mathrm{d}^{d}r}{(2\pi)^{d}}\,
\tilde{f}(q)\tilde{g}(r)K(p,q,r)e^{ip\cdot x}\\
&=\int\frac{\mathrm{d}^{d}p}{(2\pi)^{d}}\frac{\mathrm{d}^{d}q}{(2\pi)^{d}}\frac{\mathrm{d}^{d}r}{(2\pi)^{d}}\,
\tilde{g}(q)\tilde{f}(r)K(p,r,q)e^{ip\cdot x}=(g\star f)(x)
\end{align}
we get the commutativity condition
\begin{equation}\label{CC}
K(p,q,r)=K(p,r,q).
\end{equation}
This condition means that the $2$-cochain $c$
given by \eqref{2-cochain} is a $2$-coboundary. That is,
\begin{equation}\label{coboundary}
c(f,g)=\partial b(f,g)=f\star b(g)+g\star b(f)-b(f\star g)
\end{equation}
where the $1$-cochain $b$ is simply given by the identity map.
In other words, the commutativity condition \eqref{CC} is a coboundary condition in the Hochschild cohomology.

We now proceed to the discussion of translation invariance of the product \eqref{GP}.
We recall that the translation by a vector $a$ is def\mbox{}ined by
\begin{equation}
\mathcal{T}_{a}(f)(x)=f(x+a).
\end{equation}
Since in Fourier transform
\begin{equation}
\int\frac{\mathrm{d}^{d}p}{(2\pi)^{d}}\,\widetilde{\mathcal{T}_{a}(f)}(p)e^{ip\cdot x}=
\int\frac{\mathrm{d}^{d}p}{(2\pi)^{d}}\,\tilde{f}(p)e^{ip\cdot(x+a)},
\end{equation}
we have
\begin{equation}
\widetilde{\mathcal{T}_{a}(f)}(p)=e^{ia\cdot p}\tilde{f}(p).
\end{equation}
By translation invariant product we mean as usual a product which satisf\mbox{}ies the property
\begin{equation}
\mathcal{T}_{a}(f)\star\mathcal{T}_{a}(g)=\mathcal{T}_{a}(f\star g).
\end{equation}
For the translational invariance of the product \eqref{GP} we have to impose
\begin{equation}
\mathcal{T}_{a}(f\star g)=
\int\frac{\mathrm{d}^{d}p}{(2\pi)^{d}}\frac{\mathrm{d}^{d}q}{(2\pi)^{d}}\frac{\mathrm{d}^{d}r}{(2\pi)^{d}}\,
e^{ip\cdot(x+a)}\tilde{f}(q)\tilde{g}(r)K(p,q,r)
\end{equation}
is equal to
\begin{align}
\nonumber
\mathcal{T}_{a}(f)\star\mathcal{T}_{a}(g)&=
\int\frac{\mathrm{d}^{d}p}{(2\pi)^{d}}\frac{\mathrm{d}^{d}q}{(2\pi)^{d}}\frac{\mathrm{d}^{d}r}{(2\pi)^{d}}\,
e^{ip\cdot x}\widetilde{\mathcal{T}_{a}(f)}(q)\widetilde{\mathcal{T}_{a}(g)}(r)K(p,q,r)\\
&=\int\mathrm{d}p\,\mathrm{d}q\,\mathrm{d}r\,e^{ip\cdot x}
e^{ia\cdot q}\tilde{f}(q)e^{ia\cdot r}\tilde{g}(r)K(p,q,r).
\end{align}
This is achieved by setting
\begin{equation}\label{K}
K(p,q,r)=e^{\alpha(p,q)}\delta^{(d)}(r-p+q)
\end{equation}
where $\alpha$ is a generic function. Therefore, because of translation invariance,
the product \eqref{GP} takes the form
\begin{equation}\label{GTIP}
(f\star g)(x)=\int\frac{\mathrm{d}^{d}p}{(2\pi)^{d}}\frac{\mathrm{d}^{d}q}{(2\pi)^{d}}\,
\tilde{f}(q)\tilde{g}(p-q)e^{\alpha(p,q)}e^{ip\cdot x}.
\end{equation}
The ordinary product is given by $\alpha=0$, the Moyal product \eqref{mp} by
\begin{equation}\label{alphaM}
\alpha_{M}(p,q)=-\frac{i}{2}\theta^{ij}q_{i}(p_{j}-q_{j})=\frac{i}{2}\theta p\wedge q
\end{equation}
and the Wick-Voros product \eqref{wvp} by
\begin{equation}\label{alphaV}
\alpha_{V}(p,q)=-\theta q_{-}(p_{+}-q_{+})=\alpha_{M}(p,q)-\frac{\theta}{2}(p-q)\cdot q.
\end{equation}
We can express the associativity condition \eqref{AC} in terms of $\alpha$.
Indeed, from \eqref{AC} and \eqref{K} follows that
\begin{multline}
\int\mathrm{d}^{d}q\,e^{\alpha(p,q)}\delta^{(d)}(t-p+q)e^{\alpha(q,r)}\delta^{(d)}(s-q+r)=\\
\int\mathrm{d}^{d}q\,e^{\alpha(p,r)}\delta^{(d)}(q-p+r)e^{\alpha(q,s)}\delta^{(d)}(t-q+s)
\end{multline}
That is,
\begin{equation}
\int\mathrm{d}^{d}p\,e^{\alpha(p,r+s)+\alpha(r+s,r)}\delta^{(d)}(r+s+t-p)=
\int\mathrm{d}^{d}p\,e^{\alpha(p,r)+\alpha(p-r,s)}\delta^{(d)}(r+s+t-p).
\end{equation}
Therefore, $\alpha$ has to satisfy the condition
\begin{equation}
\alpha(p,r+s)+\alpha(r+s,r)=\alpha(p,r)+\alpha(p-r,s)
\end{equation}
which can be rewritten as
\begin{equation}\label{ac}
\alpha(p,q)=\alpha(p,r)-\alpha(q,r)+\alpha(p-r,q-r).
\end{equation}
Note that we can get the associativity condition by starting directly from \eqref{GTIP}. Indeed, imposing
\begin{align}
\nonumber
((f\star g)\star h)(x)=&
\int\frac{\mathrm{d}^{d}p}{(2\pi)^{d}}\frac{\mathrm{d}^{d}q}{(2\pi)^{d}}\,
(\widetilde{f\star g})(q)\tilde{h}(p-q)e^{\alpha(p,q)}e^{ip\cdot x}\\
=&\int\frac{\mathrm{d}^{d}p}{(2\pi)^{d}}\frac{\mathrm{d}^{d}q}{(2\pi)^{d}}\frac{\mathrm{d}^{d}r}{(2\pi)^{d}}\,
\tilde{f}(r)\tilde{g}(q-r)\tilde{h}(p-q)e^{\alpha(p,q)+\alpha(q,r)}e^{ip\cdot x}
\end{align}
is equal to
\begin{align}
\nonumber
(f\star(g\star h))(x)=&
\int\frac{\mathrm{d}^{d}p}{(2\pi)^{d}}\frac{\mathrm{d}^{d}q}{(2\pi)^{d}}\,
\tilde{f}(q)(\widetilde{g\star h})(p-q)e^{\alpha(p,q)}e^{ip\cdot x}\\
\nonumber
=&\int\frac{\mathrm{d}^{d}p}{(2\pi)^{d}}\frac{\mathrm{d}^{d}q}{(2\pi)^{d}}\frac{\mathrm{d}^{d}r}{(2\pi)^{d}}\,
\tilde{f}(q)\tilde{g}(r)\tilde{h}(p-q-r)e^{\alpha(p,q)+\alpha(p-q,r)}e^{ip\cdot x}\\
=&\int\frac{\mathrm{d}^{d}p}{(2\pi)^{d}}\frac{\mathrm{d}^{d}q}{(2\pi)^{d}}\frac{\mathrm{d}^{d}r}{(2\pi)^{d}}\,
\tilde{f}(r)\tilde{g}(q-r)\tilde{h}(p-q)e^{\alpha(p,r)+\alpha(p-r,q-r)}e^{ip\cdot x}
\end{align}
we reobtain the condition \eqref{ac}. We also require
\begin{equation}
f\star 1=\int\frac{\mathrm{d}^{d}p}{(2\pi)^{d}}\frac{\mathrm{d}^{d}q}{(2\pi)^{d}}\,
\tilde{f}(q)\delta^{(d)}(p-q)e^{\alpha(p,q)}e^{ip\cdot x}
=\int\frac{\mathrm{d}^{d}p}{(2\pi)^{d}}\,\tilde{f}(p)e^{\alpha(p,p)}e^{ip\cdot x}=f
\end{equation}
and
\begin{equation}
1\star f=\int\frac{\mathrm{d}^{d}p}{(2\pi)^{d}}\frac{\mathrm{d}^{d}q}{(2\pi)^{d}}\,
\delta^{(d)}(q)\tilde{f}(p-q)e^{\alpha(p,q)}e^{ip\cdot x}
=\int\frac{\mathrm{d}^{d}p}{(2\pi)^{d}}\,\tilde{f}(p)e^{\alpha(p,0)}e^{ip\cdot x}=f.
\end{equation}
That is, the identity of the algebra of functions with the product \eqref{GTIP}
is the constant function with value $1$. These conditions impose
\begin{align}
\label{alphaa}
\alpha(p,p)&=0\\
\label{alphab}
\alpha(p,0)&=0.
\end{align}
In particular,
\begin{equation}\label{alphac}
\alpha(0,0)=0
\end{equation}
Moreover, we require the algebra to be a $*$-algebra. That is, there must be a map
which satisf\mbox{}ies the following conditions~\cite{Landsman}:
\begin{align}
(f^{*})^{*}&=f\\
(\lambda f+\mu g)^{*}&=\bar{\lambda}f^{*}+\bar{\mu}g^{*}\\
(f\star g)^{*}&=g^{*}\star f^{*}
\end{align}
for any complex numbers $\lambda$ and $\mu$ where the bar dentes complex conjugation.
In our case the involution $*$ is given by complex conjugation
and in particular the last relation imposes a constrain on $\alpha$.
Indeed, since
\begin{equation}
(f\star g)^{*}=\int\frac{\mathrm{d}^{d}p}{(2\pi)^{d}}\frac{\mathrm{d}^{d}q}{(2\pi)^{d}}\,
\tilde{f}(q)^{*}\tilde{g}(p-q)^{*}e^{\alpha(p,q)^{*}}e^{-ip\cdot x}
\end{equation}
has to be equal to
\begin{align}
\nonumber
g^{*}\star f^{*}&=\int\frac{\mathrm{d}^{d}p}{(2\pi)^{d}}\frac{\mathrm{d}^{d}q}{(2\pi)^{d}}\,
\widetilde{g^{*}}(q)\widetilde{f^{*}}(p-q)e^{\alpha(p,q)}e^{ip\cdot x}\\
\nonumber
&=\int\frac{\mathrm{d}^{d}p}{(2\pi)^{d}}\frac{\mathrm{d}^{d}q}{(2\pi)^{d}}\,
\tilde{g}(-q)^{*}\tilde{f}(q-p)^{*}e^{\alpha(p,q)}e^{ip\cdot x}\\
&=\int\frac{\mathrm{d}^{d}p}{(2\pi)^{d}}\frac{\mathrm{d}^{d}q}{(2\pi)^{d}}\,
\tilde{f}(q)^{*}\tilde{g}(p-q)^{*}e^{\alpha(-p,q-p)}e^{-ip\cdot x}
\end{align}
then $\alpha$ has to satisfy the condition
\begin{equation}\label{starcondition}
\alpha(p,q)^{*}=\alpha(-p,q-p).
\end{equation}
We can express the commutativity condition \eqref{CC} in terms of $\alpha$ as well.
Indeed, from \eqref{CC} and \eqref{K} follows that
\begin{align}
\nonumber
e^{\alpha(p,q)}\delta^{(d)}(r-p+q)&=e^{\alpha(p,r)}\delta^{(d)}(q-p+r)\\
\nonumber
&=e^{\alpha(p,r)}\delta^{(d)}(r-p+q)\\
&=e^{\alpha(p,p-q)}\delta^{(d)}(r-p+q).
\end{align}
That is,
\begin{equation}\label{cc}
\alpha(p,q)=\alpha(p,p-q).
\end{equation}
Note that we can get the commutativity condition by starting directly from \eqref{GTIP}. Indeed, imposing
\begin{align}
\nonumber
(f\star g)(x)=&
\int\frac{\mathrm{d}^{d}p}{(2\pi)^{d}}\frac{\mathrm{d}^{d}q}{(2\pi)^{d}}\,
\tilde{f}(q)\tilde{g}(p-q)e^{\alpha(p,q)}e^{ip\cdot x}\\
=&\int\frac{\mathrm{d}^{d}p}{(2\pi)^{d}}\frac{\mathrm{d}^{d}q}{(2\pi)^{d}}\,
\tilde{g}(q)\tilde{f}(p-q)e^{\alpha(p,p-q)}e^{ip\cdot x}=(g\star f)(x)
\end{align}
we reobtain the condition \eqref{cc}.
From the associativity condition \eqref{ac}, we can derive some very useful relations.
For $q=r=p$ we have
\begin{equation}
\alpha(p,p)=\alpha(0,0)
\end{equation}
and for $q=r=0$ we have
\begin{equation}
\alpha(p,0)=\alpha(0,0)
\end{equation}
in agreement with \eqref{alphaa} and \eqref{alphab} respectively. For $q=0$ and $r=p$ we have
\begin{equation}\label{alpha1}
\alpha(0,-p)=\alpha(0,p).
\end{equation}
For $r=p$ we have
\begin{equation}\label{alpha2}
\alpha(p,q)=-\alpha(q,p)+\alpha(0,q-p).
\end{equation}
Moreover, from \eqref{ac} we have
\begin{equation}
\alpha(0,q)=\alpha(0,p)-\alpha(q,p)+\alpha(-p,q-p)
\end{equation}
and by using \eqref{alpha2} we obtain a very important relation%
\begin{equation}\label{alpha3}
\alpha(p,q)=-\alpha(0,p)+\alpha(0,q)+\alpha(0,p-q)-\alpha(-p,q-p).
\end{equation}
With a symbolic manipulation programme and a little work is not dif\mbox{}f\mbox{}icult
to construct polynomial expression for $\alpha$. For example, the following expression in two-dimensions
\begin{align}
\notag
\alpha&=Ap_{2}q_{1}+Bp_{1}q_{2}
-(A+B)q_{1}q_{2}+C\left(p_{2}q_{2}^{2}-p_{2}^{2}q_{2}\right)\\
&+D\left(\frac{p_{2}^{2}q_{1}-p_{1}q_{2}^{2}}{2}+p_{1}p_{2}q_{2}-p_{2}q_{1}q_{2}\right)
\end{align}
gives rise to an associative product for any complex numbers $A,B,C$ and $D$.
However, the condition \eqref{starcondition} imposes some conditions on these coef\mbox{}f\mbox{}icients.
Indeed, it is easy to see that
\begin{equation}
B^{*}=A,\quad C^{*}=-C\quad\mathrm{and}\quad D^{*}=-D.
\end{equation}
Note that the integral of the product of two functions can be written as
\begin{align}\label{ISP}
\nonumber
\int\mathrm{d}^{d}x\,f\star g=&
\int\mathrm{d}^{d}x\,\frac{\mathrm{d}^{d}p}{(2\pi)^{d}}\frac{\mathrm{d}^{d}q}{(2\pi)^{d}}\,
\tilde{f}(q)\tilde{g}(p-q)e^{\alpha(p,q)}e^{ip\cdot x}\\
=&\int\frac{\mathrm{d}q}{(2\pi)^{d}}\,\tilde{f}(q)\tilde{g}(-q)e^{\alpha(0,q)}.
\end{align}
Therefore, the integral of the product of two functions is not equal to the integral
of the ordinary product of the two functions.
%That is,
%\begin{equation}
%\int\mathrm{d}^{d}x\,f\star g\neq\int\mathrm{d}^{d}x\,fg.
%\end{equation}
However, from \eqref{alpha1} follows the trace property
\begin{equation}
\int\mathrm{d}^{d}x\,f\star g=\int\mathrm{d}^{d}x\,g\star f.
\end{equation}

\section{Cohomology}

We now proceed to show that it is possible to def\mbox{}ine
an ``$\alpha$-cohomology'' with respect to which $\alpha$
is a $2$-cocycle in the associative case, while it is a $2$-coboundary in the commutative one.
Let $\alpha\in A^{2}(\tilde{\mathcal{A}})$ be the map
\begin{equation}
\alpha:(p,q)\in\tilde{\mathcal{A}}\otimes\tilde{\mathcal{A}}\longrightarrow\tilde{\mathcal{A}}\label{coct}
\end{equation}
with $\tilde{\mathcal{A}}$ the algebra of Fourier transforms
(to be more precise $\alpha$ is def\mbox{}ined on translations, realised as linear functions in $\tilde{\mathcal{A}}$)
and the coboundary operator
\begin{equation}
\partial: A^{k}(\tilde{\mathcal{A}})\to A^{k+1}(\tilde{\mathcal{A}})
\end{equation}
def\mbox{}ined by
\begin{align}
\notag
\partial\gamma(p_{0},\ldots,p_{k})&=
\sum_{i=0}^{k}(-1)^{i}\gamma(p_{0},\ldots,p_{i-1},p_{\hat{i}},p_{i+1},\ldots,p_{k})\\
&-(-1)^{k}\gamma(p_{0}-p_{k}, p_{i}-p_{k},\ldots, p_{k-1}-p_{k}).
\end{align}
Note that a straightforward calculation verif\mbox{}ies that
\begin{equation}\label{CBO}
\partial^{2}=0.
\end{equation}
In order for $\alpha$ in \eqref{coct} to be a  $2$-cocycle in the $\alpha$-cohomology, it has to be
\begin{align}
\notag
0=\partial\alpha(p,q,r)&=\alpha(q,r)-\alpha(p,r)+\alpha(p,q)-\alpha(p-r,q-r)\\
&=2\left(f\star(g\star h)-(f\star g)\star h\right)
\end{align}
that is \eqref{ac}.
Therefore, the associativity condition \eqref{ac} is a cocycle condition in the $\alpha$-cohomology.
Analogously the commutativity condition \eqref{cc} is a coboundary condition.
Indeed, for $\alpha$ to be a $2$-coboundary in the $\alpha$-cohomology, it has to be
\begin{equation}
\alpha(p,q)=\partial\beta(p,q)=\beta(q)-\beta(p)+\beta(p-q)\label{cobt}
\end{equation}
which implies the commutativity condition \eqref{cc}. If $\alpha$ is a $2$-coboundary
in the $\alpha$-cohomology, then it is a $2$-cocycle because of \eqref{CBO}
so that the product \eqref{GTIP} is associative and commutative.
However, the coboundary condition  in the $\alpha$-cohomology \eqref{cobt}
is not equivalent to the commutative condition \eqref{cc}. Indeed, the function
\begin{equation}
\alpha(p,q)=A\beta(p)+\beta(q)+\beta(p-q)
\end{equation}
with $A$ a complex number and $A\neq-1$ is not a $2$-coboundary, but it makes the product \eqref{GTIP}
commutative and, of course, non-associative.
As we have already seen, the Moyal and Wick-Voros products, both non-commutative, are respectively given by
\eqref{alphaM} and \eqref{alphaV} which are both $2$-cocycles in the $\alpha$-cohomology and more interestingly
they dif\mbox{}fer by a term which is a $\alpha$-coboundary according to \eqref{cobt} with $\beta$ so def\mbox{}ined
\begin{equation}
\beta(q)=q^2
\end{equation}
Indeed, it is easy to verify that
\begin{equation}
\alpha_{V}(p,q)=\alpha_{M}(p,q)+\frac{\theta}{4}\partial\beta(p,q).
\end{equation}

\section{Dif\mbox{}ferential form of a general product}

Another way to get a general star product is generalizing the dif\mbox{}ferential form of the Moyal product \eqref{MP}.
To this end, consider a product which is a series in a deformation parameter which we call again $\theta$:
\begin{equation}\label{starexpans}
f\star g=\sum_{r=0}^{\infty}C_{r}(f,g)\theta^{r}
\end{equation}
where $C$'s are in general bidif\mbox{}ferential operators i.e.~bilinear maps which are
dif\mbox{}ferential operators with respect to each argument of globally bounded order.
To recover the ordinary product in the limit $\theta\to0$ we need to impose
\begin{equation}
C_{0}(f,g)=fg.
\end{equation}
Therefore, we can rewritten the product which we are considering as
\begin{equation}
f\star g=fg+\sum_{r=1}^{\infty}C_{r}(f,g)\theta^{r}.
\end{equation}
Instead, in order to ensure associativity the remaining $C_{r}$'s have to satisfy the following conditions
\begin{multline}\label{Crcond}
fC_{r}(g,h)-C_{r}(fg,h)+C_{r}(f,gh)-C_{r}(f,g)h=\\
=\sum_{j+k=r}\left(C_{j}(C_{k}(f,g),h)-C_{j}(f,C_{k}(g,h))\right)
\end{multline}
for all $r>0$. Note that a problem with the product \eqref{starexpans} is that it is def\mbox{}ined
on the space of formal series in the coordinates and then there is in general no control on the convergence
of the series after the product has been taken.
This kind of product is considered in the very general framework
of deformation quantization~\cite{DitoSternheimer,Sternheimer}.
This approach consists in f\mbox{}inding a deformation
of the algebra of functions on a Poisson manifold\footnote{
In general, a Poisson manifold is a manifold endowed with a Poisson bracket that is,
a bilinear and antisymmetric map $\{\cdot,\cdot\}$ which satisf\mbox{}ies the Jacobi identity
\begin{equation*}
\{f,\{g,h\}\}+\{g,\{h,f\}\}+\{h,\{f,g\}\}=0
\end{equation*}
and the Leibniz rule
\begin{equation*}
\{f,gh\}=\{f,g\}h+g\{f,h\}.
\end{equation*}}
with the additional property that to f\mbox{}irst order in the deformation parameter $\theta$
the star commutator of two functions:
\begin{equation}
[f,g]_{\star}=f\star g-g\star f
\end{equation}
reduces, or better is proportional, to the Poisson bracket of the two functions.
This requires that
\begin{equation}
\{f,g\}=C_{1}(f,g)-C_{1}(g,f).
\end{equation}
Since in general the Poisson bracket of two function can be written as
\begin{equation}
\{f,g\}=\Lambda^{ij}\partial_{i}f\partial_{j}g
\end{equation}
we have
\begin{equation}
\Lambda^{ij}=C_{1}(x^{i},x^{j})-C_{1}(x^{j},x^{i})
\end{equation}
which can be easily proved by setting $f=x^{i}$ and $g=x^{j}$.
Note that if
\begin{equation}
C_{1}(f,g)=C_{1}(g,f)
\end{equation}
then the product \eqref{starexpans} is commutative.
The proof is the following. First consider \eqref{Crcond} for $r=2$, $f=h=x^{m}$ and $g=x^{n}$.
Then relation \eqref{Crcond} becomes
\begin{multline}\label{c2cancel}
f C_{2}(g,f)-C_{2}(f g,f)+C_{2}(f,g f)-C_{2}(f,g)f=\\
=x^{m}(C_{2}(x^{n},x^{m})-C_{2}(x^{m},x^{n}))-C_{2}(x^{m+n},x^{m})+C_{2}(x^{m},x^{m+n})=\\
=\left(C_{1}(C_{1}(f,g),f)-C_{1}(f,C_{1}(g,f))\right)=0
\end{multline}
because of the symmetry of $C_1$.
The second line of the above equation has to hold for all $x$'s and therefore it must be
\begin{equation}
C_{2}(x^{m+n},x^{m})=C_{2}(x^{m},x^{m+n})
\end{equation}
for generic $m$ and $n$. This implies that
\begin{equation}
C_{2}(f,g)=C_{2}(g,f).
\end{equation}
It is then possible to prove exactly in the same way that if
\begin{equation}
C_{l}(f,g)=C_{l}(g,f)
\end{equation}
for $l<r$ then all the terms in the right hand side of \eqref{Crcond} pairwise cancel and
we are left to the equivalent of \eqref{c2cancel} with a generic $r$ proving that
\begin{equation}
C_{r}(f,g)=C_{r}(g,f).
\end{equation}
In general, given a Poisson manifold it is not easy to prove that it is always possible to f\mbox{}ind
a star product whose commutator reduces, to f\mbox{}irst order in the deformation parameter $\theta$,
to the Poisson bracket because the associativity conditions \eqref{Crcond} are dif\mbox{}f\mbox{}icult to satisfy.
A general result of Kontsevich~\cite{Kontsevich} in the context of formal series solves this problem
for a generic Poisson manifold. Moreover, he proves that two products with the same Poisson structure
are equivalent in sense that there exists an invertible map $T$ such that
\begin{equation}
T(f\star g)=T(f)\star' T(g).
\end{equation}
We have seen an instance of such an equivalence for the Moyal and Wick-Voros products
for which the equivalence map is given by \eqref{EM}.

To conclude this section, we brief\mbox{}ly discuss the translation invariance of the product \eqref{starexpans}.
Since the $C_{r}$'s are bidif\mbox{}ferential operators, the product \eqref{starexpans} becomes translation invariant
if and only if the $C_{r}$'s are combinations of derivatives only.
Therefore, it can be written as
\begin{equation}
f\star g=fg+\sum_{r=1}^{\infty}
\sum_{i_{1},\ldots i_{r},j_{1},\ldots j_{r}}
\theta^{r}c^{i_{1}\ldots i_{r}j_{1}\ldots j_{r}}
(\partial_{i_{1}}\ldots\partial_{i_{r}}f)(\partial_{j_{1}}\ldots\partial_{j_{r}}g)
\end{equation}
where $c$'s are complex constant coef\mbox{}f\mbox{}icients.
By using \eqref{starexpans} and the fact that the $C_{r}$'s are combinations of derivatives only, we easily get
\begin{equation}
[x^{i},x^{j}]_{\star}=x^{i}\star x^{j}-x^{i}\star x^{j}=C_{1}(x^{i},x^{j})-C_{1}(x^{j},x^{i}).
\end{equation}
That is, the commutator of coordinates reads
\begin{equation}
[x^{i},x^{j}]_{\star}=\Lambda^{ij}.
\end{equation}
Therefore, it reproduces the Poisson structure of the underlying space.

\section{Field theory with a general translation invariant product}

We now proceed to the discussion of a non-commutative f\mbox{}ield theory with the general translation invariant product \eqref{GTIP} in order to investigate its ultraviolet behaviour.
So let us consider a f\mbox{}ield theory described by the action
\begin{equation}
S=S_{0}+S_{\mathrm{int}}
\end{equation}
where $S_{0}$ is the free action given by
\begin{equation}
S_{0}=\int\mathrm{d}^{d}x\,
\frac{1}{2}\left(\partial_{\mu}\phi\star\partial^{\mu}\phi-m^{2}\phi\star\phi\right)
\end{equation}
and $S_{\mathrm{int}}$ is the interacting one given by
\begin{equation}
S_{\mathrm{int}}=\frac{g}{4!}\int\mathrm{d}^{d}x\,\phi\star\phi\star\phi\star\phi.
\end{equation}
To begin with, we calculate the equation of motion. To this end, we consider
an inf\mbox{}initesimal variation of $\phi$:
\begin{equation}
\phi\to\phi+\delta\phi.
\end{equation}
The corresponding inf\mbox{}initesimal variation of the action $S_{0}$ is given by
\begin{equation}
\delta S_{0}=\int\mathrm{d}^{d}x
\left(\partial_{\mu}\phi\star\partial^{\mu}\delta\phi-m^{2}\phi\star\delta\phi\right).
\end{equation}
By integrating by parts we obtain, up to boundary terms,
\begin{equation}
\delta S_{0}=-\int\mathrm{d}^{d}x
\left(\Box+m^{2}\right)\phi\star\delta\phi
\end{equation}
and using the relation \eqref{ISP} we have
\begin{equation}
\delta S_{0}=-\int\mathrm{d}^{d}p\,e^{\alpha(0,p)}(-p^{2}+m^{2})
\tilde{\phi}(p)\widetilde{\delta\phi}(-p).
\end{equation}
Since the variation of the action $\delta S_{0}$ have to be vanishing for every
variation of the f\mbox{}ield $\delta\phi$, we get the equation of motion in Fourier transform
\begin{equation}
e^{\alpha(0,p)}(p^{2}-m^{2})\tilde{\phi}(p)=0
\end{equation}
which reduces to the same ordinary equation of motion in Fourier transform due to the invertibility
of the exponential factor. Therefore, at classical level the free non-commutative f\mbox{}ield theory
with the general translation invariant product \eqref{GTIP} is the same as the commutative one.

We now move on to the quantum case and proceed to the calculation of Green's functions.
The propagator can be easily obtained by solving the equation
\begin{equation}
e^{\alpha(0,p)}(p^{2}-m^{2})\tilde{G}^{2}(p)=1
\end{equation}
which gives
\begin{equation}\label{TPGF}
\tilde{G}^{2}(p)=\frac{e^{-\alpha(0,p)}}{p^{2}-m^{2}}.
\end{equation}
Note that in general the presence of the exponential
in the propagator \eqref{TPGF} modif\mbox{}ies its properties.
In order to calculate the vertex, let us write down
the interacting term of the action in momentum space.
Using the relations \eqref{GTIP} and \eqref{ISP} we have
\begin{multline}
S_{\mathrm{int}}=\frac{g}{4!}
\int\mathrm{d}^{d}x\!
\frac{\mathrm{d}^{d}k_{1}}{(2\pi)^{d}}\frac{\mathrm{d}^{d}k_{2}}{(2\pi)^{d}}
\frac{\mathrm{d}^{d}k_{3}}{(2\pi)^{d}}\frac{\mathrm{d}^{d}k_{4}}{(2\pi)^{d}}\,
\tilde{\phi}(k_{2})\tilde{\phi}(k_{1}-k_{2})\tilde{\phi}(k_{4})\tilde{\phi}(k_{3}-k_{4})\\
e^{\alpha(k_{1},k_{2})}e^{\alpha(k_{3},k_{4})}e^{ik_{1}\cdot x}\star e^{ik_{3}\cdot x}\\
=\frac{g}{4!}(2\pi)^{d}\!
\int\frac{\mathrm{d}^{d}k_{1}}{(2\pi)^{d}}\frac{\mathrm{d}^{d}k_{2}}{(2\pi)^{d}}
\frac{\mathrm{d}^{d}k_{3}}{(2\pi)^{d}}\frac{\mathrm{d}^{d}k_{4}}{(2\pi)^{d}}\,
\tilde{\phi}(k_{2})\tilde{\phi}(k_{1}-k_{2})\tilde{\phi}(k_{4})\tilde{\phi}(k_{3}-k_{4})\\
e^{\alpha(k_{1},k_{2})}e^{\alpha(k_{3},k_{4})}
\int\frac{\mathrm{d}^{d}k}{(2\pi)^{d}}\,e^{\alpha(0,k)}\delta^{(d)}(k_{1}-k)\delta^{(d)}(k_{3}+k)\\
=\frac{g}{4!}(2\pi)^{d}\!
\int\frac{\mathrm{d}^{d}k_{1}}{(2\pi)^{d}}\frac{\mathrm{d}^{d}k_{2}}{(2\pi)^{d}}
\frac{\mathrm{d}^{d}k_{3}}{(2\pi)^{d}}\frac{\mathrm{d}^{d}k_{4}}{(2\pi)^{d}}\,
\tilde{\phi}(k_{2})\tilde{\phi}(k_{1}-k_{2})\tilde{\phi}(k_{4})\tilde{\phi}(k_{3}-k_{4})\\
e^{\alpha(k_{1},k_{2})+\alpha(k_{3},k_{4})+\alpha(0,k_{1})}\delta^{(d)}(k_{1}+k_{3})
\end{multline}
which can be rewritten as well as
\begin{multline}
S_{\mathrm{int}}=
\frac{g}{4!}(2\pi)^{d}\!
\int\frac{\mathrm{d}^{d}k_{1}}{(2\pi)^{d}}\frac{\mathrm{d}^{d}k_{2}}{(2\pi)^{d}}
\frac{\mathrm{d}^{d}k_{3}}{(2\pi)^{d}}\frac{\mathrm{d}^{d}k_{4}}{(2\pi)^{d}}\,
\tilde{\phi}(k_{1})\tilde{\phi}(k_{2})\tilde{\phi}(k_{3})\tilde{\phi}(k_{4})\\
e^{\alpha(k_{1}+k_{2},k_{1})+\alpha(k_{3}+k_{4},k_{3})+\alpha(0,k_{1}+k_{2})}
\delta^{(d)}(k_{1}+k_{2}+k_{3}+k_{4}).
\end{multline}
Therefore, the vertex is given by
\begin{equation}\label{V}
V_{\star}=V
e^{\alpha(k_{1}+k_{2},k_{1})+\alpha(k_{3}+k_{4},k_{3})+\alpha(0,k_{1}+k_{2})}
\end{equation}
where
\begin{equation}
V=-i\frac{g}{4!}(2\pi)^{d}\delta^{(d)}\!\left(\sum_{a=1}^{4}k_{a}\right)
\end{equation}
is the ordinary $d$-dimensional vertex.
%
%Another way to write the vertex
%
%Using \eqref{alpha3} we can rewritten the vertex \eqref{V} as
%\begin{equation}
%V_{\star}=Ve^{\sum_{a=1}^{4}\alpha(0,k_{a})
%-\alpha(-k_{1}-k_{2},-k_{2})-\alpha(0,k_{3}+k_{4})-\alpha(-k_{3}-k_{4},-k_{4})}
%\end{equation}
%or equivalently as
%\begin{align}
%\notag
%V_{\star}&=Ve^{\frac{1}{2}\left(\sum_{a=1}^{4}\alpha(0,k_{a})
%+\alpha(k_{1}+k_{2},k_{1})-\alpha(-k_{1}-k_{2},-k_{2})+\alpha(k_{3}+k_{4},k_{3})-\alpha(-k_{3}-k_{4},-k_{4})\right)}\\
%&=Ve^{\frac{1}{2}\left(\sum_{a=1}^{4}\alpha(0,k_{a})
%+\alpha(k_{1}+k_{2},k_{1})-\alpha(k_{3}+k_{4},-k_{2})+\alpha(k_{3}+k_{4},k_{3})-\alpha(k_{1}+k_{2},-k_{4})\right)}.
%\end{align}
%
%
%
It is possible to show that the vertex loses the invariance for arbitrary exchanges of the external momenta,
but it maintains invariance for cyclic permutations.
Finally, to calculate the four-point Green's function to the tree level,
we must attach to the vertex \eqref{V} four propagators \eqref{TPGF}. We have
\begin{align}
\nonumber
\tilde{G}^{(4)}=&-ig(2\pi)^{d}\,
e^{\alpha(k_{1}+k_{2},k_{1})+\alpha(k_{3}+k_{4},k_{3})+\alpha(0,k_{1}+k_{2})}
\prod_{a=1}^{4}\frac{e^{-\alpha(0,k_{a})}}{k_{a}^{2}-m^{2}}\delta^{(d)}\!\left(\sum_{a=1}^{4}k_{a}\right)\\
=&-ig(2\pi)^{d}\,
\frac{e^{\alpha(k_{1}+k_{2},k_{1})+\alpha(k_{3}+k_{4},k_{3})+\alpha(0,k_{1}+k_{2})-\sum_{a=1}^{4}\alpha(0,k_{a})}}
{\prod_{a=1}^{4}(k_{a}^{2}-m^{2})}\delta^{(d)}\!\left(\sum_{a=1}^{4}k_{a}\right).
\end{align}

\section{UV/IR mixing for a general translation invariant product}

We now proceed to the calculation of the one-loop corrections to the propagator
in order to investigate the ultraviolet behaviour of the theory.
Consider then both diagrams of f\mbox{}igure~\ref{TPD}.
The correction for the planar case (a) is obtained using three propagators \eqref{TPGF},
one with momentum $p$, one with momentum $-p$, one with momentum $q$
and the vertex \eqref{V} with assignments given by \eqref{PA}
and the integration in $q$. We have up to a constant
\begin{align}
\nonumber
G^{(2)}_{P}=&\int\mathrm{d}^{d}q\,\frac{e^{-\alpha(0,p)-\alpha(0,-p)-\alpha(0,q)}}
{(p^{2}-m^{2})^{2}(q^{2}-m^{2})}e^{\alpha(p+q,p)+\alpha(-p-q,-q)+\alpha(0,p+q)}\\
\nonumber
%=&\int\mathrm{d}^{d}q\,\frac{e^{-\alpha(0,p)-\alpha(0,-p)-\alpha(0,q)+\alpha(p+q,p)+\alpha(-p-q,-q)+\alpha(0,p+q)}}
%{(p^{2}-m^{2})^{2}(q^{2}-m^{2})}\\
%\nonumber
=&\int\mathrm{d}^{d}q\,\frac{e^{-2\alpha(0,p)-\alpha(0,q)+\alpha(p+q,p)+\alpha(-p-q,-q)+\alpha(0,p+q)}}
{(p^{2}-m^{2})^{2}(q^{2}-m^{2})}\\
=&\int\mathrm{d}^{d}q\,\frac{e^{-\alpha(0,p)}}
{(p^{2}-m^{2})^{2}(q^{2}-m^{2})}
\end{align}
since by using \eqref{alpha3} we have
\begin{equation}
\alpha(p+q,p)=-\alpha(0,p+q)+\alpha(0,p)+\alpha(0,q)-\alpha(-p-q,-q).
\end{equation}
We see that with respect to the commutative case
the only correction is in the factor $e^{-\alpha(0,p)}$ which is the correction of the free propagator.
Therefore, the ultraviolet divergence of the planar diagram is the same as the ordinary one.
Consider now the non-planar case in f\mbox{}igure~\ref{TPD}(b).
The structure is the same as in the planar case, but this time the assignments are given by \eqref{NPA}.
We have up to a constant
\begin{align}
\nonumber
G^{(2)}_{NP}=&\int\mathrm{d}^{d}q\,\frac{e^{-\alpha(0,p)-\alpha(0,-p)-\alpha(0,q)}}
{(p^{2}-m^{2})^{2}(q^{2}-m^{2})}e^{\alpha(p+q,p)+\alpha(-p-q,-p)+\alpha(0,p+q)}\\
\nonumber
%=&\int\mathrm{d}^{d}q\,\frac{e^{-\alpha(0,p)-\alpha(0,-p)-\alpha(0,q)+
%\alpha(p+q,p)+\alpha(-p-q,-p)+\alpha(0,p+q)}}{(p^{2}-m^{2})^{2}(q^{2}-m^{2})}\\
%\nonumber
=&\int\mathrm{d}^{d}q\,\frac{e^{-2\alpha(0,p)-\alpha(0,q)+
\alpha(p+q,p)+\alpha(-p-q,-p)+\alpha(0,p+q)}}
{(p^{2}-m^{2})^{2}(q^{2}-m^{2})}\\
=&\int\mathrm{d}^{d}q\,\frac{e^{-\alpha(0,p)+\alpha(p+q,p)-\alpha(p+q,q)}}
{(p^{2}-m^{2})^{2}(q^{2}-m^{2})}
\end{align}
since by using again \eqref{alpha3} we have
\begin{align}
\nonumber
\alpha(-p-q,-p)&=-\alpha(0,-p-q)+\alpha(0,-p)+\alpha(0,-q)-\alpha(p+q,q)\\
&=-\alpha(0,p+q)+\alpha(0,p)+\alpha(0,q)-\alpha(p+q,q).
\end{align}
The one-loop correction to the propagator in the non-planar case can be rewritten as
\begin{equation}
G^{(2)}_{NP}=\int\mathrm{d}^{d}q\,\frac{e^{-\alpha(0,p)+\omega(p,q)}}
{(p^{2}-m^{2})^{2}(q^{2}-m^{2})}
\end{equation}
where we have set
\begin{equation}
\omega(p,q)=\alpha(p+q,p)-\alpha(p+q,q).
\end{equation}
Note that for both the Moyal and Wick-Voros products this term is given by
\begin{equation}
\omega_{M}(p,q)=\omega_{V}(p,q)=-i\theta^{ij}p_{i}p_{j}.
\end{equation}
The function $\omega$ has some important properties which will be useful
in the discussion of the ultraviolet behaviour of the non-planar diagram.
Using \eqref{alpha3} it is easy to get the following relations:
\begin{align}
\omega(p,p)&=0\\
\label{omegaa}
\omega(p,0)&=0\\
\label{omegab}
\omega(0,q)&=0\\
\label{omega1}
\omega(p,q)&=-\omega(q,p)&\quad\mathrm{antisymmetry}\\
\label{omega2}
\omega(-p,-q)&=\omega(p,q)&\quad\mathrm{parity}\\
\omega(-p,q)&=\omega(p,-q).
\end{align}
%
%Another relation involving $\omega$
%
%From \eqref{alpha3} we have
%\begin{multline}
%\alpha(p,q)=-\alpha(0,p)+\alpha(0,q)+\alpha(0,p-q)+\alpha(-p,-q)-\alpha(-p,q-p)-\alpha(-p,-q)
%\end{multline}
%Since
%\begin{equation}
%\alpha(-p,-q)-\alpha(-p,q-p)=\alpha(p'+q',p')-\alpha(p'+q',q')=\omega(p',q')=\omega(-q,q-p)
%\end{equation}
%with a suitable change of variables, we get
%\begin{equation}
%\alpha(p,q)=-\alpha(0,p)+\alpha(0,q)+\alpha(0,p-q)+\omega(p,p-q)-\alpha(-p,-q).
%\end{equation}
%
%
%
Moreover, $\omega$ satisf\mbox{}ies the associativity condition \eqref{ac}
and then it satisf\mbox{}ies the condition \eqref{alpha3} from which follows
\begin{equation}
\label{omega3}
\omega(p,q)=\omega(p-q,p),
\end{equation}
where we have used antisymmetry \eqref{omega1} and parity \eqref{omega2}. It can be written as well as
\begin{equation}
\label{omega4}
\omega(p,q)=\omega(q,q-p),
\end{equation}
where we have exchanged $p$ and $q$ and used once again antisymmetry \eqref{omega1}.
Finally, from \eqref{ac} we have
\begin{equation}
\alpha(p+q,p)=\alpha(p+q,r)-\alpha(p,r)+\alpha(p+q-r,p-r)
\end{equation}
and by setting $r=q$ we get
\begin{equation}
\omega(p,q)=\alpha(p,p-q)-\alpha(p,q).
\end{equation}
This quantity vanishes if the product is commutative because of \eqref{cc}.
This means that the non-planar diagram captures the non-commutativity of the product.
In other words, no change in the ultraviolet can come from a commutative product.

We now prove that the contribution to the correction to the non-planar one-loop two-point diagram
must necessarily be of the form
\begin{equation}
\omega(p,q)=-i\theta^{ij}p_{i}p_{j}
\end{equation}
exactly like in the Moyal and Wick-Voros cases. We only need the extra assumption
that $\alpha$ and so $\omega$ can be expanded in a power series of $p$ and $q$.
The parity relation \eqref{omega2} requires the series to be composed only of even monomials.
Let us express the function $\omega$ with a multi-index notation as
\begin{equation}
\omega(p,q)=\sum_{\boldsymbol{ij}}a_{\boldsymbol{ij}}p^{\boldsymbol{i}}q^{\boldsymbol{j}}
\end{equation}
with $\boldsymbol{i}=(i_{1}, i_{2},\ldots, i_{d})$ and
\begin{equation}
p^{\boldsymbol{i}}=p^{i_{1}}_{1}p^{i_{2}}_{2}\ldots p^{i_{d}}_{d}.
\end{equation}
Note that from \eqref{omega2} follows that
\begin{equation}
\sum_{\boldsymbol{ij}}a_{\boldsymbol{ij}}p^{\boldsymbol{i}}q^{\boldsymbol{j}}=
-\sum_{\boldsymbol{ij}}a_{\boldsymbol{ij}}q^{\boldsymbol{i}}p^{\boldsymbol{j}}=
-\sum_{\boldsymbol{ij}}a_{\boldsymbol{ji}}p^{\boldsymbol{i}}q^{\boldsymbol{j}}.
\end{equation}
That is,
\begin{equation}
\sum_{\boldsymbol{ij}}(a_{\boldsymbol{ij}}+a_{\boldsymbol{ji}})p^{\boldsymbol{i}}q^{\boldsymbol{j}}=0
\end{equation}
Because of the independence of $p$ and $q$, we have
\begin{equation}
a_{\boldsymbol{ij}}=-a_{\boldsymbol{ji}}.
\end{equation}
Therefore, the coef\mbox{}f\mbox{}icients $a$'s are antisymmetric
for the exchange of the multi-indices $\boldsymbol{i}$ and $\boldsymbol{j}$.
Now from \eqref{omega4} follows that
\begin{equation}
\sum_{\boldsymbol{ij}}a_{\boldsymbol{ij}}p^{\boldsymbol{i}}q^{\boldsymbol{j}}
=\sum_{\boldsymbol{ij}}a_{\boldsymbol{ij}}q^{\boldsymbol{i}}(q-p)^{\boldsymbol{j}}
=\sum_{\boldsymbol{ij}}a_{\boldsymbol{ji}}(q-p)^{\boldsymbol{i}}q^{\boldsymbol{j}}
=-\sum_{\boldsymbol{ij}}a_{\boldsymbol{ij}}(q-p)^{\boldsymbol{i}}q^{\boldsymbol{j}}
\end{equation}
That is,
\begin{equation}\label{azero}
\sum_{\boldsymbol{ij}}a_{\boldsymbol{ij}}
[p^{\boldsymbol{i}}+(q-p)^{\boldsymbol{i}}]q^{\boldsymbol{j}}=0.
\end{equation}
This condition implies, because of the independence of $p$ and $q$,
that the coef\mbox{}f\mbox{}icients $a$'s must vanish
except in the case in which all of the $j_{a}$'s but one vanish.
In this case the antisymmetry of the $a$'s ensures that \eqref{azero} vanishes
without putting further constraints on the coef\mbox{}f\mbox{}icients.
Using antisymmetry the same reasoning can be done for the f\mbox{}irst multi-index
and this shows that $\omega$ is of the kind
\begin{equation}
\omega(p,q)=A\varepsilon^{ij}p_{i}p_{j}
\end{equation}
where $A$ is in general a complex number. Moreover, since from \eqref{starcondition} follows that $\bar{A}=-A$,
we can set $A=-i\theta$ with $\theta$ real number and this concludes the proof.
Therefore, like in the Moyal and Wick-Voros cases, the non-planar diagram presents
the phenomenon of ultraviolet/infrared mixing. In other words,
for high internal momentum the ultraviolet divergences are damped by a phase,
but these divergences reappear in the infrared namely for low incoming momenta.

We now want to highlight that $\omega$ is related to the commutator of the coordinates.
To this end, we must f\mbox{}irst derive the commutation relations and for this
it is more useful to rewrite \eqref{GTIP} as
\begin{equation}
(f\star g)(x)=\int\frac{\mathrm{d}^{d}p}{(2\pi)^{d}}\frac{\mathrm{d}^{d}q}{(2\pi)^{d}}\,
\tilde{f}(p)\tilde{g}(q)e^{\alpha(p+q,p)}e^{i(p+q)\cdot x}
\end{equation}
with a change of variables. We have
\begin{align}
\nonumber
x^{i}\star x^{j}&=-\int\frac{\mathrm{d}^{d}p}{(2\pi)^{d}}\frac{\mathrm{d}^{d}q}{(2\pi)^{d}}\,
\left(\frac{\partial}{\partial p_{i}}\delta^{(d)}(p)\right)
\left(\frac{\partial}{\partial q_{j}}\delta^{(d)}(q)\right)
e^{\alpha(p+q,p)}e^{i(p+q)\cdot x}\\
\nonumber
&=\int\frac{\mathrm{d}^{d}p}{(2\pi)^{d}}\frac{\mathrm{d}^{d}q}{(2\pi)^{d}}\,
\delta^{(d)}(p)\left(\frac{\partial}{\partial q_{j}}\delta^{(d)}(q)\right)
\frac{\partial}{\partial p_{i}}\left[e^{\alpha(p+q,p)}e^{i(p+q)\cdot x}\right]\\
\nonumber
&=\int\frac{\mathrm{d}^{d}p}{(2\pi)^{d}}\frac{\mathrm{d}^{d}q}{(2\pi)^{d}}\,
\delta^{(d)}(p)\left(\frac{\partial}{\partial q_{j}}\delta^{(d)}(q)\right)
\left(\frac{\partial\alpha}{\partial p_{i}}(p+q,p)+ix^{i}\right)e^{\alpha(p+q,p)}e^{i(p+q)\cdot x}\\
%\nonumber
%&=\int\frac{\mathrm{d}^{d}q}{(2\pi)^{d}}\,
%\left(\frac{\partial}{\partial q_{j}}\delta^{(d)}(q)\right)
%\left(\frac{\partial\alpha}{\partial p_{i}}(q,0)+ix^{i}\right)e^{\alpha(q,0)}e^{iq\cdot x}\\
\nonumber
&=\int\frac{\mathrm{d}^{d}q}{(2\pi)^{d}}\,
\left(\frac{\partial}{\partial q_{j}}\delta^{(d)}(q)\right)
\left(\frac{\partial\alpha}{\partial p_{i}}(q,0)+ix^{i}\right)e^{iq\cdot x}\\
\nonumber
&=-\int\frac{\mathrm{d}^{d}q}{(2\pi)^{d}}\,
\delta^{(d)}(q)
\frac{\partial}{\partial q_{j}}\left[\left(\frac{\partial\alpha}{\partial p_{i}}(q,0)+ix^{i}\right)
e^{iq\cdot x}\right]\\
\nonumber
&=-\int\frac{\mathrm{d}^{d}q}{(2\pi)^{d}}\,
\delta^{(d)}(q)
\left[\frac{\partial^{2}\alpha}{\partial p_{i}\partial q_{j}}(q,0)+
\left(\frac{\partial\alpha}{\partial p_{i}}(q,0)+ix^{i}\right)ix^{j}\right]e^{iq\cdot x}\\
%\nonumber
%&=-\left[\frac{\partial^{2}\alpha}{\partial p_{i}\partial q_{j}}(0,0)+
%\left(\frac{\partial\alpha}{\partial p_{i}}(0,0)+ix^{i}\right)ix^{j}\right]\\
&=x^{i}x^{j}-i\frac{\partial\alpha}{\partial p_{i}}(0,0)x^{j}-
\frac{\partial^{2}\alpha}{\partial p_{i}\partial q_{j}}(0,0).
\end{align}
In a similar way we get
\begin{equation}
x^{j}\star x^{i}=x^{i}x^{j}-i\frac{\partial\alpha}{\partial p_{j}}(0,0)x^{i}-
\frac{\partial^{2}\alpha}{\partial p_{j}\partial q_{i}}(0,0).
\end{equation}
Therefore, the commutator of the coordinates is given by
\begin{equation}
[x^{i},x^{j}]_{\star}=
-i\frac{\partial\alpha}{\partial p_{i}}(0,0)x^{j}+
i\frac{\partial\alpha}{\partial p_{j}}(0,0)x^{i}-
\frac{\partial^{2}\alpha}{\partial p_{i}\partial q_{j}}(0,0)+
\frac{\partial^{2}\alpha}{\partial p_{j}\partial q_{i}}(0,0).
\end{equation}
The f\mbox{}irst two terms of this expression vanish because $\alpha$ has no linear term because of
\eqref{alphab} and \eqref{alpha1}. So the commutator of the coordinates is simply given by
\begin{equation}
[x^{i},x^{j}]_{\star}=
-\frac{\partial^{2}\alpha}{\partial p_{i}\partial q_{j}}(0,0)
+\frac{\partial^{2}\alpha}{\partial p_{j}\partial q_{i}}(0,0).
\end{equation}
Expanded $\alpha$ in a power series of $p$ and $q$ as\footnote{
Note that $\alpha$ has no constant term because of \eqref{alphac}.}
\begin{equation}
\alpha(p,q)=\alpha^{ij}p_{i}q_{j}+\ldots
\end{equation}
we have
\begin{equation}
[x^{i},x^{j}]_{\star}=-\alpha^{ij}+\alpha^{ji}.
\end{equation}
Moreover, from the def\mbox{}inition of $\omega$ we have
\begin{equation}
\omega(p,q)=\alpha^{ij}(p_{i}+q_{i})p_{j}-\alpha^{ij}(p_{i}+q_{i})q_{j}=\alpha^{ij}(p_{i}+q_{i})(p_{j}-q_{j})
\end{equation}
in which survive only the mixed terms because of \eqref{omegaa} and \eqref{omegab}. That is,
\begin{equation}
\omega(p,q)=-\alpha^{ij}p_{i}q_{j}+\alpha^{ij}q_{i}p_{j}=\left(-\alpha^{ij}+\alpha^{ji}\right)p_{i}q_{j}.
\end{equation}
Therefore, the term appearing in the exponent of the one-loop correction to the propagator
in the non-planar case is just the commutator of the $x$'s multiplied by the external and internal momenta
like in the Moyal and Wick-Voros cases. In other words, the Moyal and Wick-Voros cases are generic and
their behaviour is replicated by any translation invariant associative product.
More in general we can conclude that star products with the same commutator and hence the same Poisson structure,
which are equivalent in the sense of Kontsevich, have the same structure of ultraviolet/infrared mixing
(see also~\cite{TanasaVitale}).

\chapter{The Moyal and Wick-Voros products as twisted products}

\emph{In this chapter we present a comparison of the non-commutative f\mbox{}ield theories
built using the Moyal and Wick-Voros products in the context of the twisted non-commutativity.
The comparison is made at the level of Green's functions and $S$-matrix and
we show that while the Green's functions are dif\mbox{}ferent for the two theories,
the $S$-matrix is the same in both cases and is dif\mbox{}ferent from the commutative case.}

\section{Towards twisted non-commutativity}

We have seen that the vertex and Green's functions of the Wick-Voros f\mbox{}ield theory
are dif\mbox{}ferent from the Moyal ones.
This leads to a contradiction. Indeed, with the introduction of another star product
which gives the same commutation relation, one can heuristically reason as follows.
The presence of the non-commutativity described by \eqref{CNC} gives the non-commutative structure of space-time
regardless of the realization of the product one uses. Therefore, the star product is just a way
to express the structure of space-time and so the results should be the same.

The element that we need consider to solve this puzzle is symmetries.
The commutation relation \eqref{CNC} breaks the Poincar\'{e} symmetry
and this is not a desirable feature for a fundamental theory. In particular, it breaks the Lorentz symmetry,
but retain the translational one. However, the symmetry can be reinstated at a deformed level
since both products can be seen as coming from a Drinfeld twist~\cite{Drinfeld1,Drinfeld2}.
The non-commutativity described by the two star products
is therefore a twisted non-commutativity.

The presence of a twist forces us to reconsider all of the steps
in a f\mbox{}ield theory which has to be built in a coherent twisted way.
We will see that there is equivalence between the Moyal and Wick-Voros f\mbox{}ield theories
at the level of $S$-matrix.
This is in agreement with our physical intuition since Green's functions are not observable quantities,
while $S$-matrix is. Furthermore, the equivalence is only obtained
if a consistent procedure of twisting all products is applied.
In this way the Poincar\'{e} symmetry, which appears to be broken in \eqref{CNC},
is preserved, though in a deformed way, as a non-commutative and non-cocommutative Hopf algebra.
However, there is some ambiguity in the issue of twisting%
~\cite{Vassilevich,BalachandranManganoPinzulVaidya,BalachandranPinzulQureshi,Tureanu,ChaichianTureanu,
BuKimLeeVacYee,Zhan,FioreWess,Fiore,ChaichianNishijimaSalminenTureanu}
and what we do in this chapter is just to use the f\mbox{}ield theories built
with the Moyal and Wick-Voros products to check each other.
This gives us an indication on the procedure to follow
for non-commutative theories coming from a twist%
~\cite{ChaichianKulishNishijimaTureanu,Wess,AschieriBlohmannDimitrijevicMeyerSchuppWess,
AschieriDimitrijevicKulishLizziWess}.

\section{The Moyal and Wick-Voros products as twisted products}

Given the Poincar\'{e} algebra $\mathcal{P}$ and its universal enveloping algebra $U(\mathcal{P})$,
a twist $\mathcal{F}$ is an invertible element of $U(\mathcal{P})\otimes U(\mathcal{P})$
such that
\begin{align}
\label{cocyclecondition}
\mathcal{F}_{12}(\Delta\otimes\mathrm{id})\mathcal{F}&=\mathcal{F}_{23}(\mathrm{id}\otimes\Delta)\mathcal{F}\\
(\epsilon\otimes\mathrm{id})\mathcal{F}&=(\mathrm{id}\otimes\epsilon)\mathcal{F}=\mathbbm{1}
\end{align}
where
\begin{equation}
\mathcal{F}_{12}=\mathcal{F}\otimes\mathbbm{1}\quad\mathrm{and}\quad\mathcal{F}_{23}=\mathbbm{1}\otimes\mathcal{F}.
\end{equation}
Observe that the condition \eqref{cocyclecondition} nothing other than
a cocycle condition\footnote{See appendix~\ref{Hopf algebras}.}.
Before to proceed further, we recall that the Poincar\'{e} algebra $\mathcal{P}$ describes the symmetries
of ordinary Minkowski space-time $\mathcal{M}$ and it is characterized by the following commutation relations:
\begin{align}
[M_{\mu\nu},M_{\rho\sigma}]&=i(\eta_{\mu\sigma}M_{\nu\rho}-\eta_{\nu\sigma}M_{\mu\rho}+
\eta_{\nu\rho}M_{\mu\sigma}-\eta_{\mu\rho}M_{\nu\sigma})\\
[M_{\mu\nu},P_{\rho}]&=i(\eta_{\nu\rho}P_{\mu}-\eta_{\mu\rho}P_{\nu})\\
[P_{\mu},P_{\nu}]&=0
\end{align}
where $M_{\mu\nu}$ and $P_\mu$ are respectively the Lorentz generators and translation generators
which are represented on the algebra of functions on Minkowski space-time $\mathcal{M}$ by
\begin{align}
P_{\mu}&=-i\partial_{\mu}\\
M_{\mu\nu}&=i(x_{\mu}\partial_{\nu}-x_{\nu}\partial_{\mu}).
\end{align}
Moreover, its universal enveloping algebra $U(\mathcal{P})$ has a non-commutative,
but cocommutative Hopf algebra structure with the coproduct, the counit and the antipode given respectively by
\begin{align}
\label{Coproduct}
\Delta_{0}(X)&=X\otimes\mathbbm{1}+\mathbbm{1}\otimes X\\
\varepsilon_{0}(X)&=0\\
S_{0}(X)&=-X
\end{align}
where $X$ stands for $M_{\mu\nu}$ and $P_{\mu}$.
For the Moyal and Wick-Voros cases the twist is given respectively by
\begin{align}
\mathcal{F}_{M}&=e^{-\frac{i}{2}\theta^{ij}\partial_{i}\otimes\partial_{j}}\label{moytwist}\\
\mathcal{F}_{V}&=e^{-\theta\partial_{+}\otimes\partial_{-}}\label{vortwist}.
\end{align}
We now assume the following point of view in agreement with~\cite{Wess,AschieriDimitrijevicMeyerWess,Aschieri,AschieriLizziVitale}.
The non-commutativity can be seen a consequence of twisting of all products.
Thus, for example, the ordinary commutative product between functions\footnote{
At this level we need not specify which kind of algebra of functions we are considering.
The algebra of formal series of the generators is adequate, but more restricted algebras
such as algebra of Schwarzian functions can also be considered.} on space-time $\mathcal{M}$
\begin{equation}
m_{0}:\mathrm{Fun}(\mathcal{M})\otimes\mathrm{Fun}(\mathcal{M})\to\mathrm{Fun}(\mathcal{M})
\end{equation}
def\mbox{}ined by
\begin{equation}
m_{0}(f\otimes g)=fg
\end{equation}
is consistently deformed by composing it with the twist $\mathcal{F}$ which can be viewed as well as a map
\begin{equation}
\mathcal{F}:\mathrm{Fun}(\mathcal{M})\otimes\mathrm{Fun}(\mathcal{M})\to
\mathrm{Fun}(\mathcal{M})\otimes\mathrm{Fun}(\mathcal{M}).
\end{equation}
obtaining a deformed version $m_{\star}$ of the ordinary product $m_{0}$.
In other words, the star product can be seen as the composition of the ordinary product $m_{0}$ with the twist $\mathcal{F}$:
\begin{equation}
m_{\star}=m_{0}\circ\mathcal{F}^{-1}.
\end{equation}
In particular, the Moyal and Wick-Voros products are given respectively by
\begin{align}
f\star_{M}g&=m\circ\mathcal{F}^{-1}_{\star_{M}}(f\otimes g)\\
f\star_{V}g&=m\circ\mathcal{F}^{-1}_{\star_{V}}(f\otimes g).
\end{align}
Note the the associativity of the product is ensured by the cocycle condition \eqref{cocyclecondition}~\cite{AschieriDimitrijevicMeyerWess,Aschieri}.
Moreover, notice that  the twist $\mathcal{F}$ deforms
the structure of the universal enveloping algebra of Poincar\'{e} algebra $U(\mathcal{P})$
which becomes a non-cocommutative Hopf algebra.
Indeed, the twist $\mathcal{F}$ changes the coproduct of $U(\mathcal{P})$ according to~\cite{Drinfeld2}
\begin{equation}\label{TwistedCoproduct}
\Delta_{\star}(X)=\mathcal{F}\Delta_{0}(X)\mathcal{F}^{-1}.
\end{equation}
Since the translation generators $P_{\mu}$ are commutative, their coproduct is not deformed:
\begin{equation}
\Delta_{\mathcal{F}}(P_{\mu})=\Delta_{0}(P_{\mu}).
\end{equation}
However, the coproduct of the Lorentz generators is changed~\cite{ChaichianKulishNishijimaTureanu}:
\begin{equation}
\Delta_{\mathcal{F}}(M_{\mu\nu})=\Delta_{0}(M_{\mu\nu})
-\frac{1}{2}\theta^{\rho\sigma}\left[(\eta_{\mu\rho}P_{\nu}-\eta_{\nu\rho}P_{\mu})\otimes P_{\sigma}+
P_{\rho}\otimes(\eta_{\mu\sigma}P_{\nu}-\eta_{\nu\sigma}P_{\mu})\right].
\end{equation}
In the untwisted case the action of the Poincar\'{e} generators $X$ on the ordinary product of two functions $f$ and $g$ is given by
\begin{equation}
X(fg)=(Xf)g+fXg
\end{equation}
which can be rewritten as
\begin{equation}
X(fg)=m_{0}(\Delta(X)(f\otimes g)).
\end{equation}
In other words, their action is applied through the original coproduct \eqref{Coproduct}. Instead, in the twisted case
their action  has to be applied through the twisted coproduct \eqref{TwistedCoproduct}. That is,
\begin{equation}
X(f\star g)=m_{\star}(\Delta_{\star}(X)(f\otimes g)).
\end{equation}
It is now possible to show that the Poincar\'{e} symmetry is retained as a twisted symmetry
(see~\cite{Matlock} for a discussion of twisted conformal symmetry).
Indeed, it is not dif\mbox{}f\mbox{}icult to see that the canonical non-commutative relation \eqref{CNC}
is a twisted Poincar\'{e} invariant~\cite{ChaichianKulishNishijimaTureanu}:
\begin{equation}
X([x^{\mu},x^{\nu}]_{\star})=0.
\end{equation}
Eventually, we also introduce the universal ${\mathcal R}$-matrix which represents
the permutation group in non-commutative case
\begin{equation}
\mathcal{R}=\mathcal{F}_{21}\mathcal{F}^{-1}
\end{equation}
with
\begin{equation}
\mathcal{F}_{21}(a\otimes b) = \tau\circ\mathcal{F}\circ\tau(a\otimes b)
\end{equation}
and $\tau$ the usual exchange operator
\begin{equation}
\tau(a\otimes b)=b\otimes a.
\end{equation}
For the cases at hand, it is easy to see that
\begin{equation}
\mathcal{R}_{\star_{V}}=\mathcal{R}_{\star_{M}}=\mathcal{F}_{\star_{M}}^{-2}.
\end{equation}
Therefore, the exchange operator and so the statistics are the same in both the Moyal and Wick-Voros cases.

We conclude this section by nothing that it is possible to introduce an unif\mbox{}ied notation
for the Moyal and Wick-Voros products which will be very useful in the following.
Indeed, we can write the vertex in both cases as
\begin{equation}
V_{\star}=V\prod_{a<b}e^{k_{a}\bullet k_{b}}
\end{equation}
and the four-point Green's functions as
\begin{equation}
\tilde{G}_{0}^{(4)}=-ig(2\pi)^{3}\frac{e^{\sum_{a\leq b}k_{a}\bullet k_{b}}}
{\prod_{a=1}^{4}\left(k_{a}^{2}-m_{a}^{2}\right)}\delta^{(3)}\!\left(\sum_{a=1}^{4}k_{a}\right)
\end{equation}
where
\begin{equation}
k_{a}\bullet k_{b}=
\left\{\begin{array}{ll}
-\frac{i}{2}\theta^{ij}k_{ai}k_{bj}\quad&\mbox{Moyal}\\
-\theta k_{a-}k_{b+}\quad&\mbox{Wick-Voros.}
\end{array}\right.
\end{equation}
The one-loop corrections to the propagator can be written for the planar case as
\begin{equation}
\tilde{G}_{\mathrm{P}}^{(2)}=-i\frac{g}{3}\int\frac{\mathrm{d}^{3}q}{(2\pi)^{3}}
\frac{e^{p\bullet p}}{(p^{2}-m^{2})^{2}(q^{2}-m^{2})}
\end{equation}
and for the non-planar one as
\begin{equation}
\tilde{G}_{\mathrm{NP}}^{(2)}=-i\frac{g}{6}\int\frac{\mathrm{d}^{3}q}{(2\pi)^{3}}
\frac{e^{p\bullet p+p\bullet q-q\bullet p}}{(p^{2}-m^{2})^{2}(q^{2}-m^{2})}.
\end{equation}
Furthermore, one-loop correction to the four-point Green's function can be written for the planar case as
\begin{equation}\label{PFPGF}
\tilde{G}_{\mathrm{P}}^{(4)}=\frac{(-ig)^{2}}{8}(2\pi)^{3}\int\frac{\mathrm{d}^{3}q}{(2\pi)^{3}}
\frac{e^{\sum_{a\leq b}k_{a}\bullet k_{b}}\delta^{(3)}\!\left(\sum_{a=1}^{4}k_{a}\right)}
{(q^{2}-m^{2})\left[(k_{1}+k_{2}-q)^{2}-m^{2}\right]\prod_{a=1}^{4}(k_{a}^{2}-m^{2})}
\end{equation}
and for the non-planar one as
\begin{equation}\label{NPFPGF}
\tilde{G}_{\mathrm{NP}_{a}}^{(4)}=\frac{(-ig)^{2}}{8}(2\pi)^{3}\int\frac{\mathrm{d}^{3}q}{(2\pi)^{3}}
\frac{e^{\sum_{a\leq b}k_{a}\bullet k_{b}+E_{a}}
\delta^{(3)}\!\left(\sum_{a=1}^{4}k_{a}\right)}
{(q^{2}-m^{2})\left[(k_{1}+k_{2}-q)^{2}-m^{2}\right]\prod_{a=1}^{4}(k_{a}^{2}-m^{2})}
\end{equation}
where in unif\mbox{}ied notation
\begin{align}
\notag
E_{1}&=q\bullet k_{1}- k_{1}\bullet q\\
\notag
E_{2}&= k_{2}\bullet q- q\bullet k_{2}+k_{3}\bullet q-q\bullet k_{3}\\
E_{3}&=k_{1}\bullet q-q\bullet k_{1}+k_{2}\bullet q-q\bullet k_{2}.
\end{align}

\section{Twist-deformed products}

We have now the necessary ingredients to calculate a physical process,
like the $S$-matrix for the elastic scattering of two particles.
We recall that one of the crucial ingredients in the study of the $S$-matrix
is the issue of Poincar\'{e} invariance.
If we na\"{\i}vely insert the Green's functions found for the Moyal and Wick-Voros cases
into the calculation of the $S$-matrix, we f\mbox{}ind a dependence of it from the external momenta,
something like a momentum dependence of the coupling constant.
Furthermore, we f\mbox{}ind that the result is dif\mbox{}ferent for the two cases
in contradiction with the heuristic reasoning we made in the introduction.
We also f\mbox{}ind a breaking of Poincar\'{e} invariance.
The reason for the breaking of Poincar\'{e} invariance is that
the commutator \eqref{CNC} breaks this invariance\footnote{
We are considering $\theta^{ij}$ to be constant. Another possibility which preserves Poincar\'{e} invariance
is to have it a tensor~\cite{DoplicherFredenhagenRoberts,BahnsDoplicherFredenhagenPiacitelli}
or to have it transform together with the product~\cite{Gracia-BondiaLizziRuizRuizVitale}.}.

The invariance can be reinstated considering it as a twisted symmetry
i.e.~as a symmetry described by a non-commutative and non-cocommutative Hopf algebra~\cite{ChaichianKulishNishijimaTureanu,Wess,AschieriBlohmannDimitrijevicMeyerSchuppWess}.
Our purpose is therefore to show, with an explicit calculation of scattering amplitudes,
that the na\"{\i}ve procedure which leads to a dif\mbox{}ference among the two cases
can be corrected by a coherent twisting procedure.
We see that if the twisted symmetry is properly implemented, then the f\mbox{}inal ``physical'' result
will be the same in the Wick-Voros and Moyal cases, despite the presence of dif\mbox{}ferent propagators and vertices.

Consider the elastic scattering of two particles as described in f\mbox{}igure~\ref{TPES}.
\begin{figure}[htb]
\begin{center}
\unitlength 1.5mm
\linethickness{0.5pt}
\ifx\plotpoint\undefined\newsavebox{\plotpoint}\fi
\begin{picture}(28.938,19.813)(0,0)
\put(21.92,11.932){\circle*{2.358}}
\thicklines
\multiput(20.948,12.993)(-.034114826,.033597935){171}{\line(-1,0){.034114826}}
\put(17.854,15.026){\line(1,0){1.061}}
\put(19.092,16.087){\line(0,-1){1.061}}
\put(21.063,10.938){\line(-1,-1){4.875}}
\put(19,8.938){\line(-1,0){.938}}
\put(19,8.875){\line(0,-1){.75}}
\multiput(23,12.875)(.033735821,.034090935){176}{\line(0,1){.034090935}}
\put(25.063,15.938){\line(1,0){.875}}
\put(25.938,15.938){\line(0,-1){.875}}
\multiput(23,12.938)(-.03125,-.0625){4}{\line(0,-1){.0625}}
\multiput(22.813,11.063)(.03435433,-.033526515){151}{\line(1,0){.03435433}}
\put(26,8.875){\line(0,-1){.75}}
\put(25,7.938){\line(1,0){.875}}
\put(18.688,19.625){\makebox(0,0)[cc]{\large $k_1$}}
\put(26.25,19.813){\makebox(0,0)[cc]{\large $k_3$}}
\put(15.688,8.75){\makebox(0,0)[cc]{\large $k_2$}}
\put(28.438,9.063){\makebox(0,0)[cc]{\large $k_4$}}
\end{picture}
\end{center}
\caption{The two-particles elastic scattering.}
\label{TPES}
\end{figure}
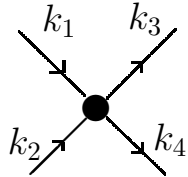
The f\mbox{}irst consequence of non-commutativity is that,
since the vertex is not invariant for noncyclic exchange of the particles,
we have to twist-symmetrize the incoming and outgoing states using the $\mathcal{R}$-matrix.
Several aspects of this twist-symmetrization and the consequences for spin and statistics
have been discussed in~\cite{BalachandranManganoPinzulVaidya,BalachandranJosephPadmanabhan,FioreSchupp1,FioreSchupp2}.
In the commutative case the order of the propagators into the vertex is irrelevant,
but in our case there are several twists at work and we must be careful in considering all of them.
Therefore, let us become with the def\mbox{}inition of multiparticle states as twisted tensor products.
Consider f\mbox{}irst the one-particle state. It is def\mbox{}ined as usual as
\begin{equation}\label{ketk}
\ket{k}=a^{\dagger}_{k}\ket{0}
\end{equation}
where the operators $a_{k}$ and $a_{k}^{\dag}$ can be expressed in terms of
the free f\mbox{}ield of Klein-Gordon equation
\begin{equation}
\phi(x)=\int\frac{\mathrm{d}^{3}k}{\sqrt{(2\pi)^{3}2\omega_{k}}}
\left(a_{k}e^{-ik\cdot x}+a^{\dag}_{k}e^{ik\cdot x}\right)
\end{equation}
respectively as
\begin{align}
\notag
a_{k}&=\frac{i}{\sqrt{(2\pi)^{3}2\omega_{k}}}\int\mathrm{d}^{3}x\,
e^{ik\cdot x}\stackrel{\leftrightarrow}{\partial_{0}}\phi(x)\\
\label{aexpansion}
a^{\dag}_{k}&=-\frac{i}{\sqrt{(2\pi)^{3}2\omega_{k}}}\int\mathrm{d}^{3}x\,
e^{-ik\cdot x}\stackrel{\leftrightarrow}{\partial_{0}}\phi(x)
\end{align}
with
\begin{equation}
f\stackrel{\leftrightarrow}{\partial_{0}}g=f\partial_{0}g-(\partial_{0}f)g.
\end{equation}
Since $a_{k}$ and $a_{k}^{\dag}$ may be regarded, for any f\mbox{}ixed $k$, as functionals of the f\mbox{}ields,
their star product can be obtained as in~\cite{AschieriLizziVitale} obtaining in the Moyal case
\begin{align}
a(k)\star_{M}a(k')&=e^{-\frac{i}{2}\theta^{ij}k_{i}k'_{j}}a(k)a(k')\\
a(k)\star_{M}a^{\dag}(k')&=e^{\frac{i}{2}\theta^{ij}k_{i}k'_{j}}a(k)a^{\dag}(k')\\
a^{\dag}(k)\star_{M}a(k')&=e^{\frac{i}{2}\theta^{ij}k_{i}k'_{j}}a^{\dag}(k)a(k')\\
a^{\dag}(k)\star_{M}a^{\dag}(k')&=e^{-\frac{i}{2}\theta^{ij}k_{i}k'_{j}}a^{\dag}(k)a^{\dag}(k')
\end{align}
and in the Wick-Voros one
\begin{align}
a(k)\star_{V}a(k')&=e^{-\theta k_{-}k'_{+}}a(k)a(k')\\
a(k)\star_{V}a^{\dag}(k')&=e^{\theta k_{-}k'_{+}}a(k)a^{\dag}(k')\\
a^{\dag}(k)\star_{V}a(k')&=e^{\theta k_{-}k'_{+}}a^{\dag}(k)a(k')\\
a^{\dag}(k)\star_{V}a^{\dag}(k')&=e^{-\theta k_{-}k'_{+}}a^{\dag}(k)a^{\dag}(k').
\end{align}
Consider now the two-particle state. In the ordinary case it is def\mbox{}ined by
\begin{equation}\label{ketkakb}
\ket{k_{a},k_{b}}=\ket{k_{a}}\otimes\ket{k_{b}}
\end{equation}
and the symmetrized state, which is an eigenstate of the exchange operator
\begin{equation}
\tau\ket{k_{a}}\otimes\ket{k_{b}}=\ket{k_{b}}\otimes\ket{k_{a}}
\end{equation}
with eigenvalue $1$, is
\begin{equation}\label{ketkakbsimm}
\ket{k_{a},k_{b}}_{\mathrm{simm}}=\frac{\ket{k_{a}}\otimes\ket{k_{b}}+\ket{k_{b}}\otimes\ket{k_{a}}}{2}.
\end{equation}
Note however that for the comparison we are going to make later we will not
actually use the fact that the state has to be symmetrized. In fact, inserting the two expressions
\eqref{ketkakb} or \eqref{ketkakbsimm} in the calculation of the $S$-matrix
does not make a dif\mbox{}ference (for the connected diagrams)
because of the invariance for exchange on the incoming momenta
and we are discussing the issue of symmetrization of states just for completeness.
The symmetries for identical particles change for the non-commutative
case~\cite{BalachandranManganoPinzulVaidya,FioreSchupp1,FioreSchupp2}
and we must take into account the fact that the tensor product is twisted
as well as the exchange of particles. Therefore, we def\mbox{}ine
\begin{equation}\label{twist2ket}
\ket{k_{a},k_{b}}_{\star}=\tilde{\mathcal{F}}^{-1}\ket{k_{a}}\otimes\ket{k_{b}}=
\tilde{\mathcal{F}}^{-1}\ket{k_{a},k_{b}}
\end{equation}
where by $\tilde{\mathcal{F}}$ we indicate the twist acting in momentum space:
\begin{align}
\tilde{\mathcal{F}}_{\star_{M}}^{-1}\ket{k_{a}}\otimes\ket{k_{b}}&=
e^{-\frac{i}{2}\theta^{ij} k_{ai}\otimes k_{bj}}\ket{k_{a}}\otimes\ket{k_{b}}\\
\tilde{\mathcal{F}}_{\star_{V}}^{-1}\ket{k_{a}}\otimes\ket{k_{b}}&=
e^{\theta k_{a-}\otimes k_{b+}}\ket{k_{a}}\otimes\ket{k_{b}}.
\end{align}
This is not the only change we have to make to the state \eqref{ketkakb}.
Indeed, it has to be eigenstate of the twist-exchange, given by the $\mathcal{R}$-matrix acting in momentum space.
The properly symmetrized state is
\begin{align}
\notag
\ket{k_{a},k_{b}}_{\mathrm{simm}_{\star}}&=
\frac{1}{2}\left(\tilde{\mathcal{F}}^{-1}\ket{k_{a}}\otimes\ket{k_{b}}+
\tilde{\mathcal{F}}^{-1}\tilde{\mathcal{R}}^{-1}\ket{k_{a}}\otimes\ket{k_{b}}\right)\\
&=\frac{1}{2}\left(\tilde{\mathcal{F}}^{-1}\ket{k_{a}}\otimes\ket{k_{b}}
+\tilde{\mathcal{F}}^{-1}\tilde{\mathcal{F}}\tilde{\mathcal{F}}^{-1}_{21}
\ket{k_{a}}\otimes\ket{k_{b}}\right)\label{twist2sym}.
\end{align}
We can reexpress equation \eqref{twist2ket} as
\begin{equation}\label{tensorkakb}
\ket{k_a,k_b}_{\star}=a^{\dag}_{k_{a}}\star a^{\dag}_{k_b}\ket{0}
\end{equation}
and equation \eqref{twist2sym} as
\begin{equation}\label{tensorkakbsym}
\ket{k_a,k_b}_{\mathrm{simm}_{\star}}=\frac{a^{\dag}_{k_{a}}\star a^{\dag}_{k_{b}}+
a^{\dag}_{k_{b}}\star a^{\dag}_{k_{a}}}2\ket{0}.
\end{equation}
The next step is to twist the inner product among one-particle states.
In the commutative case we have
\begin{equation}
\langle k|k'\rangle=\bra{0}a_{k}a^{\dag}_{k'}\ket{0}=\delta^{(3)}(k-k').
\end{equation}
We twist this product in the usual way composing it with the twist
\begin{equation}\label{twistinner1part}
\left\langle k_{1}\stackrel{\star}{\big|} k_{2}\right\rangle=
\langle\cdot|\cdot\rangle\circ\mathcal{F}^{-1}(\ket{k}\otimes\ket{k'})=
\tilde{\mathcal{F}}^{-1}(k,k')\langle k|k'\rangle=\bra{0}a_k\star a^\dagger_{k'}\ket{0}
\end{equation}
where we have set
\begin{align}
\tilde{\mathcal{F}}_{\star_{M}}^{-1}(k,k')&=e^{-\frac{i}{2}\theta^{ij}k_{i}k'_{j}}\\
\tilde{\mathcal{F}}_{\star_{V}}^{-1}(k,k')&=e^{-\theta k_{-}k'_{+}}.
\end{align}
for the Moyal and Wick-Voros products respectively.
We f\mbox{}inally have to twist the inner product among two-particle states. In the commutative case:
\begin{equation}
\langle k_{1},k_{2}|k_{3},k_{4}\rangle=\delta^{(3)}(k_{1}-k_{3})\delta^{(3)}(k_{2}-k_{4}).
\end{equation}
Instead, in the non-commutative case we have to twist the two-particle state according to \eqref{twist2ket}
and then we have to twist the inner product according to the two-particle generalization of \eqref{twistinner1part}.
To this end, we must consider the action of the twist on two-particle states which is given
by the coproduct of the Hope algebra. Given a representation of an element of the Hope algebra on a space,
the representation of the element on the product of states is given in the undeformed case by
\begin{equation}\label{coprod0}
\Delta_{0}(u)(f\otimes g)=(\mathbbm{1}\otimes u+u\otimes\mathbbm{1})(f\otimes g)
\end{equation}
For the twisted Hope algebra the coproduct is deformed according to the fact that
it is the $\mathcal{R}$-matrix which realizes the permutations:
\begin{equation}\label{coprodstar}
\Delta_{\star}(u)(f\otimes g)=\left(\mathbbm{1}\otimes u +\mathcal{R}^{-1}(u\otimes\mathbbm{1})\right)(f\otimes g)
\end{equation}
However, the twists we are considering are built out of translations whose coproduct is undeformed
\begin{equation}
\Delta_{\star_{M}}(\partial_{i})=\Delta_{\star_{V}}(\partial_{i})=
\Delta_{0}(\partial_{i})=\mathbbm{1}\otimes\partial_{i}+\partial_{i}\otimes\mathbbm{1}.
\end{equation}
Since we are acting on two-particle states we need to def\mbox{}ine also
\begin{equation}
\Delta_{\star}(\partial_{i}\otimes\partial_{j})=\Delta_{0}(\partial_{i}\otimes\partial_{j})=
\mathbbm{1}\otimes\mathbbm{1}\otimes\partial_{i}\otimes\partial_{j}+
\partial_{i}\otimes\partial_{j}\otimes\mathbbm{1}\otimes\mathbbm{1}.
\end{equation}
Therefore, the twisted inner product among two-particle states
\begin{equation}
\left\langle{k_{1},k_{2}\stackrel{\star}{\big|}k_{3},k_{4}}\right\rangle=
\langle\cdot|\cdot\rangle\circ\Delta_{\star}\left({\mathcal{F}}^{-1}\right)(\ket{k_{1},k_{2}}\otimes\ket{k_{3},k_{4}})
\end{equation}
may be easily computed to be
\begin{align}
\notag
\left\langle{k_{1},k_{2}\stackrel{\star_{M}}{\big|}k_{3},k_{4}}\right\rangle&=
e^{\frac{i}{2}\theta^{ij}(k_{1i}+k_{2i})(k_{3j}+k_{4j})}
\langle k_{1},k_{2}|k_{3},k_{4}\rangle\\
\label{twistinner2}
\left\langle{k_{1},k_{2}\stackrel{\star_{V}}{\big|}k_{3},k_{4}}\right\rangle&=
e^{\theta(k_{1-}+k_{2-})(k_{3+}+k_{4+})}\langle k_{1},k_{2}|k_{3},k_{4}\rangle.
\end{align}
We can now calculate the twisted inner product of twisted states.
Combining \eqref{twistinner2} with \eqref{twist2ket} we obtain the simple expressions
\begin{align}
{\phantom{\bigg\rangle}}_{\star_{M}}\!\!\!\hsstarmoy{k_{1},k_{2}}{k_{3},k_{4}}_{\!\!\!\star_{M}}&=
e^{\frac{i}{2}\theta^{ij}\sum_{a<b}k_{ai}k_{bj}}\langle k_{1},k_{2}|k_{3},k_{4}\rangle\\
{\phantom{\bigg\rangle}}_{\star_{V}}\!\!\!\hsstarvor{k_{1},k_{2}}{k_{3},k_{4}}_{\!\!\!\star_{V}}&=
e^{\theta\sum_{a<b}k_{a-}k_{b+}}\langle k_{1},k_{2}|k_{3},k_{4}\rangle.
\end{align}
That is, in unif\mbox{}ied notation
\begin{equation}
{\phantom{\bigg\rangle}}_{\star}\!\!\!\hsstar{k_1,k_2}{k_3,k_4}_{\!\!\!\star}=
e^{-\sum_{a<b}k_{a}\bullet k_{b}}\langle k_{1},k_{2}|k_{3},k_{4}\rangle
\end{equation}
which can be cast in the form
\begin{equation}
{\phantom{\bigg\rangle}}_{\star}\!\!\!\hsstar{k_{1},k_{2}}{k_{3},k_{4}}_{\!\!\!\star}
=\bra{0}a_{k_{1}}\star a_{k_{2}}\star a_{k_{3}}^{\dag}\star a_{k_{4}}^{\dag}\ket{0}.
\end{equation}
This is in some sense also a consistency check. We could have
started with the commutative expression
\begin{equation}
\hs{k_{1},k_{2}}{k_{3},k_{4}}=\bra{0}a_{k_{1}}a_{k_{2}}a_{k_{3}}^{\dag}a_{k_{4}}^{\dag}\ket{0}
\end{equation}
and twisted the product among the creation and annihilation operators $a_{k}$ and $a_{k}^{\dag}$
obtaining the above result. We decided to follow a longer procedure to highlight the appearance of the various twists.

\section[The twisted $S$-matrix]{The twisted $\boldsymbol{S}$-matrix}

Let $\ket{f}$ and $\ket{i}$ denote a collection of free asymptotic states at $t=\pm\infty$ respectively.
We also assume that we can def\mbox{}ine in some way the one-particle incoming and outgoing states.
This is a very nontrivial assumption since in a theory in which localization is impossible,
the concept of asymptotic state may not be well def\mbox{}ined.
Nevertheless, is it reasonable to expect that also in this theory
for small $\theta$ and for large distances and times, it is possible
to talk on incoming and outgoing states expandable in terms of momentum eigenstates $\ket{k}$.

As in standard books in quantum f\mbox{}ield theory, we def\mbox{}ine the $S$-matrix
as the matrix which describes the scattering of the initial $\ket{i}$ states into the f\mbox{}inal $\ket{f}$ states
\begin{equation}
S_{fi}={\phantom{\bigg\rangle}}_{\mathrm{in}\star}\left\langle{f}\stackrel{\star}{\big|}{i}
\right\rangle_{\star\mathrm{out}}=
{\phantom{\bigg\rangle}}_{\mathrm{out}\star}\left\langle {f}\stackrel{\star}{\big|}S\stackrel{\star}{\big|}{ i}
\right\rangle_{\star\mathrm{out}}=
{\phantom{\bigg\rangle}}_{\mathrm{in}\star}\left\langle{f}\stackrel{\star}{\big|}S\stackrel{\star}{\big|}{i}
\right\rangle_{\star\mathrm{in}}.
\end{equation}
The one-particle asymptotic state is def\mbox{}ined as in \eqref{ketk} to be
\begin{equation}\label{kek}
\ket{k}_{\mathrm{in}}=N_{\star}(k)a^{\dag}_{k}\ket{0}_{\mathrm{in}}=
-N_{\star}(k)\frac{i}{\sqrt{(2\pi)^{3}2\omega_{k}}}\int\mathrm{d}^{3}x\,
e^{-ik\cdot x}\stackrel{\leftrightarrow}{\partial_{0}}\phi_{\mathrm{in}}(x)\ket{0}_{\mathrm{in}}
\end{equation}
for the in states and an analogous formula for the out states
where $N_{\star}(k)$ is a normalization factor to be determined for the Moyal and Wick-Voros cases separately.
Moreover, we assume, as in the commutative case, that the matrix elements
of the interacting f\mbox{}ield $\phi(x)$ approaches those of the free asymptotic f\mbox{}ield
as time goes to $\pm\infty$. That is,
\begin{equation}\label{freint}
\lim_{x^{0}\to\pm\infty}\langle f|\phi(x)|i\rangle=
Z^{1/2}\langle f|\phi_{\stackrel{\mathrm{out}}{\mathrm{in}}}(x)|i\rangle
\end{equation}
with $Z$ a renormalization factor. To be def\mbox{}inite, let us consider an elastic process
of two particles in two particles. According to the previous section we have
\begin{equation}
{S_{fi}}_{\star}(k_{1},\ldots,k_{4})=
{\phantom{\bigg\rangle}}_{\mathrm{in}\star}\hstar{k_{1},k_{2}}{k_{3},k_{4}}_{\mathrm{in}\star}=
e^{\sum_{a< b}k_{a}\bullet k_{b}}
{\phantom{\rangle}}_{\mathrm{in}}\hs{k_{1},k_{2}}{k_{3},k_{4}}_{\mathrm{out}}
\end{equation}
which can be expressed in terms of Green's functions
following the same procedure as in the commutative case~\cite{Kaku}.
On repeatedly using \eqref{kek} and \eqref{freint} we arrive at
\begin{align}
\notag
S_{fi}&={\phantom{\bigg\rangle}}_{\mathrm{in}\star}\hstar{k_{1},k_{2}}{k_{3},k_{4}}_{\mathrm{out}\star}=
\mathrm{disconnected}\,\,\mathrm{graphs}\\
\notag
&+\bar{N}_{\star}(k_{1})\bar{N}_{\star}(k_{2})N_{\star}(k_{3})N_{\star}(k_{4})
\left(iZ^{-1/2}\right)^{2}e^{-\sum_{a<b}k_{a}\bullet k_{b}}\\
\label{ourproc}
&\int\prod_{a=1}^{4}\frac{\mathrm{d}^{3}x^{a}}{\sqrt{(2\pi)^{3}2\omega_{k_{a}}}}\,
e^{-ik_{a}\cdot x^{a}}\left(\partial_{\mu}^{2}+m^{2}\right)_{a}G^{(4)}(x_{1},x_{2},x_{3},x_{4})
\end{align}
where $G^{(4)}(x_{1},x_{2},x_{3},x_{4})$ is the four-point Green's function.
In order to f\mbox{}ix the normalization of the asymptotic states,
let us compute the scattering amplitude for one particle going into one particle at zeroth order.
Up to the undeformed normalization factors $N(p_{a})$, this has to give a delta function
\begin{align}
\notag
\bar{N}(k)N(p)\delta^{(3)}(k-p)&=N_{\star}^{*}(k)
N_{\star}(p){\phantom{\bigg\rangle}}_{\mathrm{in}\star}\hstar{k}{p}_{\mathrm{out}\star}\\
\notag
&=N_{\star}^{*}(k)N_{\star}(p)e^{-k\bullet p}{\phantom{\rangle}}_{\mathrm{in}}\hs{k}{p}_{\mathrm{out}}\\
&=N_{\star}^{*}(k)N_{\star}(p)e^{-k\bullet p}\delta^{(3)}(k-p)
\end{align}
from which follows
\begin{align}
N_{\star_{M}}(p)&=N(p)\\
N_{\star_{V}}(p)&=e^{-\frac{\theta}{4}\boldsymbol{p}^{2}}N(p).
\end{align}
Let us now compute the scattering amplitude for the process above
that is, the scattering of two-particles in two particles at one loop.
We have two kinds of contribution to \eqref{ourproc}, one coming from the planar term \eqref{PFPGF}
which in spatial coordinates reads
\begin{equation}
G^{(4)}_{\mathrm{P}}(x_{1},x_{2},x_{3},x_{4})=
\int\prod_{a=1}^{4}\frac{\mathrm{d}^{3}k_{a}}{\sqrt{(2\pi)^{3}2\omega_{k_{a}}}}\,
e^{ik_{a}\cdot x^{a}}\tilde{G}^{(4)}_{\mathrm{P}}(k_{1},k_{2},k_{3},k_{4})
\end{equation}
and the other coming from non-planar terms \eqref{NPFPGF}
\begin{equation}
G^{(4)}_{\mathrm{NP}}(x_{1},x_{2},x_{3},x_{4})=
\int\prod_{a=1}^{4}\frac{\mathrm{d}^{3}k_{a}}{\sqrt{(2\pi)^{3}2\omega_{k_{a}}}}\,
e^{ik_{a}\cdot x^{a}}\tilde{G}^{(4)}_{\mathrm{NP}}(k_{1},k_{2},k_{3},k_{4}).
\end{equation}
In the planar case we f\mbox{}ind the same result in the Moyal and Wick-Voros cases
which coincide with the ordinary result:
\begin{multline}
{S_{fi}}_{\star P}(k_{1},\ldots,k_{4})=\frac{(-ig)^{2}}{8}(2\pi)^{3}\bar{N}(k_{1})\bar{N}(k_{2})N(k_{3})N(k_{4})
\prod_{a=1}^{4}e^{\frac{\theta}{4}\boldsymbol{k}_{a}^{2}}\\
e^{-\sum_{a<b}k_{a}\bullet k_{b}}
\int\prod_{a=1}^{4}\frac{\mathrm{d}^{3}x^{a}}{\sqrt{(2\pi)^{3}2\omega_{k_{a}}}}\,e^{-ik_{a}\cdot x^{a}}
\int\prod_{a=1}^{4}\frac{\mathrm{d}^{3}p_{a}}{\sqrt{(2\pi)^{3}2\omega_{p_{a}}}}\,e^{ip_{a}\cdot x^{a}}
\left(-p_{a}^{2}+m^{2}\right)\\
\int\frac{\mathrm{d}^{3}q}{(2\pi)^{3}}
\frac{e^{\sum_{a\leq b}p_{a}\bullet p_{b}}\delta^{(3)}\!\left(\sum_{a=1}^{4}p_{a}\right)}
{\left(q^{2}-m^{2}\right)\left[(p_{1}+p_{2}-q)^{2}-m^{2}\right]\prod_{a=1}^{4}{\left(p_{a}^{2}-m^{2}\right)}}.
\end{multline}
The integration over the $x^a$ variables yields factors of $(2\pi)^{3}\delta^{(3)}(k_{a}-p_{a})$
and so the propagators of the external legs cancel as in the standard case; as well as the factor
\begin{equation}\label{canc}
\prod_{a=1}^{4}e^{\frac{\theta}{4}\boldsymbol{k}_{a}^{2}}e^{-\sum_{a<b}k_{a}\bullet k_{b}}
e^{\sum_{a\leq b}p_{a}\bullet p_{b}}\delta^{(3)}(k_{a}-p_{a})\to1
\end{equation}
we are left with the usual commutative expression so that
\begin{equation}
{S_{fi}}_{\star P}(k_{1},\ldots,k_{4})=S_{fi}(k_{1},\ldots,k_{4}).
\end{equation}
In the non-planar case instead we f\mbox{}ind
\begin{multline}
{S_{fi}}_{\star NP}(k_{1},\ldots,k_{4})=\frac{(-ig)^{2}}{8}(2\pi)^{3}\bar{N}(k_{1})\bar{N}(k_{2})N(k_{3})N(k_{4})
\prod_{a=1}^{4}e^{\frac{\theta}{4}\boldsymbol{k}_{a}^{2}}\\
e^{-\sum_{a<b}k_{a}\bullet k_{b}}
\int\prod_{a=1}^{4}\frac{\mathrm{d}^{3}x^{a}}{\sqrt{(2\pi)^{3}2\omega_{k_{a}}}}\,e^{-ik_{a}\cdot x^{a}}
\int\prod_{a=1}^{4}\frac{\mathrm{d}^{3}p_{a}}{\sqrt{(2\pi)^{3}2\omega_{p_{a}}}}\,e^{ip_{a}\cdot x^{a}}
\left(-p_{a}^{2}+m^{2}\right)\\
\int\frac{\mathrm{d}^{3}q}{(2\pi)^{3}}\frac{e^{\sum_{a\leq b}p_{a}\bullet p_{b}+E_{a}}
\delta^{(3)}\!\left(\sum_{a=1}^{4}p_{a}\right)}
{\left(q^{2}-m^{2}\right)\left[(p_{1}+p_{2}-q)^{2}-m^{2}\right]\prod_{a=1}^{4}{\left(p_{a}^{2}-m^{2}\right)}}.
\end{multline}
After integrating over $x^{a}$ the propagators of the external legs cancel and the simplif\mbox{}ication \eqref{canc} continues to hold, but we are left with the exponential of $E_{a}$ which does not simplify.
This factor is an imaginary phase and it has the same expression in the Moyal and Wick-Voros cases.
It depends on the $q$ so that it gets integrated and modif\mbox{}ies the ultraviolet behaviour of the loop.
Furthermore, it is responsible for the UV/IR mixing~\cite{MVRS}. Therefore, we can conclude that
\begin{equation}
{S_{fi}}_{\star_{M}NP}(k_{1},\ldots,k_{4})={S_{fi}}_{\star_{V}NP}(k_{1},\ldots,k_{4})
\ne{S_{fi}}(k_{1},\ldots,k_{4}).
\end{equation}

\chapter*{Conclusions}
\addcontentsline{toc}{chapter}{Conclusions}

Throughout this thesis we have investigated the ultraviolet behaviour of a non-commutative f\mbox{}ield theory
obtained from an ordinary one substituting the commutative product with a non-commutative one that is, a star product.
In particular, we have considered the scalar $\phi^{4}$ f\mbox{}ield theory deformed with the Wick-Voros product,
a variant of the well-known Moyal product. We have discussed both the classical and the quantum f\mbox{}ield theory
and calculated the vertex, the two- and four-point Green's functions and their corrections up to one-loop.
We have found that the vertex, like in the Moyal case,
is not anymore invariant for the exchange of the external momenta,
but it maintains invariance for cyclic permutations. Thus the planar and non-planar diagrams
for the calculation of the one-loop corrections to the Green's functions behave dif\mbox{}ferently.
Indeed, the planar diagrams have the same behaviour as the commutative ones,
while the non-planar diagrams present the phenomenon of ultraviolet/infrared mixing, like in the Moyal case~\cite{MVRS}. That is, for high internal momentum the ultraviolet divergences are damped by a phase,
but these divergences reappear in the infrared (for low incoming momenta).
This is to be expected because heuristically this is consequence of commutation relation
which is, of course, the same in both theories.

More in general, we have shown that the ultraviolet/infrared mixing found for the
Moyal and Wick-Voros products is a generic feature of any translation invariant associative product.
To this end, we have introduced a general associative product
and then discussed its translational invariance properties.
We have found that the vertex is changed by an exponential which maintains invariance for cyclic permutation
of the external momenta, but not for any arbitrary exchange.
So, like in the Moyal and Wick-Voros cases, the planar and non-planar diagrams behave dif\mbox{}ferently.
In particular, the non-planar diagram present the same kind of ultraviolet/infrared mixing.
Moreover, we have showed that the phase appearing in the exponent
in the non-planar diagram is related to the commutator of the coordinates so that we can state that the mixing
is given by the Poisson structure of the underlying space.

Going back to the discussion about the Moyal product versus the Wick-Voros one,
the two products are not equivalent at f\mbox{}irst sight.
Indeed, we have found dif\mbox{}ferent Green's functions despite both the physical intuition
and the fact that the two star products are algebraically equivalent,
in the sense that they def\mbox{}ine exactly the same deformed algebra~\cite{Zachos,AlexanianPinzulStern}
and as such describe the same non-commutative geometry.

The element we have used in order to solve this puzzle is symmetries.
Indeed, the commutation relation \eqref{CNC} breaks the Poincar\'{e} symmetry,
but it can be easily reinstated at a deformed level, as a non-commutative and non-cocommutative Hopf algebra
as described in%
~\cite{ChaichianKulishNishijimaTureanu,Wess,AschieriBlohmannDimitrijevicMeyerSchuppWess},
since both products can be seen as coming from a Drinfeld twist~\cite{Drinfeld1,Drinfeld2}.
We have showed how the presence of a twist forces us to reconsider all of the steps
in a f\mbox{}ield theory which has to be built in a coherent twisted way. We have found that
there is equivalence between the Moyal and Wick-Voros f\mbox{}ield theories at the level of $S$-matrix
in agreement with our physical intuition, since Green's functions are not observable quantities, while $S$-matrix is.
Moreover, this equivalence is obtained only if a consistent procedure of twisting all products is applied.
Therefore, we have used the f\mbox{}ield theories built with the Moyal and Wick-Voros products to check each other and
to obtain an indication on the procedure to follow for non-commutative theories coming from a twist.

\appendix

\chapter{An elementary introduction to the Hopf algebras}\label{Hopf algebras}

\emph{In the following appendix we present a very elementary introduction to the theory of Hopf algebras.
We just collect some very essential def\mbox{}initions of Hopf algebras~\cite{Majid}
that we have used throughout the thesis, mainly in the last chapter.}

\section{Algebras and coalgebras}

We begin with the notion of algebra for completeness.
A complex algebra is a complex vector space $\mathcal{A}$
equipped with a linear map called multiplication
\begin{equation*}
m: a\otimes b\in\mathcal{A}\otimes\mathcal{A}\to
m(a\otimes b)=ab\in\mathcal{A}
\end{equation*}
associative namely which satisf\mbox{}ies the condition
\begin{equation}
m\circ(m\otimes\mathrm{id})=m\circ(\mathrm{id}\otimes m)
\end{equation}
or equivalently the condition
\begin{equation}
(ab)c=a(bc)
\end{equation}
for any $a,b,c\in\mathcal{A}$, where $\mathrm{id}$ denotes the identity map of $\mathcal{A}$.
An algebra $\mathcal{A}$ is commutative if
\begin{equation}
m\circ\tau=m
\end{equation}
or equivalently if
\begin{equation}
ab=ba
\end{equation}
for any $a,b\in\mathcal{A}$, where
$$\tau:\mathcal{A}\otimes\mathcal{A}\to\mathcal{A}\otimes\mathcal{A}$$
def\mbox{}ined by
\begin{equation}
\tau(a\otimes b)=b\otimes a
\end{equation}
for any $a,b\in\mathcal{A}$ is the exchange map. An algebra $\mathcal{A}$
is unitary if there exists a linear map called unit
$$\eta:\mathbbm{C}\to\mathcal{A}$$
which satisf\mbox{}ies the condition
\begin{equation}
m\circ(\eta\otimes\mathrm{id})=m\circ(\mathrm{id}\otimes\eta)=\mathrm{id}
\end{equation}
or equivalently the condition
\begin{equation}
\eta(1)a=a\eta(1)=a
\end{equation}
for any $a\in\mathcal{A}$. In what follows we set
\begin{equation}
\eta(1)=\mathbbm{1}.
\end{equation}
Finally, a homomorphisms between two algebras $\mathcal{A}$ and $\mathcal{A}'$ is a linear map
$$ f:\mathcal{A}\to\mathcal{A'}$$
such that
\begin{equation}
f\circ m=m'\circ(f\otimes f)
\end{equation}
or equivalently such that
\begin{equation}
f(ab)=f(a)f(b)
\end{equation}
for any $a,b\in\mathcal{A}$. Moreover, if $\mathcal{A}$ and $\mathcal{A'}$ are unitary algebras,
we assume that
\begin{equation}
f\circ\eta=\eta'
\end{equation}
or equivalently that
\begin{equation}
f(\mathbbm{1})=\mathbbm{1}'.
\end{equation}

A complex coalgebra is a complex vector space $\mathcal{A}$ equipped
with a linear map called comultiplication
$$\Delta:\mathcal{A}\to\mathcal{A}\otimes\mathcal{A}$$
coassociative namely which satisf\mbox{}ies the condition
\begin{equation}
(\Delta\otimes\mathrm{id})\circ\Delta=(\mathrm{id}\otimes\Delta)\circ\Delta
\end{equation}
or equivalently the condition
\begin{equation}
a_{(1)(1)}\otimes a_{(1)(2)}\otimes a_{(2)}= a_{(1)}\otimes
a_{(2)(1)}\otimes a_{(2)(2)}
\end{equation}
for any $a\in\mathcal{A}$, where we have used the generalized Sweedler's notation\footnote{For any
$a\in\mathcal{A}$, the Sweedler's notation consists of writing
\begin{equation*}
\Delta(a)=\sum_{i}a_{(1)}^i\otimes a_{(2)}^i
\end{equation*}
with $a_{(1)}^i,a_{(2)}^i\in\mathcal{A}$ or more simply
\begin{equation*}
\Delta(a)=\sum_{(a)}a_{(1)}\otimes a_{(2)}.
\end{equation*}}:
\begin{equation}
\Delta(a)=a_{(1)}\otimes a_{(2)}
\end{equation}
for any $a\in\mathcal{A}$. A coalgebra $\mathcal{A}$ is cocommutative if
\begin{equation}
\tau\circ\Delta=\Delta
\end{equation}
or equivalently if
\begin{equation}
a_{(1)}\otimes a_{(2)}=a_{(2)}\otimes a_{(1)}
\end{equation}
for any $a\in\mathcal{A}$. A coalgebra $\mathcal{A}$ is counitary if
there exists a linear map called counit
$$\varepsilon:\mathcal{A}\to\mathbb{C}$$
which satisf\mbox{}ies the condition
\begin{equation}
(\varepsilon\otimes\mathrm{id})\circ\Delta=(\mathrm{id}\otimes\varepsilon)\circ\Delta=\mathrm{id}
\end{equation}
or equivalently the condition
\begin{equation}
\varepsilon\left(a_{(1)}\right)a_{(2)}=
a_{(1)}\varepsilon\left(a_{(2)}\right)=a
\end{equation}
for any $a\in\mathcal{A}$. Finally, a homomorphism between two coalgebras $\mathcal{A}$ and $\mathcal{A'}$
is a linear map
$$f:\mathcal{A}\to\mathcal{A}'$$
such that
\begin{equation}
\Delta'\circ f=(f\otimes f)\circ\Delta
\end{equation}
or equivalently such that
\begin{equation}
\Delta'(f(a))=f\left(a_{(1)}\right)\otimes f\left(a_{(2)}\right)
\end{equation}
for any $a\in\mathcal{A}$. Moreover, if $\mathcal{A}$ and $\mathcal{A'}$ are counitary coalgebras, we assume as well
\begin{equation}
\varepsilon'\circ f=\varepsilon.
\end{equation}

\section{Bialgebras and Hopf algebras}

A complex bialgebra is a complex vector space $\mathcal{A}$
that is at the same time a complex unitary algebra and a complex counitary coalgebra
in a compatible way namely the multiplication, the comultiplication, the unit and the counit
satisfy the conditions
\begin{align}
\nonumber
\Delta\circ m&=
(m\otimes m)\circ(\mathrm{id}\otimes\tau\otimes\mathrm{id})\circ(\Delta\otimes\Delta)\\
\nonumber
\varepsilon\circ m&=\varepsilon\otimes\varepsilon\\
\nonumber
\Delta\circ\eta&=\eta\otimes\eta\\
\varepsilon\circ\eta&=\mathrm{id}_\mathbb{C}
\end{align}
where $\mathrm{id}_\mathbb{C}$ denotes the identity map of $\mathbb{C}$.
Equivalently the compatibility relations between the two structures can be written as\footnote{
$\Delta(a)\Delta(b)=a_{(1)}b_{(1)}\otimes a_{(2)}b_{(2)}$
for any $a,b\in\mathcal{A}$.}
\begin{align}
\nonumber
\Delta(ab)&=\Delta(a)\Delta(b)\\
\nonumber
\varepsilon(ab)&=\varepsilon(a)\varepsilon(b)\\
\nonumber
\Delta(\mathbbm{1})&=\mathbbm{1}\otimes\mathbbm{1}\\
\varepsilon(\mathbbm{1})&=1
\end{align}
for any $a,b\in\mathcal{A}$. A bialgebra $\mathcal{A}$ is
commutative if it is commutative like an algebra and it is
cocommutative if it is cocommutative like a coalgebra.
Finally, a homomorphism between two bialgebras $\mathcal{A}$ and $\mathcal{A}'$
is linear map
$$f:\mathcal{A}\to\mathcal{A}'$$
which is both an unitary algebra and counitary coalgebra homomorphism.

A complex Hopf algebra is a complex bialgebra $\mathcal{A}$
equipped with a linear map called antipode
$$S:\mathcal{A}\to\mathcal{A}$$
such that
\begin{equation}
m\circ(S\otimes\mathrm{id})\circ\Delta=m\circ(\mathrm{id}\otimes S)\circ\Delta=
\eta\circ\varepsilon
\end{equation}
or equivalently such that
\begin{equation}
S\left(a_{(1)}\right)a_{(2)}=a_{(1)}S\left(a_{(2)}\right)=
\varepsilon(a)\mathbbm{1}
\end{equation}
for any $a\in\mathcal{A}$. The role of the antipode is like that of an inverse.
However, we do not demand that $S^{2}=\mathrm{id}$. The antipode is unique and satisf\mbox{}ies the conditions
\begin{align}
\nonumber
S\circ m&=m\circ\tau\circ(S\otimes S)\\
\nonumber
\Delta\circ S&=(S\otimes S)\circ\tau\circ\Delta\\
\nonumber
S\circ\eta&=\eta\\
\varepsilon\circ S&=\varepsilon
\end{align}
or equivalently the conditions
\begin{align}
\nonumber
S(ab)&=S(b)S(a)\\
\nonumber
\Delta(S(a))&=S\left(a_{(2)}\right)\otimes S\left(a_{(1)}\right)\\
\nonumber
S(\mathbbm{1})&=\mathbbm{1}\\
\varepsilon(S(a))&=\varepsilon(a)
\end{align}
for any $a,b\in\mathcal{A}$. Therefore, the antipode is an unitary algebra and counitary coalgebra antihomomorphism.
Like for bialgebras, a Hopf algebra is commutative if it is commutative like an algebra
and it is cocommutative if it is cocommutative like a coalgebra.
Notice that if $\mathcal{A}$ is a commutative or cocommutative Hopf algebra, then
\begin{equation}
S^2=\mathrm{id}.
\end{equation}
Moreover, a bialgebra homomorphism between two Hopf algebras $\mathcal{A}$ e $\mathcal{A}'$
$$f:\mathcal{A}\to\mathcal{A}'$$
is automatically a Hopf algebra homomorphism i.e~it satisf\mbox{}ies the condition
\begin{equation}
f\circ S=S'\circ f.
\end{equation}
An example of Hopf algebra is  given by universal enveloping algebra $U(g)$
of a Lie algebra $g$, where the comultiplication is def\mbox{}ined by
\begin{equation}
\Delta(\xi)=\xi\otimes\mathbbm{1}+\mathbbm{1}\otimes\xi
\end{equation}
the counit is def\mbox{}ined by
\begin{equation}
\varepsilon(\xi)=0
\end{equation}
and the antipode is def\mbox{}ined by
\begin{equation}
S(\xi)=-\xi
\end{equation}
for any $\xi\in U(g)$. Furthermore, the comultiplication and the counit are extended
as unitary algebra homomorphisms, while the antipode is extended as a counitary coalgebra antihomomorphism.

\section{Cocycles and twists}

Let $\mathcal{A}$ be a Hopf algebra. Consider the maps
\begin{equation}
\Delta_{i}:\underbrace{\mathcal{A}\otimes\mathcal{A}\ldots\otimes\mathcal{A}}_{n-\mathrm{times}}\to
\underbrace{\mathcal{A}\otimes\mathcal{A}\ldots\otimes\mathcal{A}}_{(n+1)-\mathrm{times}}
\end{equation}
def\mbox{}ined by
\begin{equation}
\Delta_{i}=\mathrm{id}\otimes\mathrm{id}\ldots\otimes\mathrm{id}\otimes\Delta
\otimes\mathrm{id}\otimes\mathrm{id}\ldots\otimes\mathrm{id}
\end{equation}
with $\Delta$ is in the ith position and $i=1,2,\ldots,n$. Moreover, we def\mbox{}ine
\begin{equation}
\Delta_{0}=\mathbbm{1}\otimes()\quad\mathrm{and}\quad\Delta_{n+1}=()\otimes\mathbbm{1}.
\end{equation}
An $n$-cochain is an invertible element
\begin{equation}
\chi\in\underbrace{\mathcal{A}\otimes\mathcal{A}\ldots\otimes\mathcal{A}}_{n-\mathrm{times}}
\end{equation}
and its coboundary as the $(n+1)$-cochain
\begin{equation}
\partial\chi=\left(\prod_{i\,\mathrm{even}}\Delta_{i}\chi\right)
\left(\prod_{i\,\mathrm{odd}}\Delta_{i}\chi\right)
\end{equation}
and the products are each taken in increasing order.
An $n$-cochain $\chi$ is an $n$-cocycle if
\begin{equation}
\partial\chi=\mathbbm{1}
\end{equation}
and it is counitary if
\begin{equation}
\varepsilon_{i}\chi=\mathbbm{1}
\end{equation}
for all
\begin{equation}
\varepsilon_{i}=\mathrm{id}\otimes\mathrm{id}\ldots\otimes\mathrm{id}\otimes\varepsilon
\otimes\mathrm{id}\otimes\mathrm{id}\ldots\otimes\mathrm{id}
\end{equation}
with $\varepsilon$ is in the ith position.
For example, a $1$-cocycle is an invertible element $\chi\in\mathcal{A}$ such that
\begin{equation}
\chi\otimes\chi=\Delta\chi
\end{equation}
and it is automatically counitary. Instead,
a $2$-cocycle is an invertible element $\chi\in\mathcal{A}\otimes\mathcal{A}$ such that
\begin{equation}
(\mathbbm{1}\otimes\chi)(\mathrm{id}\otimes\Delta)\chi=(\chi\otimes\mathbbm{1})(\Delta\otimes\mathrm{id})\chi
\end{equation}
and it is counitary if
\begin{equation}
(\varepsilon\otimes\mathrm{id})\chi=(\mathrm{id}\otimes\varepsilon)\chi=\mathbbm{1}.
\end{equation}
A twist is a counitary $2$-cocycle which is usually denoted by $\mathcal{F}$
and with the generalized Sweedler's notation it can be written as
\begin{equation}
\mathcal{F}=\mathcal{F}_{(1)}\otimes\mathcal{F}_{(2)}.
\end{equation}

\section{Quasi-triangular Hopf algebras}

A Hopf algebra $\mathcal{A}$ is quasi-triangular if there exists an invertible element $\mathcal{R}\in\mathcal{A}\otimes\mathcal{A}$ called a quasi-triangular structure or universal $\mathcal{R}$-matrix
which satisf\mbox{}ies the following conditions:
\begin{align}
(\Delta\otimes\mathrm{id})\mathcal{R}&=\mathcal{R}_{13}\mathcal{R}_{23}\\
(\mathrm{id}\otimes\Delta)\mathcal{R}&=\mathcal{R}_{13}\mathcal{R}_{12}\\
\tau\circ\Delta(a)&=\mathcal{R}\Delta(a)\mathcal{R}^{-1}\quad\forall a\in\mathcal{A}
\end{align}
where with the generalized Sweedler's notation
\begin{equation}
\mathcal{R}=\mathcal{R}_{(1)}\otimes\mathcal{R}_{(2)}
\end{equation}
and
\begin{align}
\mathcal{R}_{12}&=\mathcal{R}_{(1)}\otimes\mathcal{R}_{(2)}\otimes\mathbbm{1}\\
\mathcal{R}_{13}&=\mathcal{R}_{(1)}\otimes\mathbbm{1}\otimes\mathcal{R}_{(2)}\\
\mathcal{R}_{23}&=\mathbbm{1}\otimes\mathcal{R}_{(1)}\otimes\mathcal{R}_{(2)}.
\end{align}
Note that it is possible to show that
\begin{equation}
(\varepsilon\otimes\mathrm{id})\mathcal{R}=(\mathrm{id}\otimes\varepsilon)\mathcal{R}=\mathbbm{1}.
\end{equation}
Moreover,
\begin{align}
(S\otimes\mathrm{id})\mathcal{R}&=\mathcal{R}^{-1}\\
(\mathrm{id}\otimes S)\mathcal{R}^{-1}&=\mathcal{R}
\end{align}
and hence
\begin{equation}
(S\otimes S)\mathcal{R}=\mathcal{R}.
\end{equation}
Finally, it is easy to see that $\mathcal{R}$ obeys the abstract Yang-Baxter equation:
\begin{equation}
\mathcal{R}_{12}\mathcal{R}_{13}\mathcal{R}_{23}=\mathcal{R}_{23}\mathcal{R}_{13}\mathcal{R}_{12}.
\end{equation}
To conclude this section, we recall that given a Hopf algebra $\mathcal{A}$,
we can get a new Hopf algebra by means of a twist $\mathcal{F}$ that is,
by twisting the initial Hopf algebra $\mathcal{A}$.
Indeed, it is not dif\mbox{}f\mbox{}icult to see that there is a new Hopf algebra $\mathcal{A}_{\mathcal{F}}$
with the same algebra structure and counit of $\mathcal{A}$
and the comultiplication, the antipode and the universal $\mathcal{R}$-matrix given respectively by
\begin{align}
\Delta_{\mathcal{F}}(a)&=\mathcal{F}\Delta(a)\mathcal{F}^{-1}\\
S_{\mathcal{F}}(a)&=US(a)U^{-1}\\
\mathcal{R}_{\mathcal{F}}&=\mathcal{F}_{21}\mathcal{R}\mathcal{F}^{-1}
\end{align}
for any $a\in\mathcal{A}_{\mathcal{F}}$ where $U$ is invertible and given by
\begin{equation}
U=\sum\mathcal{F}^{(1)}S\left(\mathcal{F}^{(2)}\right)
\end{equation}
and
\begin{equation}
\mathcal{F}_{21}=\mathcal{F}_{(2)}\otimes\mathcal{F}_{(1)}.
\end{equation}
Notice that if $\mathcal{A}$ is just a Hopf algebra, then so is $\mathcal{A}_{\mathcal{F}}.$


\begin{thebibliography}{99}
\addcontentsline{toc}{chapter}{Bibliography}






\bibitem{Connes}
A.~Connes,
\emph{Noncommutative Geometry},
Academic Press, (1994).

\bibitem{Madore}
J.~Madore,
\emph{An Introduction to Noncommutative Dif\mbox{}ferential Geometry and its Physical Applications},
Cambridge University Press (1995).

\bibitem{Landi}
G.~Landi,
\emph{An Introduction to Noncommutative Spaces and Their Geometry},
Springer (1997)
[arXiv:hep-th/9701.078].
%%CITATION = HEP-TH/9701078;%%

\bibitem{Gracia-BondiaVarillyFigueroa}
J.~M.~Gracia-Bond\'{\i}a, J.~C.~V\'{a}rilly and H.~Figueroa,
\emph{Elements of Noncommutative Geometry},
Birkhaeuser (2001)

\bibitem{Heisenberg}
Heisenberg,
\emph{Letter of Heisenberg to Peierls} (1930) in:
Wolfgang Pauli,
\emph{Scientif\mbox{}ic Correspondence},
Vol. II, 15, Ed. Karl von Meyenn, Springer-Verlag (1985).

\bibitem{Snyder}
H.~S.~Snyder,
\emph{Quantized space-time},
Phys.\ Rev.\ \textbf{71}, 38 (1947).
%%CITATION = PHRVA,71,38;%%

\bibitem{DoplicherFredenhagenRoberts}
S.~Doplicher, K.~Fredenhagen and J.~E.~Roberts,
\emph{The Quantum structure of space-time at the Planck scale and quantum f\mbox{}ields},
Commun.\ Math.\ Phys.\ \textbf{172}, 187 (1995)
[arXiv:hep-th/0303.037].
%%CITATION = CMPHA,172,187;%%

\bibitem{Doplicher}
S.~Doplicher,
\emph{Spacetime and f\mbox{}ields, a quantum texture},
[arXiv:hep-th/0105.251].
%%CITATION = HEP-TH/0105251;%%

\bibitem{SeibergWitten}
N.~Seiberg and E.~Witten,
\emph{String theory and noncommutative geometry},
JHEP \textbf{9909}, 032 (1999)
[arXiv:hep-th/9908.142].
%%CITATION = JHEPA,9909,032;%%

\bibitem{GrosseWulkenhaar}
H.~Grosse and R.~Wulkenhaar,
\emph{Renormalisation of $\varphi^{4}$ theory on noncommutative $\mathbb{R}^{4}$ in the matrix base},
Commun.\ Math.\ Phys.\ \textbf{256}, 305 (2005)
[arXiv:hep-th/0401.128].
%%CITATION = CMPHA,256,305;%%

\bibitem{GrosseSteinacker}
H.~Grosse and H.~Steinacker,
\emph{Exact renormalization of a noncommutative $\varphi^{3}$ model in 6 dimensions},
[arXiv:hep-th/0607.235].
%%CITATION = HEP-TH/0607235;%%

\bibitem{Rivasseau}
V.~Rivasseau,
\emph{Why Renormalizable NonCommutative Quantum Field Theories?},
[arXiv:math-ph/0711.1748].
%%CITATION = ARXIV:0711.1748;%%

\bibitem{Szabo}
R.~J.~Szabo,
\emph{Quantum f\mbox{}ield theory on noncommutative spaces},
Phys.\ Rept.\ \textbf{378}, 207 (2003)
[arXiv:hep-th/0109.162].
%%CITATION = PRPLC,378,207;%%

\bibitem{DouglasNekrasov}
M.~R.~Douglas and N.~A.~Nekrasov,
\emph{Noncommutative f\mbox{}ield theory},
Rev.\ Mod.\ Phys.\ \textbf{73}, 977 (2001)
[arXiv:hep-th/0106.048].
%%CITATION = RMPHA,73,977;%%

\bibitem{Gronewold}
H.~Gr\"onewold,
\emph{On principles of quantum mechanics}
Physica \textbf{12}, 405 (1946).
%%CITATION = PHYSA,12,405;%%

\bibitem{Moyal}
J.~E.~Moyal,
\emph{Quantum mechanics as a statistical theory},
Proc.\ Cambridge Phil.\ Soc.\ \textbf{45}, 99 (1949).
%%CITATION = PCPSA,45,99;%%

\bibitem{Bayen}
F. Bayen, in \emph{Group Theoretical Methods in Physics},
Lect.\ Notes Phys.\ \textbf{94}, 260 edited by E. Beiglbock et al.
(Springer, New York, 1979).

\bibitem{Voros}
A.~Voros,
\emph{Wentzel-Kramers-Brillouin method in the Bargmann representation},
Phys.\ Rev.\ A \textbf{40}, 6814 (1989).
%%CITATION = PHRVA,A40,6814;%%

\bibitem{BordemannWaldmann1}
M.~Bordemann and S.~Waldmann,
\emph{A Fedosov Star Product of Wick Type for K\"{a}hler Manifolds},
Lett.\ Math.\ Phys.\ \textbf{41}, 243 (1997)
[arXiv:q-alg/9605.012].
%%CITATION = Q-ALG/9605012;%%

\bibitem{BordemannWaldmann2}
M.~Bordemann and S.~Waldmann,
\emph{Formal GNS Construction and States in Deformation Quantization},
Comm.\ Math.\ Phys.\ \textbf{195}, 549 (1998)
[arXiv:q-alg/9607.019].
%%CITATION = Q-ALG/9607019;%%

\bibitem{LizziVitaleZampini1}
F.~Lizzi, P.~Vitale and A.~Zampini,
\emph{The fuzzy disc},
JHEP {\bf 0308}, 057 (2003)
[arXiv:hep-th/0306.247].
%%CITATION = JHEPA,0308,057;%%

\bibitem{LizziVitaleZampini2}
F.~Lizzi, P.~Vitale and A.~Zampini,
\emph{The beat of a fuzzy drum: fuzzy Bessel functions for the disc},
JHEP {\bf 0509}, 080 (2005)
[arXiv:hep-th/0506.008].
%%CITATION = JHEPA,0509,080;%%

\bibitem{MVRS}
S.~Minwalla, M.~Van Raamsdonk and N.~Seiberg,
\emph{Noncommutative perturbative dynamics},
JHEP \textbf{0002}, 020 (2000)
[arXiv:hep-th/9912.072].
%%CITATION = JHEPA,0002,020;%%

\bibitem{GalluccioLizziVitale1}
S.~Galluccio, F.~Lizzi and P.~Vitale,
\emph{Twisted Noncommutative Field Theory with the Wick-Voros and Moyal Products},
Phys.\ Rev.\ D \textbf{78}, 085007 (2008)
[arXiv:hep-th/0810.2095].
%%CITATION = PHRVA,D78,085007;%%

\bibitem{GalluccioLizziVitale2}
S.~Galluccio, F.~Lizzi and P.~Vitale,
\emph{Translation Invariance, Commutation Relations and Ultraviolet/Infrared Mixing},
JHEP \textbf{0909}, 054 (2009)
[arXiv:hep-th/0907.3640].
%%CITATION = JHEPA,0909,054;%%

\bibitem{HammouLagraaSheikh-Jabbari}
A.~B.~Hammou, M.~Lagraa and M.~M.~Sheikh-Jabbari,
\emph{Coherent state induced star-product on R(lambda)**3 and the fuzzy sphere},
Phys.\ Rev.\ D \textbf{66}, 025025 (2002)
[arXiv:hep-th/0110.291].
%%CITATION = PHRVA,D66,025025;%%

\bibitem{Drinfeld1}
V.~G.~Drinfeld,
\emph{On constant quasiclassical solutions of the Yang-Baxter equations},
Soviet Math.\ Dokl.\ \textbf{28}, 667-671 (1983).
%%CITATION = SVMDA,28,667;%%

\bibitem{Drinfeld2}
V.~G.~Drinfeld,
\emph{Quasi-Hopf Algebras},
Alg.\ Anal.\ \textbf{1N6}, 114 (1989).
%%CITATION = 00040,1N6,114;%%

\bibitem{Oeckl}
R.~Oeckl,
\emph{Untwisting noncommutative R**d and the equivalence of quantum f\mbox{}ield theories},
Nucl.\ Phys.\ B \textbf{581}, 559 (2000)
[arXiv:hep-th/0003.018].
%%CITATION = NUPHA,B581,559;%%

\bibitem{ChaichianKulishNishijimaTureanu}
M.~Chaichian, P.~P.~Kulish, K.~Nishijima and A.~Tureanu,
\emph{On a Lorentz-invariant interpretation of noncommutative space-time and
its implications on noncommutative quantum f\mbox{}ield theory},
Phys.\ Lett.\ B \textbf{604}, 98 (2004)
[arXiv:hep-th/0408.069].
%%CITATION = HEP-TH/0408069;%%

\bibitem{Wess}
J.~Wess,
\emph{Deformed coordinate spaces: Derivatives},
[arXiv:hep-th/0408.080].
%%CITATION = HEP-TH/0408080;%%

\bibitem{AschieriBlohmannDimitrijevicMeyerSchuppWess}
P.~Aschieri, C.~Blohmann, M.~Dimitrijevi$\mathrm{\acute{c}}$, F.~Meyer, P.~Schupp and J.~Wess,
\emph{A gravity theory on noncommutative spaces},
Class.\ Quant.\ Grav.\  \textbf{22}, 3511 (2005)
[arXiv:hep-th/0504.183].
%%CITATION = CQGRD,22,3511;%%






\bibitem{Weyl}
H.~Weyl,
\emph{The theory of groups and Quantum Mechanics},
Dover (1931).

\bibitem{Zampini}
A.~Zampini,
\emph{Applications of the Weyl-Wigner formalism to noncommutative geometry},
arXiv:hep-th/0505.271.
%%CITATION = HEP-TH/0505271;%%

\bibitem{EspositoMarmoSudarshan}
G.~Esposito, G.~Marmo and E.~C.~G.~Sudarshan,
\emph{From Classical to Quantum Mechanics},
Cambridge University Press (2004).

\bibitem{Pool}%%%%%%%%%%%%%%%%%%%%%%%%%%%%%%%%%%%%%%%%%% l'anno mi sembra errato
J.~C.~T.~Pool,
\emph{Mathematical Aspects of the Weyl Correspondence},
J.\ Math.\ Phys.\ \textbf{7} (1), 66-76 (1996).

\bibitem{Gracia-BondiaLizziMarmoVitale}
J.~M.~Gracia-Bond\'{\i}a, F.~Lizzi, G.~Marmo and P.~Vitale,
\emph{Inf\mbox{}initely many star products to play with},
JHEP \textbf{0204}, 026 (2002)
[arXiv:hep-th/0112.092].
%%CITATION = JHEPA,0204,026;%%

\bibitem{EstradaVarillyGracia-Bondia}
R.~Estrada, J.~M.~Gracia-Bond\'{\i}a and J.~C.~V\'{a}rilly,
\emph{On Asymptotic expansions of twisted products},
J.\ Math.\ Phys.\ \textbf{30}, 2789 (1989).
%%CITATION = JMAPA,30,2789;%%

\bibitem{Daoud}
M.~Daoud,
\emph{Extended Voros product in the coherent states framework},
Phys.\ Lett.\ A \textbf{309} 167 (2003).
%%CITATION = PHLTA,A309,167;%%






\bibitem{CahillGlauber1}
K.~E.~Cahill and R.~L.~Glauber,
\emph{Ordered Expansions in Boson Amplitude Operators},
Phys.\ Rev.\ \textbf{177}, 5 1857-1881 (1969)
%%CITATION = PHRVA,177,1857;%%

\bibitem{MankoMankoMarmo}
O.~V.~Man'ko, V.~I.~Man'ko and G.~Marmo,
\emph{Alternative commutation relations, star products and tomography},
J.\ Phys.\ A  \textbf{35}, 699-719 (2002).
%%CITATION = JPAGB,A35,699;%%

\bibitem{MankoMarmoVitale}
V.~I.~Man'ko, G.~Marmo, and P.~Vitale,
\emph{Duality symmetry for star products},
Phys.\ Lett.\  A \textbf{334}, 1 (2005)
[arXiv:hep-th/0407.131].
%%CITATION = PHLTA,A334,1;%%

\bibitem{MankoMankoMarmoVitale}
O.~V.~Man'ko, V.~I.~Man'ko, G.~Marmo and P.~Vitale
\emph{Star products, duality and double Lie algebras},
Phys.\ Lett.\  A \textbf{360}, 522 (2007)
[arXiv:quant-th/0609.041].
%%CITATION = PHLTA,A360,522;%%






\bibitem{Filk}
T.~Filk,
\emph{Divergencies in a f\mbox{}ield theory on quantum space},
Phys.\ Lett.\ B \textbf{376}, 53 (1996).
%%CITATION = PHLTA,B376,53;%%

\bibitem{Kaku}
M.~Kaku,
\emph{Quantum f\mbox{}ield theory: A Modern introduction}
Oxford Univ. Press (1993).

\bibitem{Gracia-BondiaVarilly}
J.~M.~Gracia-Bond\'{\i}a and J.~C.~V\'{a}rilly,
\emph{On the ultraviolet behaviour of quantum f\mbox{}ields over noncommutative manifolds},
Int.\ J.\ Mod.\ Phys.\  A \textbf{14}, 1305 (1999)
[arXiv:hep-th/9804.001].
%%CITATION = IMPAE,A14,1305;%%

\bibitem{MartinGracia-BondiaVarilly}
C.~P.~Martin, J.~M.~Gracia-Bond\'{\i}a and J.~C.~V\'{a}rilly,
\emph{The Standard model as a noncommutative geometry: The Low-energy regime},
Phys.\ Rept.\  {\bf 294} (1998) 363
[arXiv:hep-th/9605.001].
%%CITATION = PRPLC,294,363;%%


\bibitem{ChaichianDemichevPresnajder}
M.~Chaichian, A.~Demichev and P.~Presnajder,
\emph{Quantum f\mbox{}ield theory on noncommutative space-times and the persistence of ultraviolet divergences},
Nucl.\ Phys.\ B \textbf{567}, 360 (2000)
[arXiv:hep-th/9812.180].
%%CITATION = NUPHA,B567,360;%%






\bibitem{Zachos}
C.~K.~Zachos,
\emph{Geometrical evaluation of star products},
J.\ Math.\ Phys.\ \textbf{41}, 5129 (2000)
[arXiv:hep-th/9912.238].
%%CITATION = JMAPA,41,5129;%%

\bibitem{AlexanianPinzulStern}
G.~Alexanian, A.~Pinzul and A.~Stern,
\emph{Generalized Coherent State Approach to Star Products and Applications to the Fuzzy Sphere},
Nucl.\ Phys.\ B \textbf{600}, 531 (2001)
[arXiv:hep-th/0010.187].
%%CITATION = NUPHA,B600,531;%%






\bibitem{BFFLS1}
F.~Bayen, M.~Flato, C.~Fronsdal, A.~Lichnerowicz and D.~Sternheimer,
\emph{Deformation Theory And Quantization 1. Deformations Of Symplectic Structures},
Annals Phys.\ \textbf{111}, 61 (1978).
%%CITATION = APNYA,111,61;%%

\bibitem{BFFLS2}
F.~Bayen, M.~Flato, C.~Fronsdal, A.~Lichnerowicz and D.~Sternheimer,
\emph{Deformation Theory And Quantization 2. Physical Applications},
Annals Phys.\ \textbf{111}, 111 (1978).
%%CITATION = APNYA,111,111;%%

\bibitem{Landsman}
N.~P.~Landsman,
\emph{Lecture Notes on $C^*$-algebras, Hilbert $C^*$-modules and Quantum Mechanics},
[arXiv:math-ph/9807.030].
%%CITATION = MATH-PH/9807030;%%

\bibitem{DitoSternheimer}
G.~Dito and D.~Sternheimer,
\emph{Deformation quantization: genesis, developments and metamorphoses},
[arXiv:math/0201.168].
%%CITATION = MATH/0201168;%%

\bibitem{Sternheimer}
D.~Sternheimer,
\emph{Deformation Quantization: Twenty Years After},
AIP Conf.\ Proc.\ \textbf{453}, 107 (1998)
[arXiv:math/9809.056].
%%CITATION = APCPC,453,107;%%

\bibitem{Kontsevich}
M.~Kontsevich,
\emph{Deformation quantization of Poisson manifolds, I},
Lett.\ Math.\ Phys.\ \textbf{66}, 157 (2003)
[arXiv:q-alg/9709.040].
%%CITATION = LMPHD,66,157;%%

\bibitem{TanasaVitale}
A.~Tanasa and P.~Vitale,
\emph{Curing the UV/IR mixing for field theories with translation-invariant $\star$ products},
Phys.\ Rev.\  D {\bf 81} (2010) 065008
[arXiv:hep-th/0912.0200].
%%CITATION = PHRVA,D81,065008;%%






\bibitem{Vassilevich}
D.~V.~Vassilevich,
\emph{Twist to close},
Mod.\ Phys.\ Lett.\ A \textbf{21}, 1279 (2006)
[arXiv:hep-th/0602.185].
%%CITATION = MPLAE,A21,1279;%%

\bibitem{BalachandranManganoPinzulVaidya}
A.~P.~Balachandran, G.~Mangano, A.~Pinzul and S.~Vaidya,
\emph{Spin and statistics on the Gronewold-Moyal plane: Pauli-forbidden levels and transitions},
Int.\ J.\ Mod.\ Phys.\ A \textbf{21}, 3111 (2006)
[arXiv:hep-th/0508.002].
%%CITATION = IMPAE,A21,3111;%%

\bibitem{BalachandranPinzulQureshi}
A.~P.~Balachandran, A.~Pinzul and B.~A.~Qureshi,
\emph{UV-IR mixing in non-commutative plane},
Phys.\ Lett.\ B \textbf{634}, 434 (2006)
[arXiv:hep-th/0508.151].
%%CITATION = PHLTA,B634,434;%%

\bibitem{Tureanu}
A.~Tureanu,
\emph{Twist and spin-statistics relation in noncommutative quantum f\mbox{}ield theory},
Phys.\ Lett.\ B \textbf{638}, 296 (2006)
[arXiv:hep-th/0603.219].
%%CITATION = PHLTA,B638,296;%%

\bibitem{ChaichianTureanu}
M.~Chaichian and A.~Tureanu,
\emph{Twist symmetry and gauge invariance},
Phys.\ Lett.\ B \textbf{637}, 199 (2006)
[arXiv:hep-th/0604.025].
%%CITATION = PHLTA,B637,199;%%

\bibitem{BuKimLeeVacYee}
J.~G.~Bu, H.~C.~Kim, Y.~Lee, C.~H.~Vac and J.~H.~Yee,
\emph{Noncommutative f\mbox{}ield theory from twisted Fock space},
Phys.\ Rev.\ D \textbf{73}, 125001 (2006)
[arXiv:hep-th/0603.251].
%%CITATION = PHRVA,D73,125001;%%

\bibitem{Zhan}
J.~Zahn,
\emph{Remarks on twisted noncommutative quantum f\mbox{}ield theory},
Phys.\ Rev.\  D \textbf{73}, 105005 (2006)
[arXiv:hep-th/0603.231].
%%CITATION = PHRVA,D73,105005;%%

\bibitem{FioreWess}
G.~Fiore and J.~Wess,
\emph{On full twisted Poincar\'{e} symmetry and QFT on Moyal-Weyl spaces},
Phys.\ Rev.\ D \textbf{75}, 105022 (2007)
[arXiv:hep-th/0701.078].
%%CITATION = PHRVA,D75,105022;%%

\bibitem{Fiore}
G.~Fiore,
\emph{Can QFT on Moyal-Weyl spaces look as on commutative ones?},
Prog.\ Theor.\ Phys.\ Suppl.\  {\bf 171}, 54 (2007)
[arXiv:hep-th/0705.1120].
%%CITATION = PTPSA,171,54;%%

\bibitem{ChaichianNishijimaSalminenTureanu}
M.~Chaichian, K.~Nishijima, T.~Salminen and A.~Tureanu,
\emph{Noncommutative Quantum Field Theory: A Confrontation of Symmetries},
JHEP \textbf{0806}, 078 (2008)
[arXiv:hep-th/0805.3500].
%%CITATION = JHEPA,0806,078;%%

\bibitem{AschieriDimitrijevicKulishLizziWess}
P.~Aschieri, M.~Dimitrijevi$\mathrm{\acute{c}}$, P.~Kulish F.~Lizzi and J.~Wess,
\emph{Noncommutative Spacetimes: Symmetries in Noncommutative Geometry and Field Theory},
Lect. Notes Phys. 774 (Springer, Berlin Heidelberg, 2009)

\bibitem{AschieriDimitrijevicMeyerWess}
P.~Aschieri, M.~Dimitrijevi$\mathrm{\acute{c}}$, F.~Meyer and J.~Wess,
\emph{Noncommutative geometry and gravity},
Class.\ Quant.\ Grav.\ \textbf{23}, 1883 (2006)
[arXiv:hep-th/0510.059].
%%CITATION = CQGRD,23,1883;%%

\bibitem{Aschieri}
P.~Aschieri,
\emph{Noncommutative symmetries and gravity},
J.\ Phys.\ Conf.\ Ser.\ \textbf{53}, 799 (2006)
[arXiv:hep-th/0608.172].
%%CITATION = 00462,53,799;%%

\bibitem{AschieriLizziVitale}
P.~Aschieri, F.~Lizzi and P.~Vitale,
\emph{Twisting all the way: from Classical Mechanics to Quantum Fields},
Phys.\ Rev.\ D \textbf{77}, 025037 (2008)
[arXiv:hep-th-0708.3002].
%%CITATION = PHRVA,D77,025037;%%

\bibitem{Matlock}
P.~Matlock,
\emph{Non-Commutative Geometry and Twisted Conformal Symmetry}
Phys.\ Rev.\  D \textbf{71}, 126007 (2005)
[arXiv:hep-th/0504.084].
%%CITATION = PHRVA,D71,126007;%%

\bibitem{BahnsDoplicherFredenhagenPiacitelli}
D.~Bahns, S.~Doplicher, K.~Fredenhagen and G.~Piacitelli,
\emph{On the unitarity problem in space/time noncommutative theories},
Phys.\ Lett.\ B \textbf{533}, 178 (2002)
[arXiv:hep-th/0201.222].
%%CITATION = PHLTA,B533,178;%%

\bibitem{Gracia-BondiaLizziRuizRuizVitale}
J.~M.~Gracia-Bond\'{\i}a, F.~Lizzi, F.~Ruiz Ruiz and P.~Vitale,
\emph{Noncommutative spacetime symmetries: Twist versus covariance},
Phys.\ Rev.\ D \textbf{74}, 025014 (2006)
[Erratum-ibid.\ D \textbf{74}, 029901 (2006)]
[arXiv:hep-th/0604.206].
%%CITATION = PHRVA,D74,025014;%%

\bibitem{BalachandranJosephPadmanabhan}
A.~P.~Balachandran, A.~Joseph and P.~Padmanabhan,
\emph{Causality and statistics on the Groenewold-Moyal plane},
arXiv:hep-th/0905.0876.
%%CITATION = ARXIV:0905.0876;%%

\bibitem{FioreSchupp1}
G.~Fiore and P.~Schupp,
\emph{Statistics and Quantum Group Symmetries},
[arXiv:hep-th/9605.133].
%%CITATION = HEP-TH/9605133;%%

\bibitem{FioreSchupp2}
G.~Fiore and P.~Schupp,
\emph{Identical particles and quantum symmetries},
Nucl.\ Phys.\ B \textbf{470}, 211 (1996)
[arXiv:hep-th/9508.047].
%%CITATION = NUPHA,B470,211;%%






\bibitem{Majid}
S.~Majid,
\emph{Foundations of Quantum Group Theory},
Cambridge University Press (2000).






\end{thebibliography}
\end{document}